\documentclass[journal, a4paper, onecolumn]{IEEEtran}
\usepackage{comment}
\usepackage{silence}
\usepackage{url}
\usepackage{amsmath}
\usepackage{mathtools}
\usepackage[noend]{algorithmic}
\usepackage{algorithm} 
\usepackage{enumerate}
\usepackage{tabularx}
\usepackage{rotating}
\usepackage[center]{caption}
\usepackage{subcaption}
\usepackage{tabularray}

\usepackage{multirow}
\usepackage{enumitem}
\usepackage{xcolor}
\usepackage{orcidlink}
\usepackage[inkscapeformat=png]{svg}
\usepackage{makecell}
\usepackage{todonotes}
\usepackage{amssymb}
\usepackage{booktabs}
\usepackage{amsthm}
\usepackage{textcomp}
\usepackage{pgfplots}
\usepgfplotslibrary{fillbetween}
\usepackage{pgfplotstable}
\pgfplotsset{compat=newest}
\usepackage{siunitx}
\usepackage{amsfonts}
\usepgfplotslibrary{groupplots}
\usepackage{graphicx}
\usepackage{float}
\usepackage{adjustbox}
\graphicspath{{results_100_uniform/heatmaps/}}

\definecolor{PURPLE}{RGB}{154, 28, 237}
\definecolor{BLUE}{RGB}{28, 105, 237}
\definecolor{GREEN}{RGB}{29, 176, 12}
\definecolor{ORANGE}{RGB}{242, 125, 15}
\definecolor{RED}{RGB}{230, 63, 53}

\pgfplotsset{
    ieee/.style={
        width=\columnwidth,
        height=5cm,
        xlabel style={font=\footnotesize},
        ylabel style={font=\footnotesize},
        title style={font=\footnotesize\bfseries},
        legend style={
            font=\scriptsize,
            at={(0.03,0.97)},
            anchor=north west,
            draw=none,
            fill=none
        },
        tick label style={font=\scriptsize},
        grid=major,
        grid style={line width=.1pt, draw=gray!30},
        major grid style={line width=.2pt,draw=gray!50},
        every axis plot/.append style={thick, smooth},
        xmin=0, xmax=100,
        ymin=0, ymax=100,
        tick style={draw=none},
        axis lines=left,
        axis line style={draw=none},
        tick align=outside,
        enlarge x limits=0.02,
        enlarge y limits=0.05,
        xtick={0,20,40,60,80,100},
        ytick={0,20,40,60,80,100}
    },
    ieee-wide/.style={
        width=0.8\textwidth,
        height=5cm,
        xlabel style={font=\small},
        ylabel style={font=\small},
        title style={font=\small\bfseries},
        legend style={font=\footnotesize},
        tick label style={font=\footnotesize},
        grid=major,
        grid style={line width=.1pt, draw=gray!30},
        major grid style={line width=.2pt,draw=gray!50},
        every axis plot/.append style={thick},
    }
}

\usepackage{times}
\usepackage{helvet}
\usepackage{courier}
\usepackage{graphicx}
\usepackage[numbers]{natbib}

\usepackage{arxiv}
\usepackage{amsmath}
\usepackage{graphicx}
\usepackage{xcolor}
\usepackage{amssymb}
\usepackage{yfonts}
\usepackage{hyperref}
\usepackage{textalpha}

\title{Altruistic Ride Sharing: A Framework for Fair and Sustainable Urban Mobility via Peer-to-Peer Incentives}

\author{
  Divyanshu Singh\\
  Dept. of CSIS and APPCAIR, BITS Pilani K K Birla Goa Campus, Goa, India\\
  \texttt{f20221129@goa.bits-pilani.ac.in}
\And
Ashman Mehra\\
School of Computer Science, Carnegie Mellon University, Pittsburgh, PA, USA\\
\texttt{ashmanm@andrew.cmu.edu}
\And
Kavya Makwana\\
  Dept. of CSIS and APPCAIR, BITS Pilani K K Birla Goa Campus, Goa, India\\
  \texttt{f20240634@goa.bits-pilani.ac.in}
\And
Snehanshu Saha\\
Dept. of CSIS and APPCAIR, BITS Pilani K K Birla Goa Campus, Goa, India\\
\texttt{snehanshus@goa.bits-pilani.ac.in}
\And
Santonu Sarkar\\
Dept. of CSIS and APPCAIR,  BITS Pilani K K Birla Goa Campus, Goa, India\\
\texttt{santonus@goa.bits-pilani.ac.in}
}

\begin{document}

\maketitle

\begin{abstract}
Urban mobility systems face persistent challenges of congestion,
underutilized vehicles, and rising emissions driven by private
point-to-point commuting. Although ride-sharing platforms exist,
their profit-driven incentive structures often fail to align
individual participation with broader community benefit.
We introduce \textit{Altruistic Ride Sharing} (ARS), a decentralized
peer-to-peer mobility framework in which commuters alternate between
driver and rider roles using \textit{altruism points}, a non-monetary
credit mechanism that rewards providing rides and discourages
persistent free-riding. To enable scalable coordination among agents,
ARS formulates ride-sharing as a multi-agent reinforcement learning
problem and introduces \textit{ORACLE} (One-Network Actor–Critic for
Learning in Cooperative Environments), a shared-parameter learning
architecture for decentralized rider selection.
We evaluate ARS using real-world New York City Taxi and Limousine
Commission (TLC) trajectory data under varying agent populations and
behavioral dynamics. Across simulations, ARS reduces total travel
distance and associated carbon emissions by approximately 20\%,
reduces urban traffic density by up to 30\%, and doubles vehicle
utilization relative to no-sharing baselines while maintaining
balanced participation across agents.
These results demonstrate that altruism-based incentives combined
with decentralized learning can provide a scalable and equitable
alternative to profit-driven ride-sharing systems.
\end{abstract}

\section{Introduction}
Urban mobility systems are increasingly strained as rapid urban population growth places growing pressure on road infrastructure, with point-to-point commuting remaining the dominant travel pattern. While conventional ride-sharing platforms such as UberPOOL and Lyft Line attempt to reduce inefficiencies through monetary incentives, their optimization objectives are primarily profit-driven rather than community-oriented. Carpooling adoption remains limited despite policy interventions, such as high-occupancy vehicle (HOV) lanes~\cite{shoham_multiagent_2008}. The primary reason for this is the lack of coordinated, community-driven solutions.
A growing body of work has investigated ride-sharing driven by social incentives rather than financial ones. Ma and Hanrahan~\cite{ma_unpacking_2020} analyzed peer-to-peer communities where shared needs foster more cooperative behavior. However, these efforts remained largely exploratory and lacked systematic integration with learning-based decision-making.
Recent work has begun incorporating fairness considerations into ride-sharing systems. Vlachogiannis et al.~\cite{vlachogiannis2023humanlightincentivizingridesharinghumancentric} explored the design of monetary incentives using deep reinforcement learning for ridesharing, while Zhang et al. ~\cite{10529943} focused on fairness with respect to race and income in travel demand forecasting. Zhou~\cite{Zhou2025RideSharingRec} employed Graph Attention Networks with Opinion Dynamics (OD-GAT) to improve prediction and reasoning of social relationships, enhancing service quality and ride-sharing safety.
Zhou and Roncoli~\cite{10241664} introduce a joint pricing and matching framework that uses a fairness-aware discount function to maximize platform profit while ensuring equitable fares for passengers.
Despite these advances, existing approaches largely remain focused on monetary incentives or predictive fairness, leaving the potential of community-driven cooperative ride-sharing largely unexplored. 
This paper introduces Altruistic Ride-Sharing (ARS), a decentralized, peer-to-peer mobility framework in which participants voluntarily alternate between being drivers (givers) and riders (takers). The key contributions of our work are summarized as follows:
\begin{itemize}[leftmargin=*]

\item {\em Novel Concept:} The ARS model operates using the concept of
\textit{altruism points} as a mechanism for incentivizing cooperation
and balanced participation instead of traditional monetary
transactions. ARS is designed to minimize total travel distance,
detours, and wait times while promoting fairness, sustainability, and
system stability.

\item {\em Role-Switching:} We introduce a probabilistic,
altruism-driven role assignment model where agents dynamically switch
between driver and rider roles, maintaining fairness and balance in
the system (\autoref{subsec:daily_role_assign}).

\item {\em Population Modeling:} We design a biologically inspired
birth–death–dropout process to simulate realistic participation,
dropout, and re-entry behaviors. This improves system resilience and
prevents persistent dominance of either drivers or riders.

\item {\em Reinforcement Learning Integration:} We develop a
multi-agent reinforcement learning framework based on the proposed
ORACLE (One-Network Actor--Critic for Learning in Cooperative
Environments) architecture. ORACLE employs parameter sharing and an
attention-based critic to enable scalable decentralized ride-selection
policies in dynamic urban environments (\autoref{subsec:algo}).

\item {\em Continual Adaptation:} The ORACLE policy can operate as a
human-in-the-loop recommendation system where drivers may accept,
reject, or override suggested actions. These interactions generate
off-policy experience that is incorporated into prioritized replay,
allowing continual adaptation of the policy to evolving real-world
behavior.

\item {\em Model Stability:} The system dynamics implicitly promote
fairness and long-term stability (\autoref{metric:fairness}) in role
distribution and altruism dynamics \cite{shoham_multiagent_2008}.
Stability here refers to the system's ability to maintain a balanced
driver–rider population while reintegrating agents who temporarily
leave the platform and allowing them to rebuild their altruism scores.

\item {\em Evaluation Framework:} Using real-world data from the New
York City Taxi and Limousine Commission (TLC), we evaluate ARS across
multiple metrics capturing efficiency, service quality, and fairness.
The framework is benchmarked against several baselines including
learning-based methods, optimization-based solvers, and a no-sharing
scenario, demonstrating improvements in sustainability and cooperative
participation.

\end{itemize}

The paper has been organized as follows. We present the relevant work in this area in Section~\ref{sec:relwork}. Section~\ref{sec:arsintro} introduces the ARS framework, detailing its core components including the altruism score mechanism, the multi-agent reinforcement learning model, and the population. The simulation design, including the experimental setup, performance metrics, and baseline models, is described in Section~\ref{sec:simulation}. We present and analyze the results of our experiments in Section~\ref{sec:results}. Finally, Section~\ref{sec:conclusion} concludes the paper and discusses future research directions.

\begin{figure}[ht]
    \centering
    \begin{minipage}[t]{\linewidth}
        \centering
        \includegraphics[width=\linewidth]{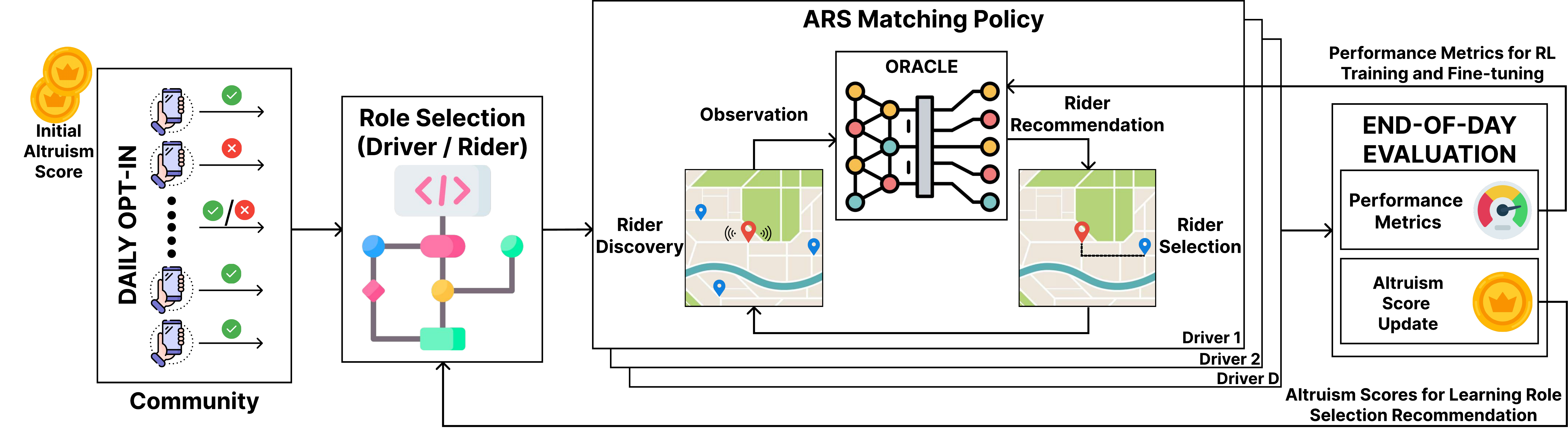}
    \end{minipage}
    \caption{Overview of the Altruistic Ride-Sharing (ARS) framework.
            Each day, community members join the system and receive role
            recommendations based on their altruism scores. Drivers use the
            ARS policy to select riders in real time using local observations.
            At the end of the day, system performance is evaluated and user
            scores are updated, creating a feedback loop that influences
            future role recommendations and policy learning.}
    \label{fig:delivery-sharing}
\end{figure}

\section{The Altruistic Ride Sharing Model}
\label{sec:arsintro}
The Altruistic Ride-Sharing (ARS) framework models short-distance urban
mobility as a decentralized peer-to-peer system in which participants
dynamically alternate between two roles: \emph{driver (giver)} and
\emph{rider (taker)}. Instead of financial transactions, participants
exchange a virtual social currency called the \emph{altruism score},
which reflects their contribution to the community.

Unlike conventional ride-hailing platforms where services can be
repeatedly consumed through monetary payments, ARS operates as a
closed-loop cooperative system. Altruism points must be earned by
providing rides before they can be spent to request rides, encouraging
reciprocal participation and discouraging persistent free-riding.
This mechanism enables decentralized coordination while maintaining a
balanced driver–rider population over time.

\subsection{Core Concepts}
\label{subsec:terminology}

\begin{enumerate}[leftmargin=*, label=\textbf{\arabic*.}]

\item \textbf{Agent.}
An \emph{agent} represents a commuter in the system. Let $A$ denote the
set of agents and $t \in \{1,\dots,T\}$ denote a simulation day. On day
$t$, agents may act as drivers $D_t \subseteq A$, riders $R_t \subseteq A$,
or remain inactive.

\item \textbf{Altruism Score.}
Each agent $a_i$ maintains an altruism score $s_i^t \in [0,1]$ that
reflects their contribution to the community. Altruism increases when
providing rides and decreases when receiving rides.

\item \textbf{Popularticipation Dynamics.}
Agents may dynamically join (\emph{birth}), temporarily leave
(\emph{dropout}), or permanently exit (\emph{death}) the system
(\autoref{subsec:popdynamics}).

\item \textbf{Detour Cost.}
Detour cost measures the additional distance a driver travels to
accommodate riders relative to their original direct route.

\end{enumerate}

A comprehensive notation reference is provided in the supplementary material.
Key variables are defined inline as they are introduced.

\subsection{Optimization Objectives}
\label{sec:optimization}

The ARS system aims to balance individual travel efficiency with
community-level benefits. We therefore define three objectives:
minimizing driver detours, maximizing rider service coverage, and
encouraging cooperative participation through altruism incentives.

Let $d_i(\cdot)$ denote the trip distance for agent $a_i$, where
$d_i(\emptyset)$ is the direct solo trip distance without ride-sharing,
and $d_i(\phi_i^t)$ is the total distance traveled by driver $a_i$ on
day $t$ when serving a sequence of riders $\phi_i^t$.

The objectives are defined as follows:

\begin{align}
\text{Minimize} ~\mathbb{O}_1 &: \quad
\sum_{t=1}^{T} \sum_{i=1}^{|D_t|}
\bigl[d_i(\phi_i^t) - d_i(\emptyset)\bigr]
\label{eq:o1} \\
\text{Maximize} ~(\mathbb{O}_2,\mathbb{O}_3) &: \quad
\sum_{t=1}^{T} \sum_{a_i \in R_t}
\mathbb{I}(a_i \text{ gets ride on day } t)\, d_i(\emptyset)
\;,\;
\quad\sum_{i=1}^{|A|} s_i^t
\label{eq:o23}
\end{align}

Objective $\mathbb{O}_1$ captures the additional distance incurred by
drivers when serving riders, encouraging efficient ride-sharing routes.
Objective $\mathbb{O}_2$ rewards serving riders with longer trips,
reflecting the greater system-wide benefit obtained when a shared ride
replaces a longer solo trip. Objective $\mathbb{O}_3$ promotes sustained
cooperative participation by maximizing the cumulative altruism scores
across the agent population.

Rather than solving a strict multi-objective optimization problem,
ARS approximates these objectives through the reward formulation used
by the reinforcement learning policy (\autoref{subsec:algo}). The reward
acts as a weighted scalarization of driver detour cost, rider trip
benefit, and altruism incentives, guiding decentralized decision-making
while implicitly balancing the objectives above.

The system operates under two practical constraints. First, drivers
have a tolerance limit on detours to ensure ride-sharing remains
feasible, restricting detours to at most $50\%$ of the direct route.
Second, each vehicle has a fixed seating capacity that limits the
number of riders served per trip.

These objectives guide the decision-making of drivers within the ARS
framework, while role assignment and altruism dynamics regulate
participation across the agent population.

\subsection{The Altruism Score (Virtual Currency)}
\label{subsec:altruismscore}

The altruism score is a non-monetary incentive mechanism that regulates
participation in the ARS community. Agents earn altruism by providing
rides as drivers and spend it when requesting rides. Scores are bounded
to the interval $[0,1]$ to prevent long-term hoarding and encourage
continued participation in both roles.

When a driver $a_i$ provides a ride to rider $a_j$ on day $t$, the
driver's altruism increases according to
\begin{align}
\Delta s_{i}^{t+1} =
\alpha_s s_j^t
\left(
1 - \frac{d_i(\phi_i^t \parallel a_j) - d_i(\phi_i^t)}{\text{max\_detour}}
\right)
\label{eq:driver_altruism}
\end{align}

where $d_i(\phi_i^t \parallel a_j)$ denotes the route after adding rider
$a_j$ and $\text{max\_detour}$ represents the maximum allowable detour.
Larger detours therefore yield smaller altruism gains.

The rider’s altruism decreases proportionally to the benefit received,
after which scores are clamped to remain within valid bounds:
\begin{align}
\Delta s_{j}^{t+1} &= -\beta_s \cdot \Delta s_{i}^{t+1},
\qquad
s_i^{t+1} = \max\left(0,\min\left(1, s_i^t + \Delta s_i^{t+1}\right)\right)
\end{align}

This mechanism ensures that agents who repeatedly request rides must
eventually contribute as drivers to replenish their altruism.

\subsection{Daily Role Assignment (Probabilistic Switching)}
\label{subsec:daily_role_assign}

At the beginning of each simulation day, agents are assigned roles
based on their current altruism score. This mechanism promotes
reciprocal participation while preventing persistent free-riding.

Agents with low altruism ($s_i^t \leq 0.2$) are required to act as
drivers, ensuring that agents who have benefited from the system
contribute back to the community. Agents with higher altruism
($s_i^t > 0.2$) select roles probabilistically according to their
normalized altruism scores. A small exploration probability allows
occasional random role selection to maintain behavioral diversity.

\begin{figure}[t]
    \centering
    \begin{minipage}{0.48\textwidth}
        \centering
        \includegraphics[width=\linewidth]{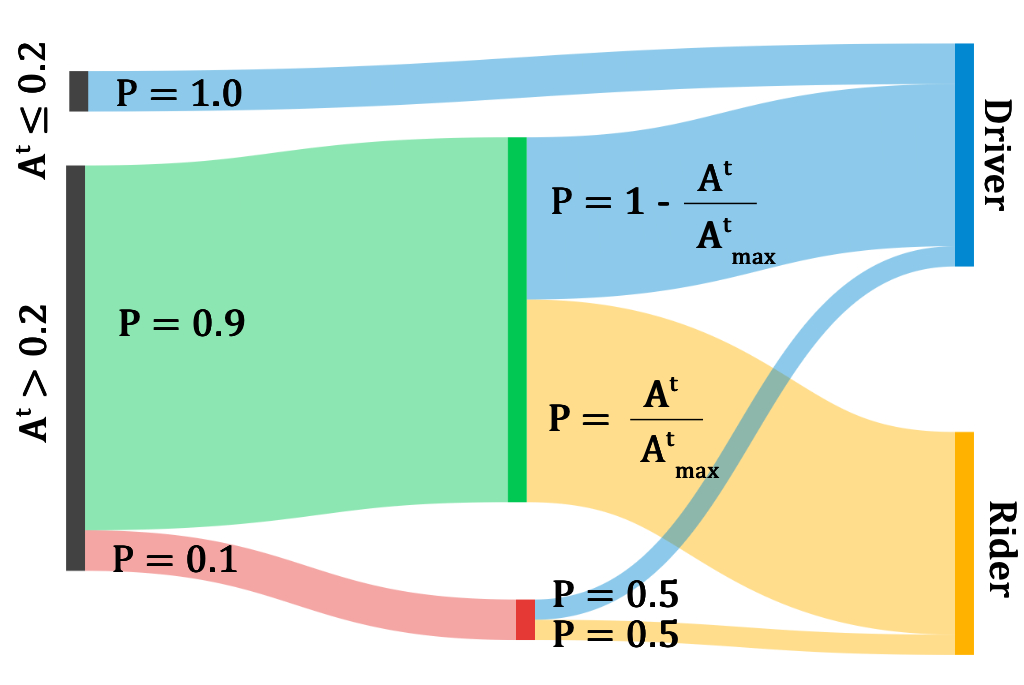}
        \caption{Sankey diagram illustrating the probabilistic role assignment
        mechanism. Agents with low altruism ($\leq 0.2$) must act as drivers,
        while agents with higher altruism probabilistically choose between
        driver and rider roles based on their normalized scores.}
        \label{fig:role_assignment_fig}
    \end{minipage}
    \hfill
    \begin{minipage}{0.43\textwidth}
        \centering
        \includegraphics[width=\linewidth]{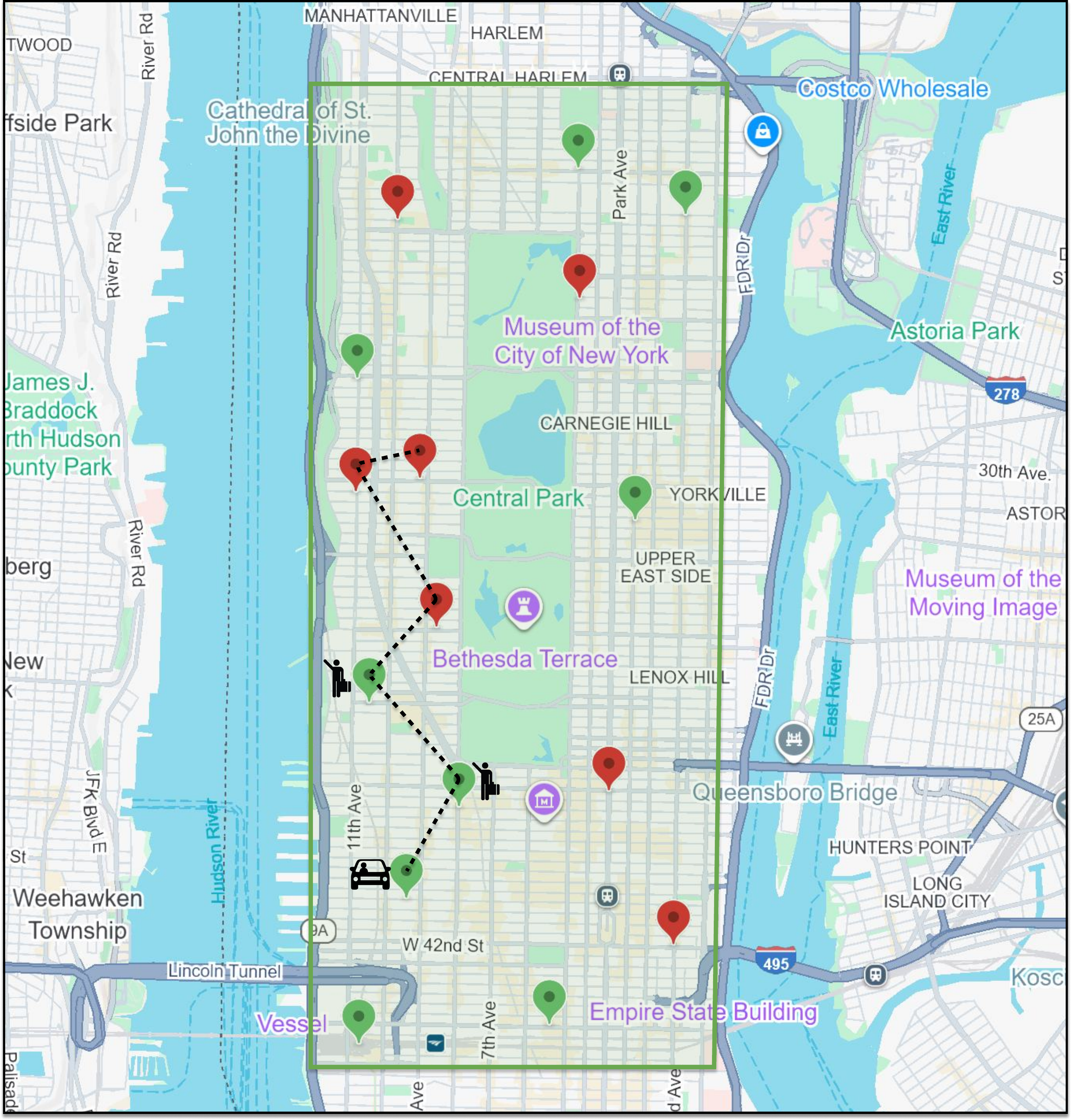}
        \caption{The \(15 \times 15\) grid overlay on the selected Manhattan
        corridor forming the simulation environment. Points represent trip
        origins and destinations sampled from the NYC TLC dataset.}
        \label{fig:dataset}
    \end{minipage}
\end{figure}

The resulting role assignment process is illustrated in
Figure~\ref{fig:role_assignment_fig}.

\subsection{Population Dynamics and Network Adoption}
\label{subsec:popdynamics}

\begin{figure}[ht]
    \centering
    \includegraphics[width=1\textwidth]{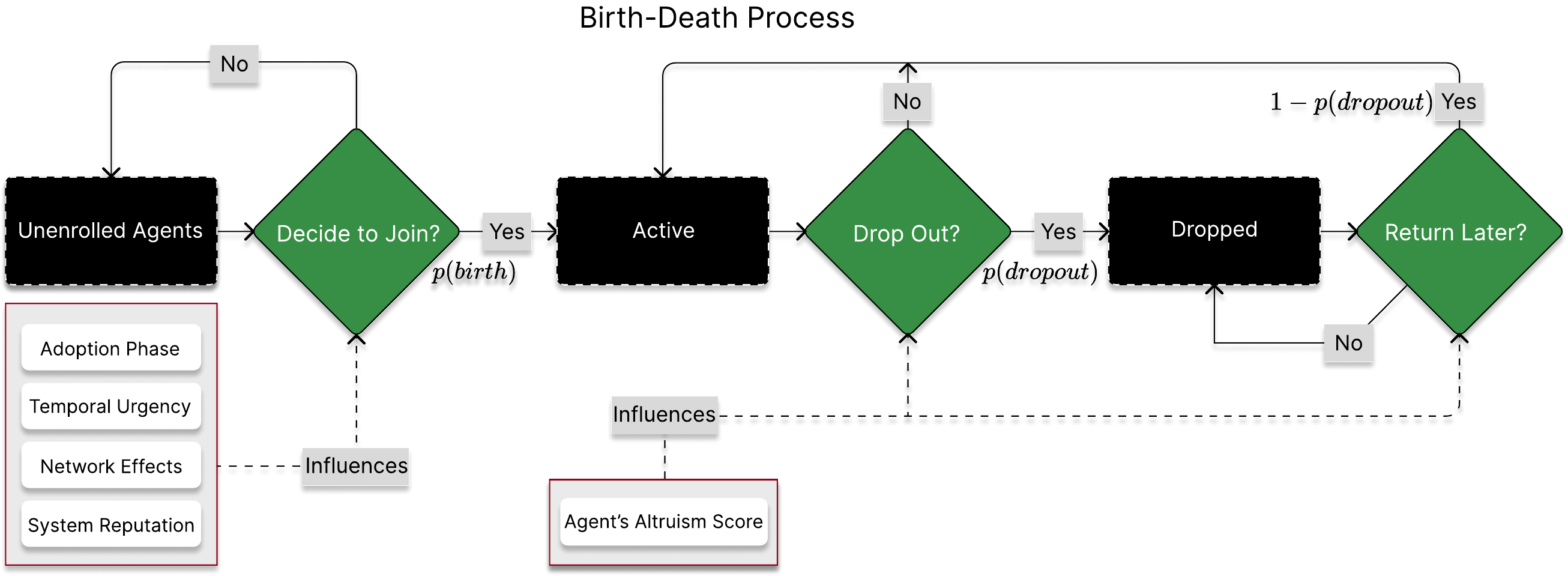}
    \caption{Population dynamics in the ARS system. Agents may join the
    platform (birth), remain active, temporarily drop out, and later
    return. Adoption is influenced by network effects and system
    reputation, while dropout probability is linked to an agent's
    altruism score.}
    \label{fig:birth_death_representation}
\end{figure}

Decentralized peer-to-peer mobility systems require sufficient
participation to remain viable. To capture realistic community growth
and retention, the ARS simulation incorporates a dynamic population
model in which agents may enter, leave, and later rejoin the system.
The overall process is illustrated in
Figure~\ref{fig:birth_death_representation}.

\textbf{Dropout Process.}
Agent retention is linked to the altruism economy. Agents with
persistently low altruism scores have a higher probability of leaving
the platform, reflecting reduced incentives when they cannot secure
rides. Conversely, agents with higher altruism tend to remain active.

\textbf{Birth Process.}
New agents join the system through a probabilistic adoption model
capturing network effects and platform reputation:
\begin{equation}
P_{\text{birth}} =
P_{\text{base}}
\cdot
F_{\text{phase}}(\rho)
\cdot
F_{\text{urgency}}(d)
\cdot
F_{\text{net}}(N_{\text{act}})
\cdot
F_{\text{rep}}(\bar{s})
\end{equation}

where $\rho$ denotes the current community adoption rate,
$N_{\text{act}}$ represents the number of active agents,
and $\bar{s}$ is the average altruism score of the population.
The multiplier $F_{\text{phase}}$ models adoption stages inspired by
diffusion-of-innovation theory, while $F_{\text{net}}$ and
$F_{\text{rep}}$ capture positive feedback from user density and
system reputation.

To mitigate cold-start challenges, newly joining agents receive a
small initialization bonus to their altruism score. This encourages
early participation and helps the system reach a cooperative critical
mass more rapidly.

Detailed definitions and mathematical formulations pro-
vided in the supplementary material.

\subsection{ORACLE: Shared-Parameter Learning Architecture}
\label{subsec:algo}

Ride-sharing environments involve large populations of agents making
structurally similar decisions. Standard multi-agent reinforcement
learning (MARL) methods such as MADDPG~\cite{lowe_multi-agent_2017}
or MAAC~\cite{iqbal2019actorattentioncriticmultiagentreinforcementlearning}
maintain independent policy networks for each agent, causing the number
of parameters and training instability to grow with population size.
To support scalable learning in ARS, we introduce ORACLE
(\emph{One-Network Actor--Critic for Learning in Cooperative
Environments}), a parameter-sharing MARL architecture designed for
homogeneous agent populations.

Instead of learning separate policies for each agent, ORACLE trains a
single shared actor network that is reused across all agents. Each agent
conditions its decisions on its own observation while sharing the same
policy parameters. This allows knowledge learned by one agent to
immediately benefit the entire population and enables the policy to
generalize to varying numbers of active agents.

During centralized training, ORACLE employs an ego-conditioned
attention-based critic to evaluate joint agent interactions. The critic
processes the ego agent’s state together with contextual information
from neighboring agents and the selected action, capturing coordination
effects between drivers competing for riders in the local environment.
The attention mechanism focuses on relevant agents while ignoring
unrelated ones, improving learning stability in dense multi-agent
settings.

\begin{figure}[ht]
    \centering
    \includegraphics[width=0.65\textwidth]{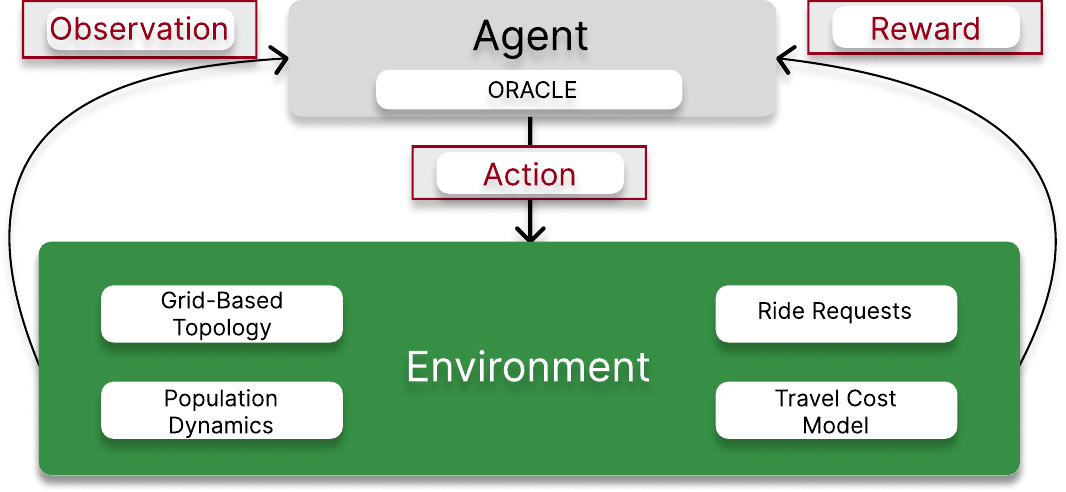}
    \caption{Interaction loop of the ORACLE policy in the ARS system.
    Drivers observe local state information and select riders using
    the shared policy. Learning is guided by rewards that balance
    driver detour cost and altruism incentives.}
\end{figure}

The ORACLE policy operates using structured observations,
discrete rider-selection actions, and a reward signal that
balances efficiency and cooperative incentives.

\paragraph{\textbf{Observation Space.}}
Each agent observes a structured state vector composed of three components:
\begin{itemize}[leftmargin=*]
    \item \textbf{Spatial coordinates}: The $(x,y)$ position of the agent
    in the discrete grid environment.

    \item \textbf{Role identifier}: A binary indicator distinguishing
    drivers (1) from riders (0).

    \item \textbf{Local perception field}: A $5\times5$ observation grid
    centered on the agent that encodes nearby rider locations and local
    environmental context.
\end{itemize}
Together these components form a 28-dimensional observation vector.

\paragraph{\textbf{Action Space.}}
Driver actions correspond to selecting riders from their local
neighborhood. The action space is defined as
$u \in \{0,1,\dots,N\}$ where $u \in \{0,\dots,N-1\}$ corresponds to
selecting a specific rider for pickup and $u=N$ represents declining
all available requests.
To ensure feasibility, ORACLE employs \emph{action masking}, which
dynamically restricts available actions to riders visible within the
local perception field. This prevents invalid assignments while
maintaining computational efficiency.

\paragraph{\textbf{Reward Mechanism.}}
To guide decentralized learning, ORACLE optimizes a scalar reward
$r_i$ that balances driver cost, rider benefit, and altruism incentives:
\begin{equation}
  r_i = \tanh\!\Big(
         \alpha_r \cdot
         \underbrace{d_j(\emptyset)}_{\substack{\text{\scriptsize rider} \\ \text{\scriptsize trip saved}}}
         -
         (1 - \alpha_r) \cdot
         \underbrace{\Delta d_i(a_j)}_{\substack{\text{\scriptsize driver} \\ \text{\scriptsize detour cost}}}
         +
         \underbrace{\Delta s_i^+}_{\substack{\text{\scriptsize altruism} \\ \text{\scriptsize gain}}}
       \Big)
  \label{eq:reward}
\end{equation}
The first term rewards serving riders with longer trips, the second
penalizes additional detour incurred by the driver, and the third
captures altruism gains from the virtual currency mechanism.
Together these terms encourage drivers to select riders that improve
overall system efficiency while maintaining fair participation across
the community.

\subsection{Continual Adaptation via Human-in-the-Loop}
\label{continual}

In real-world deployment, the ORACLE policy can operate as a
recommendation engine rather than a fully autonomous controller.
Drivers may accept, delay, or override suggested rider selections
based on factors that are difficult to model explicitly, such as
personal schedules or perceived detour tolerance.
These interactions generate off-policy transitions that are stored in a
prioritized experience replay (PER) buffer. Transitions with higher
temporal-difference (TD) error are sampled more frequently during
training, enabling the policy to focus on situations where model
predictions diverge from observed behavior.
This mechanism allows the ARS policy to continually adapt to evolving
real-world driving patterns without requiring explicit exploration that
could negatively affect service quality.

\section{Altruistic Ride Sharing Simulation Design}
\label{sec:simulation}

This section describes the simulation environment used to evaluate the
Altruistic Ride-Sharing (ARS) framework. The simulator integrates the
ORACLE multi-agent reinforcement learning model with the population
dynamics and altruism mechanisms introduced in \autoref{sec:arsintro}. The goal of
the simulation is to analyze how decentralized ride-selection policies
influence system-level outcomes such as travel efficiency, congestion
reduction, and cooperative participation.

\subsection{Environment Modeling and Evaluation Metrics}
\label{modeling}

The ARS environment is modeled as a multi-agent system operating on a
discrete \(15 \times 15\) grid representing an approximately
\(23\,\text{km}^2\) urban region derived from the New York City Taxi and
Limousine Commission (TLC) Yellow Taxi dataset (January 2016).

Trips are filtered to a one-hour window (9:00–10:00 AM on January 2,
2016) within a Manhattan corridor and mapped to the nearest grid cells,
reducing the continuous spatial space while preserving realistic travel
patterns. Inter-cell travel distances are precomputed using the
OpenStreetMap road network assuming an average speed of 25 km/h.

When a pickup is chosen, the driver’s route is
updated and the resulting detour cost and altruism updates are applied.
The environment evolves in discrete daily cycles consisting of role
assignment, ride-selection interactions, route updates, and altruism
score adjustments.

Population dynamics—including agent birth, dropout, and reintegration
mechanisms described in \autoref{subsec:popdynamics}, are also simulated to capture realistic variations in participation over time. Trips are sampled uniformly across distance ranges to ensure
representative coverage of short and medium-distance urban travel.

Figure~\ref{fig:dataset} illustrates the grid discretization and example
trip trajectories used in the simulation.

This environment forms the basis for evaluating ARS under the metrics defined next.

\paragraph{\textbf{System Efficiency Metrics}}

\begin{itemize}[leftmargin=*]

\item \textbf{Total Distance and Carbon Emissions.}
Total distance traveled by all vehicles during the simulation.
Carbon emissions are estimated as a linear function of total distance. Lower values indicate improved
environmental efficiency.

\item \textbf{Vehicle Utilization.}
Measures the average occupancy of vehicles over time.
Higher utilization indicates more efficient use of available vehicles
through ride-sharing.

\item \textbf{Road Traffic Density.}
Measures the number of vehicles present in spatial grid cells of the
city map during the simulation. Lower traffic density indicates
reduced congestion.

\item \textbf{Detour Factor.}
Ratio of actual trip distance to the corresponding direct route.

\end{itemize}

\paragraph{\textbf{Service Quality Metrics}}

\begin{itemize}[leftmargin=*]

\item \textbf{Average Trip Time.}
Average travel time experienced by agents during a simulation day.

\item \textbf{Rider Acceptance Rate.}
Fraction of rider requests successfully matched with drivers.
Higher acceptance rates indicate better service availability.

% \item \textbf{Average Agent Reward.}
% For MARL-based methods, we report the average cumulative reward
% obtained by agents during the simulation to compare the effectiveness
% of learning policies.

\end{itemize}

\paragraph{\textbf{Fairness and Participation Metrics}}
\label{metric:fairness}
\begin{itemize}[leftmargin=*]

\item \textbf{Benefit Distribution Analysis.}
Evaluates inequality in ride-sharing benefits across agents using
a 3D Lorenz surface and the Gini coefficient.

\item \textbf{Reintegration Score.}
Measures the system’s ability to reintegrate agents who temporarily
leave the platform and later return. The score aggregates multiple
aspects of user re-entry behavior:
\begin{equation}
REINT = \alpha R_{basic} + \beta R_{time} + \gamma R_{quick} + \delta R_{stable}
\end{equation}
where $R_{basic}$ is the overall return rate, $R_{time}$ measures
reintegration speed, $R_{quick}$ captures short-term returns, and
$R_{stable}$ reflects long-term participation stability.

\end{itemize}

Detailed definitions and mathematical formulations of these metrics
are provided in the supplementary material.

\subsection{Baselines}
\label{subsec:baselines}

We compare ARS against several learning-based, optimization-based,
and system-level baselines to evaluate the effectiveness of the
proposed framework. All learning-based baselines are trained using the same
environment, reward structure, and observation space
to ensure fair comparison.

\paragraph{\textbf{Learning-Based Baselines}}

\begin{itemize}[leftmargin=*]

\item \textbf{MADDPG}~\cite{lowe_multi-agent_2017}.
A centralized-training decentralized-execution MARL algorithm where
each agent maintains an independent actor–critic policy. In our
experiments, each driver is modeled as an independent MADDPG agent
trained within the same ARS environment and reward formulation.

\item \textbf{MAAC}~\cite{iqbal2019actorattentioncriticmultiagentreinforcementlearning}.
An actor–critic MARL method with an attention-based centralized critic.
Agents operate using local observations while the critic selectively
attends to relevant agents during training.

\end{itemize}

\paragraph{\textbf{Optimization Baseline}}

\begin{itemize}[leftmargin=*]

\item \textbf{Particle Swarm Optimization (PSO)}~\cite{pso}.
PSO is used as a heuristic optimization baseline where each particle
represents a rider–driver assignment configuration.

\end{itemize}

\paragraph{\textbf{System Baseline}}

\begin{itemize}[leftmargin=*]

\item \textbf{No Ride Sharing}.
Each agent travels directly to its destination without ride-sharing.
This scenario provides a lower-bound reference for evaluating the
impact of cooperative mobility.

\end{itemize}

\section{Performance Analysis and Results}
\label{sec:results}

We evaluate the Altruistic Ride-Sharing (ARS) framework through a
series of simulations designed to analyze system-level efficiency,
service quality, and cooperative participation. Experiments are
conducted with agent populations of $N \in \{100,150,200\}$ to study
system behavior under varying demand levels. For each configuration,
100 independent simulation runs are performed to ensure statistical
robustness.

Agents are initialized with altruism scores drawn from two different
distributions: a uniform distribution over $[0,1]$ and a Gaussian
distribution centered at $0.5$ with bounded support in $[0,1]$. This
allows us to examine system dynamics under both evenly distributed and
centrally concentrated cooperation levels.

We evaluate ARS under both fixed-population and dynamic population
settings. In the fixed scenario, the number of agents remains constant
throughout the simulation. In the dynamic scenario, agents may enter or
leave the system according to the birth--death processes described in
\autoref{subsec:popdynamics}, enabling evaluation of system stability
under evolving participation.

Across these experimental settings, we compare the ORACLE-based ARS
model with learning-based baselines, optimization-based solvers, and a
no-sharing scenario to quantify improvements in mobility efficiency,
traffic reduction, and community participation.

\begin{figure*}[t!]
\centering
\includegraphics[width=\textwidth]{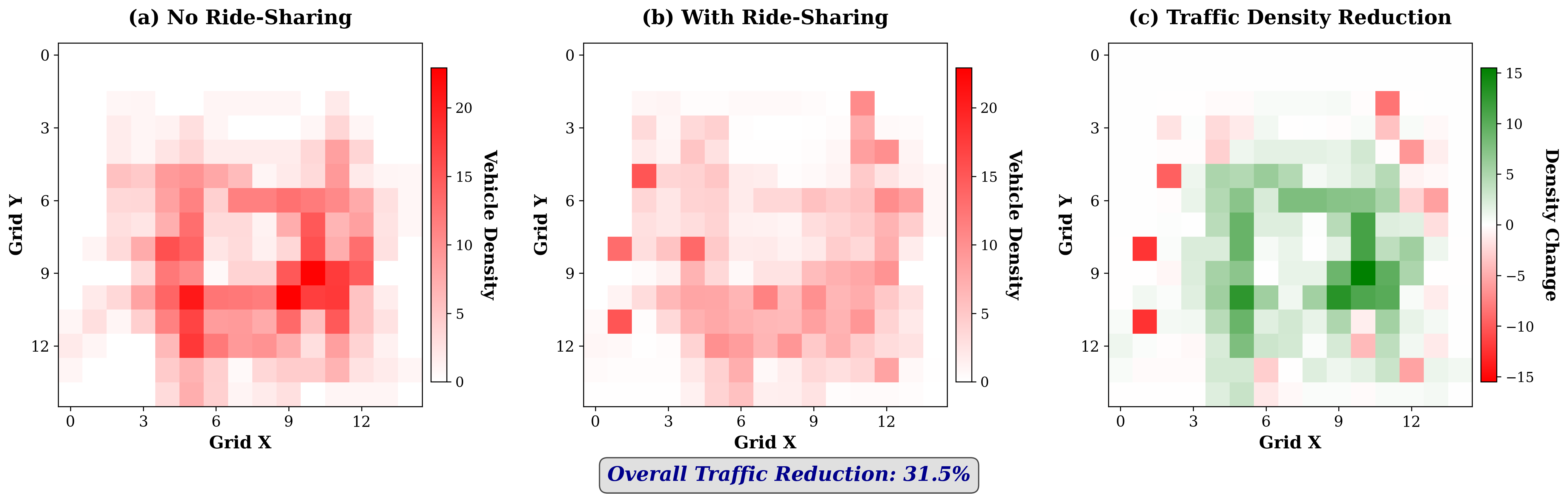}
\vspace{1mm}
\caption{Comparison of traffic density maps illustrating the reduction in urban congestion achieved by the ARS framework. The figure displays vehicle density for (a) the No Ride-Sharing baseline and (b) the With Ride-Sharing scenario. Subplot (c) visualizes the difference, quantifying a significant overall traffic reduction of 31.5\%.}
\label{fig:traffic_density}
\end{figure*}

\begin{figure*}[t!]
\centering
\includegraphics[width=\textwidth]{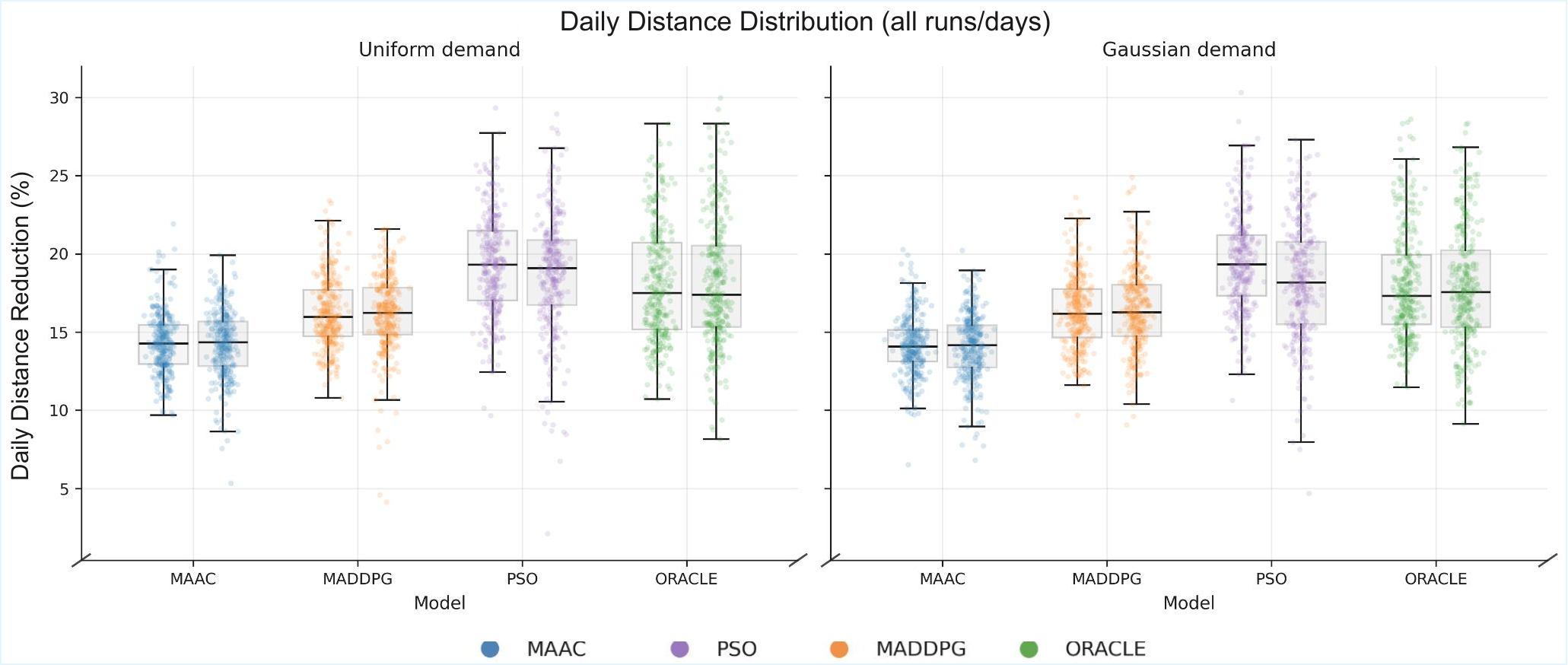}
\vspace{1mm}
\caption{
Daily travel distance reduction across simulation runs under
uniform (left) and Gaussian (right) demand distributions.
PSO achieves the largest reductions through global optimization,
while ORACLE closely competes with PSO while operating with
decentralized decision-making.
}
\label{fig:distance_reduction}
\end{figure*}

\begin{figure*}[t!]
\centering
\includegraphics[width=\textwidth]{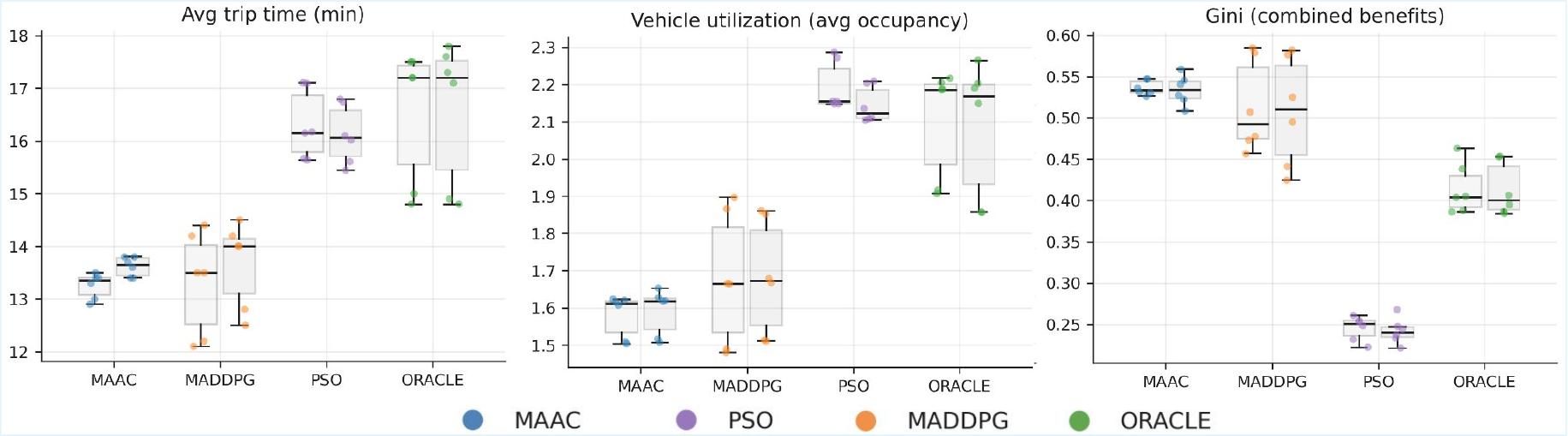}
\vspace{1mm}
\caption{
Distribution of key performance metrics across ride-sharing methods.
Box plots compare average trip time, vehicle utilization (average
occupancy), and benefit inequality (Gini coefficient) for MAAC,
MADDPG, PSO, and ORACLE. ORACLE achieves competitive utilization
while maintaining lower inequality in benefit distribution compared
to optimization-based approaches.
}
\label{fig:box_metrics}
\end{figure*}

\begin{figure*}[t!]
\centering
\includegraphics[width=\textwidth]{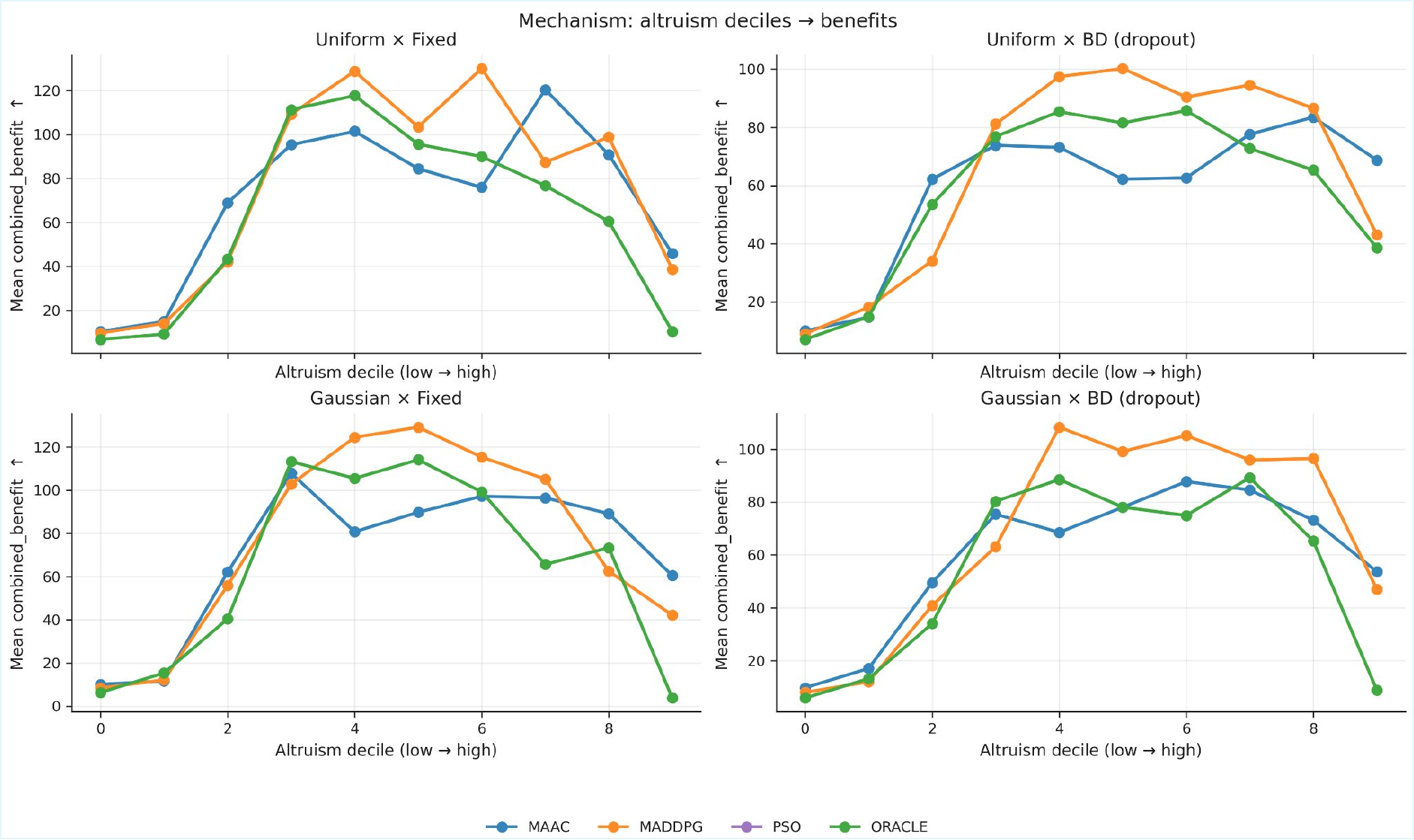}
\vspace{1mm}
\caption{
Relationship between altruism score deciles and mean community benefit
across different simulation settings. Results are shown for four
initialization scenarios: Uniform and Gaussian altruism distributions
with fixed populations and with birth--death dynamics. ORACLE maintains
stable benefit allocation across altruism levels while avoiding the
extreme concentration observed in some optimization-based assignments.
}
\label{fig:benefit_altruism}
\end{figure*}

\begin{figure*}[t!]
\centering
\includegraphics[width=\textwidth]{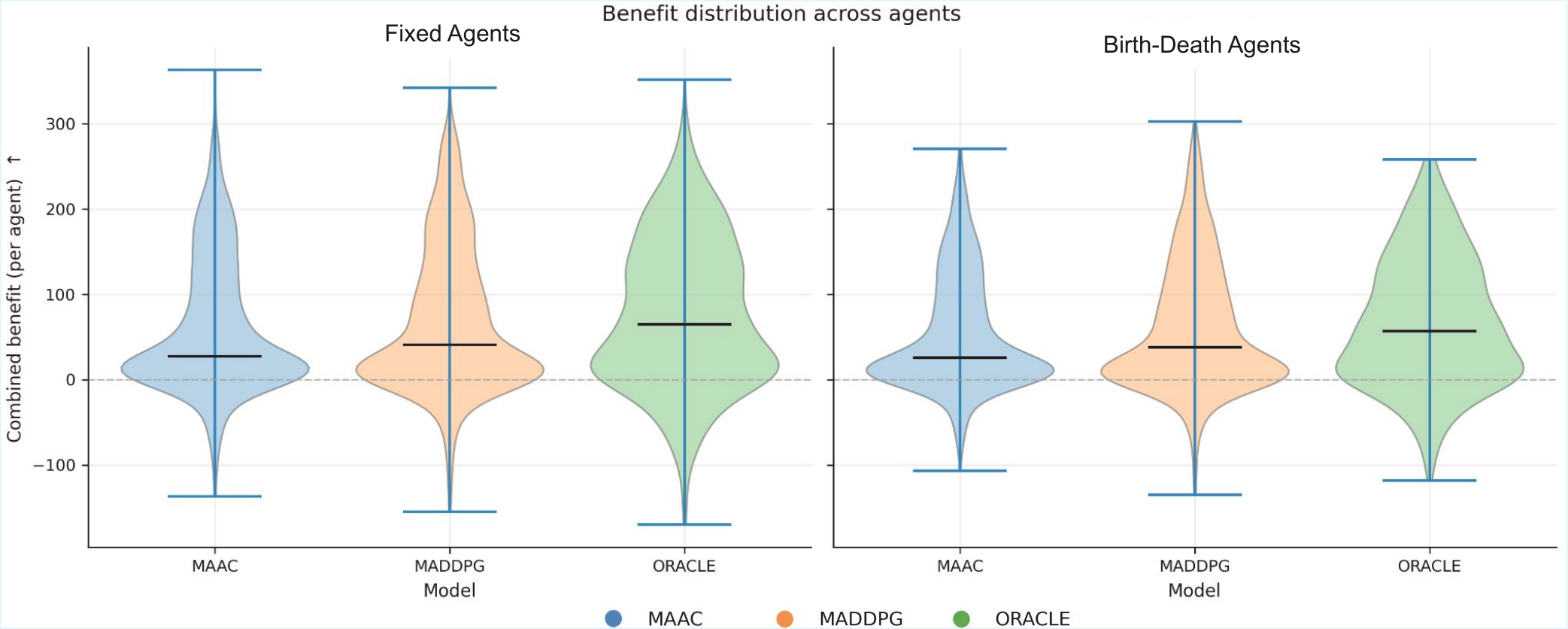}
\vspace{1mm}
\caption{
Distribution of cumulative ride-sharing benefits across agents.
Violin plots show the distribution of individual benefits for MAAC,
MADDPG, and ORACLE under two population scenarios: fixed agents
(left) and dynamic birth–death participation (right). ORACLE produces
a more balanced benefit distribution while maintaining higher median
community benefit, indicating improved fairness and sustained
participation incentives.
}
\label{fig:benefit_distribution}
\end{figure*}

\begin{figure*}[t!]
\centering
\includegraphics[width=\textwidth]{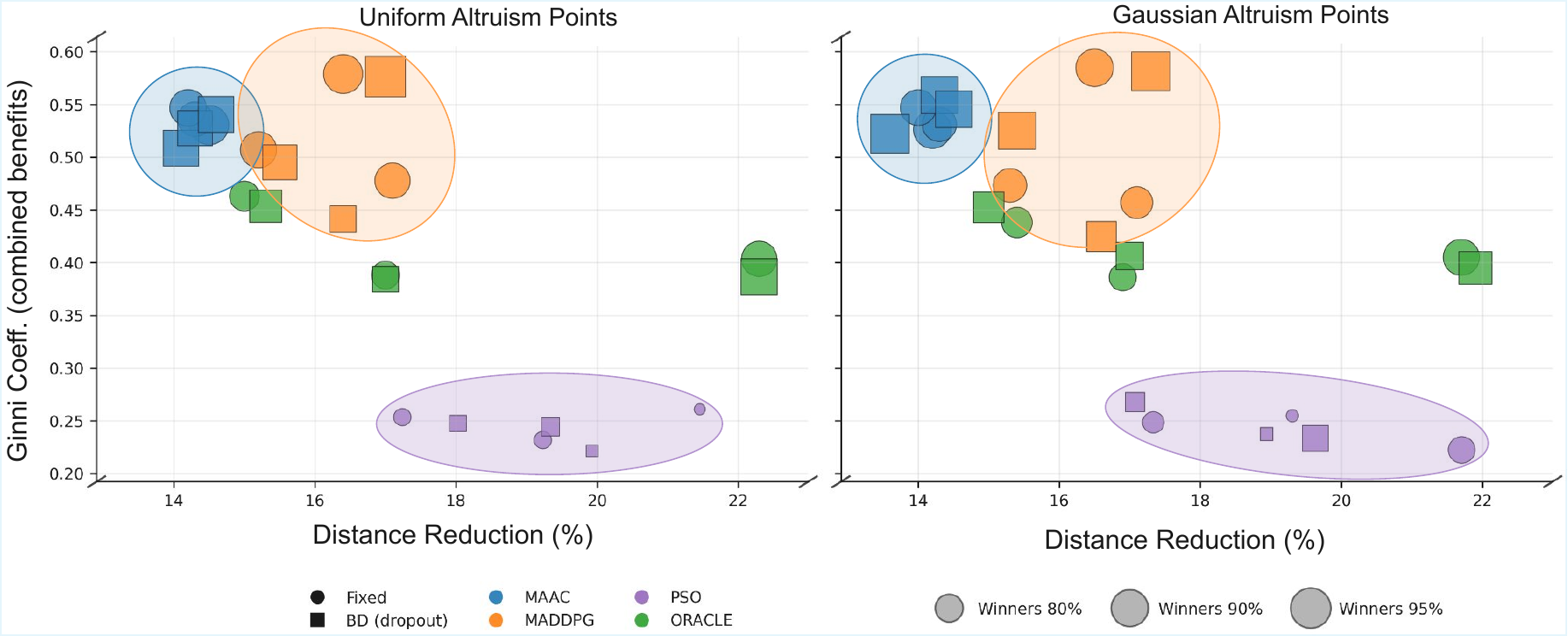}
\vspace{1mm}
\caption{
Distance reduction vs.\ benefit inequality (Gini; lower is fairer)
under uniform (left) and Gaussian (right) altruism initialization.
ORACLE achieves competitive distance reduction while maintaining
lower inequality than other learning-based methods.
}
\label{fig:gini_distance}
\end{figure*}

\subsection{ARS vs. No Sharing}

We first compare the ARS framework against a baseline scenario in which
agents travel independently without ride-sharing. Across the 100-day
simulation horizon, ARS consistently reduces the total distance
traveled by the community by approximately \textit{20\%} across
different agent populations and simulation settings
(\autoref{fig:distance_reduction}). Because vehicle emissions are
proportional to distance traveled, this reduction directly translates
into a comparable decrease in estimated carbon emissions.

In addition to environmental benefits, ARS substantially reduces road
congestion. As shown in \autoref{fig:traffic_density}, the number of
vehicles occupying grid cells during peak periods decreases
significantly under ARS, resulting in an overall traffic reduction of
up to \textit{31.5\%}. These results indicate that altruism-driven
ride-sharing can improve transportation efficiency and reduce
congestion without relying on monetary incentives. Beyond pure efficiency, the ARS framework was designed to be equitable and self-sustaining. Our results confirm its success in fostering a resilient and fair community and demonstrate its ability to recover users who drop out and maintain a stable, active user base over time. (\autoref{tab:reintegration_metrics}).

\subsection{Architectural Advantages of ORACLE and Learning-Based Methods}

Optimization-based approaches such as PSO can achieve strong
performance on certain metrics under controlled conditions,
particularly when rider requests and driver locations are
available simultaneously for global assignment. However,
these approaches treat ride matching as a static combinatorial
problem and require repeatedly solving large optimization
instances as the system state evolves.

In real-world mobility systems, ride requests arrive
asynchronously and drivers depart immediately after
matching, making static batch assignment difficult to
maintain in practice. Recomputing global assignments
as new requests appear introduces significant
computational overhead and may invalidate previously
generated routes.
In addition, evaluating candidate assignments in PSO requires solving
small multi-rider routing problems (e.g., three- or four-rider TSP
variants) for each driver, significantly increasing the computational
cost of a single global optimization step compared to policy inference
in learning-based approaches.

Learning-based methods instead operate in a sequential
decision-making setting where agents act based on
locally observable information. This allows the system
to process new ride requests incrementally while
maintaining stable assignments over time.

This sequential decision structure also enables human-in-the-loop
operation. Drivers may override recommended matches (e.g., rejecting
or selecting a different rider), and the resulting transitions are
incorporated into the prioritized experience replay buffer during
training. As summarized in Table~\ref{tab:deployment_comparison},
this makes human override support native to ORACLE. In contrast,
global optimization approaches would require recomputing the full
assignment when deviations occur, while MARL architectures such as
MADDPG and MAAC rely on separate per-agent policies, making it
difficult to consistently incorporate override behavior into a
shared learning process.

Within this class of approaches, ORACLE introduces
additional architectural advantages over standard
multi-agent reinforcement learning methods such as
MADDPG and MAAC. Rather than maintaining separate
actor and critic networks for each agent, ORACLE
uses a single shared network that evaluates the joint
state from the perspective of an ego agent using an
attention-based critic.

This shared-network formulation reduces parameter
complexity, improves training stability, and enables
continual adaptation through real-world interaction.
Empirically, ORACLE achieves stronger performance
across efficiency, fairness, and utilization metrics
while remaining computationally more efficient than
both optimization-based baselines and existing
learning-based architectures.

\begin{table}[t]
\centering
\caption{Deployment-oriented comparison of optimization and learning approaches.}
\label{tab:deployment_comparison}
\small
\setlength{\tabcolsep}{4pt}
\begin{tabular}{lccc}
\toprule
\textbf{Deployment Dimension} & \textbf{PSO} & \textbf{MADDPG / MAAC} & \textbf{ORACLE} \\
\midrule
Information requirement & Static batch & Local obs. & Local obs. \\
Computational latency & $\sim$100$\times$ slower & $\sim$5$\times$ slower & Fast \\
Live requests & No & Yes & Yes \\
Human override & Infeasible & Difficult & Native \\
Acceptance rate (batch) & Highest & Moderate & Moderate \\
Stability (Reintegration) & Not modeled & Moderate (78.20) & High (89.40) \\
\bottomrule
\end{tabular}

\vspace{2pt}
{\footnotesize Latency measured relative to ORACLE inference time.}
\end{table}

\paragraph{\textbf{Community Fairness and System Resilience.}}
Beyond pure efficiency, the ARS framework was designed to be equitable and self-sustaining. Our results confirm its success in fostering a resilient and fair community. The model's high Reintegration Score of \textit{89.40} and Stability Score of \textit{0.8316} demonstrate its ability to recover users who drop out and maintain a stable, active user base over time (\autoref{tab:reintegration_metrics}).
\begin{table}[t]
\centering

\begin{minipage}{0.47\linewidth}
\centering
\caption{Summary of Reintegration Metrics - ORACLE}
\label{tab:reintegration_metrics}
\small
\begin{tabular}{lc}
\toprule
\textbf{Metric} & \textbf{Value} \\
\midrule
Final Score $\uparrow$ & 89.40 \\
Basic Rate $\uparrow$ & 0.9888 \\
Time-Weighted Rate $\uparrow$ & 0.8039 \\
Quick Return Rate $\uparrow$ & 0.9887 \\
Stability Score $\uparrow$ & 0.8316 \\
Avg. Return Time (d) $\downarrow$ & 1.10 \\
\bottomrule
\end{tabular}
\end{minipage}
\hfill
\begin{minipage}{0.47\linewidth}
\centering
\caption{Avg. Vehicle Utilization and Detour Factor Metrics}
\label{tab:utilization_detour}
\small
\begin{tabular}{lcc}
\toprule
\textbf{Method} & \textbf{Utilization $\uparrow$} & \textbf{Detour $\downarrow$} \\
\midrule
ORACLE & 2.10 & 1.27 \\
MADDPG & 1.68 & 1.14 \\
PSO & 2.17 & 1.53 \\
MAAC & 1.59 & 1.19 \\
\bottomrule
\end{tabular}
\end{minipage}

\end{table}

\paragraph{\textbf{Closed Economy Design.}}
ARS intentionally operates as a closed incentive economy in which
altruism points can only be earned through providing rides and spent
when receiving rides. No external mechanisms such as point purchasing
or membership tiers are allowed. This restriction preserves the
interpretability of altruism scores as a measure of community
contribution and ensures that role assignment remains tied to actual
participation. Allowing external injection of altruism points would
break this signal and destabilize the driver–rider equilibrium.
A detailed analysis of these failure modes is provided in the
supplementary material.

\section{Related Work}\label{sec:relwork}

Research on ride-sharing has evolved across optimization methods,
reinforcement learning frameworks, and decentralized coordination
mechanisms. Wen et al.~\cite{10529131} review machine learning
approaches for ride-hailing and categorize planning strategies into two
key tasks: matching (assigning vehicles to riders) and repositioning
(relocating vehicles to meet anticipated demand), further distinguishing
between centralized collective planning and distributed agent-based
decision-making. Broader surveys by Qin et al.~\cite{qin_reinforcement_2022}
examine reinforcement learning applications in ride-sharing, while recent
studies explore hybrid transit systems~\cite{kumar_algorithm_2021} and
pooled routing with passenger transfers~\cite{wang_optimization_2023},
highlighting growing interest in multi-modal and socially responsible
transport systems.

\noindent\textbf{Optimization-Based Ride-Sharing.}
Traditional ride-sharing research largely focuses on centralized
optimization for system-wide efficiency. Alonso-Mora et al.~\cite{alonso-mora_-demand_2017}
introduced a real-time high-capacity ride-pooling algorithm capable of
handling large-scale urban demand. Zhou and Roncoli~\cite{10241664}
extended this paradigm with a joint pricing and matching framework
incorporating fairness-aware discounts. To address scalability,
distributed approaches have also been explored. Masoud and
Jayakrishnan~\cite{masoud_real-time_2017} proposed heuristic-based
peer-to-peer matching, while Wang et al.~\cite{Wang2023pdRide}
investigated privacy-preserving decentralized ride-sharing. CARE-Share
\cite{manjunath_et_al_care-share_nodate} further explored fully
distributed multi-objective optimization using ant colony techniques.
However, most optimization-based systems rely on static assumptions and
lack mechanisms to adapt to dynamic user behaviors or fairness
considerations.

\noindent\textbf{Reinforcement Learning in Ride-Sharing.}
Reinforcement learning has become a natural framework for dynamic
ride-matching. Tang et al.~\cite{tang2021deepvaluenetworkbasedapproach} proposed a deep
value-network dispatcher, while DeliverAI~\cite{mehra_deliverai_2024,mehra_last_2024}
extended RL-based routing to logistics systems. Deep reinforcement learning further enables scalable decision making, with frameworks such as DQN~\cite{vanhasselt2015deepreinforcementlearningdouble} inspiring
applications like DeepPool~\cite{al-abbasi_deeppool_2019} and its
distributed extensions~\cite{haliem_distributed_2021,singh_distributed_2022}.
Other works improve coordination using mean-field
approximations~\cite{qin_efficient_2020}, distribution
matching~\cite{zhang_multi-agent_2019}, hierarchical
MARL~\cite{zhou_multi-agent_2020}, transfer
learning~\cite{alberto_castagna_multi-agent_2022}, and attention
mechanisms~\cite{li_efficient_2021}. Wei et al.~\cite{Wei2024Lookahead}
introduced lookahead-based RL for real-time ride matching. Despite
these advances, most RL systems remain economically driven,
optimizing revenue, latency, or fleet utilization rather than
community-driven cooperation.

\section{Conclusion and Future Work}
\label{sec:conclusion}

This paper presented Altruistic Ride-Sharing (ARS), an innovative framework designed to enhance short-distance urban mobility through community involvement. Distinct from profit-based ride-sharing models, ARS replaces financial incentives with altruism points, a virtual currency improving cooperation and mutual support. Employing multi-agent reinforcement learning (ORACLE), ARS offers a fair, scalable, and sustainable resolution to urban transport inefficiencies. Analysis utilizing real New York City taxi data reveals that ARS substantially cuts travel distances, reduces carbon emissions and traffic congestion, while enhancing vehicle utilization and service equity. Through altruism-based incentives, the system achieves equilibrium between drivers and riders, positioning ARS as a practical alternative to profit-driven urban mobility. 

To improve system robustness, future work must incorporate more realistic and non-cooperative behaviors, such as modeling dishonest participants (e.g., no-shows, deliberate dropouts). Future work should also explore practical trust and safety mechanisms for real-world deployment, such as user verification and reputation-based accountability systems. To counter the reality that ``real users may be strategic and malicious'', this requires developing a better game-theoretic model (e.g., one based on incomplete information or adversarial search) to analyze and mitigate non-cooperative strategies. In addition to the current grid structure, future works could also explore using dynamic road graph topology (e.g., a high-performance, \textit{GraphHopper}-based structure).

\newpage

\appendices

%% ===================================================================
\section{Notation and Symbols}
\label{sec:notation}
%% ===================================================================

Tables~\ref{tab:env_variables} and~\ref{tab:model_params} collect all
symbols used in the paper.  Environment and problem-level symbols
appear in Table~\ref{tab:env_variables}; learning and population-model
parameters appear in Table~\ref{tab:model_params}.

\begin{table}[H]
    \centering
    \caption{Model and Training Parameters}
    \label{tab:model_params}
    \renewcommand{\arraystretch}{1.4}
    \begin{tabular}{l p{0.6\linewidth}}
        \hline
        \textbf{Symbol} & \textbf{Meaning} \\
        \hline \hline
        \multicolumn{2}{l}{\textit{ORACLE training parameters}} \\
        \hline
        $\pi_\theta$ & Shared actor network (parameters $\theta$) \\
        \hline
        $Q_\phi$ & Shared ego-conditioned attention critic (parameters $\phi$) \\
        \hline
        $\alpha_\pi$ & Actor learning rate \\
        \hline
        $\alpha_Q$ & Critic learning rate \\
        \hline
        $\gamma$ & Discount factor \\
        \hline
        $\tau$ & Soft (Polyak) target-network update rate \\
        \hline
        $\lambda$ & Actor weight regularisation coefficient \\
        \hline
        $\mathcal{E}$ & Exploration strategy ($\epsilon$-greedy, TAGS, or GST) \\
        \hline
        $\mathcal{D}$ & Prioritised experience replay buffer \\
        \hline
        $\alpha_{\mathcal{D}}, \beta_{\mathcal{D}}$ & PER prioritisation exponent and importance-sampling exponent \\
        \hline
        \multicolumn{2}{l}{\textit{Reward shaping}} \\
        \hline
        $\alpha_r$ & Trade-off weight between rider benefit and driver detour cost in the reward \\
        \hline
        $\alpha_s, \beta_s$ & Driver and rider altruism score scaling factors \\
        \hline
        \multicolumn{2}{l}{\textit{Population dynamics (birth/death)}} \\
        \hline
        $\alpha_{bd}$ & Base dropout probability when $s_i^d \approx 0$ \\
        \hline
        $\beta_{bd}$ & Exponential decay rate for low-altruism agents \\
        \hline
        $\gamma_{bd}$ & Base dropout value for high-altruism agents \\
        \hline
        $\delta_{bd}$ & Linear decay range above altruism threshold $s_{\text{th}}^d$ \\
        \hline
        $s_{\text{th}}^d$ & Altruism threshold separating exponential and linear dropout regimes \\
        \hline
        $P_{\text{dropout}}$ & Dropout probability (function of $s_i^d$) \\
        \hline
        $P_{\text{birth}}$ & Birth probability for unenrolled agents \\
        \hline
        \multicolumn{2}{l}{\textit{Reintegration score}} \\
        \hline
        $\alpha, \beta, \gamma, \delta$ & Weighting factors for the four reintegration components \\
        \hline
        $\lambda_{\text{re}}$ & Time-decay parameter for time-weighted return rate \\
        \hline
        $\tau_{\text{re}}$ & Quick-return threshold (number of days) \\
        \hline
    \end{tabular}
\end{table}

\begin{table}[H]
    \centering
    \caption{Environment and Problem Variables}
    \label{tab:env_variables}
    \renewcommand{\arraystretch}{1.4}
    \begin{tabular}{l l}
        \hline
        \textbf{Symbol} & \textbf{Meaning} \\
        \hline \hline
        $N$ & Total number of agents \\
        \hline
        $D$ & Total simulation days, $d \in \{1, \dots, D\}$ \\
        \hline
        $\mathcal{A}$ & Set of all agents, $\mathcal{A} = \{a_1, \dots, a_N\}$ \\
        \hline
        $\mathcal{A}_{\text{active}}$ & Set of active agents on a given day \\
        \hline
        $\mathcal{D}_d$ & Set of drivers on day $d$, \ $\mathcal{D}_d \subseteq \mathcal{A}_{\text{active}}$ \\
        \hline
        $\mathcal{R}_d$ & Set of riders on day $d$, \ $\mathcal{R}_d \subseteq \mathcal{A}_{\text{active}}$ \\
        \hline
        $\mathbf{M}$ & \makecell[l]{Role assignment map; $\mathbf{M}_i \in \{\text{driver}, \text{rider}\}$} \\
        \hline
        $s_i^d$ & Altruism score of agent $a_i$ at the end of day $d$, \ $s_i^d \in [0,1]$ \\
        \hline
        $s_{\max}^d$ & Maximum altruism score on day $d$, \  $s_{\max}^d = \max_i s_i^d$ \\
        \hline
        $\Phi^d$ & \makecell[l]{Set of all rider pick-up permutations on day $d$, \\
        $\Phi^d = \bigcup_{k=0}^{|\mathcal{R}_d|} \mathrm{Perm}_k(\mathcal{R}_d)$} \\
        \hline
        $\phi_i^d$ & Ordered rider pick-up sequence for driver $a_i$ on day $d$, \ $\phi_i^d \in \Phi^d$ \\
        \hline
        $l_i$ & Vehicle capacity of driver $a_i$, \ $l_i \in \{1,\dots,4\}$ \\
        \hline
        $d_i(\cdot)$ & Trip distance function for $a_i$; \ $d_i: \Phi^d \to \mathbb{R}$; \ $d_i(\emptyset)$ = solo trip \\
        \hline
        $s_i$ & Observation of agent $a_i$ at the current time-step \\
        \hline
        $u_i$ & \makecell[l]{Action of $a_i$; \ $u_i \in \{0,\dots,N\}$ \\ $u_i = j$: pick up $a_j$; \ $u_i = N$: no pick-up (no-op)} \\
        \hline
        $r_i$ & Scalar reward received by $a_i$ from the environment \\
        \hline
        $\mathbf{m}_i$ & Binary valid-action mask for agent $a_i$ \\
        \hline
        $\rho(d)$ & Community adoption rate at day $d$: fraction of the population that has joined \\
        \hline
    \end{tabular}
\end{table}

\section{Closed Economy Design}
\label{sec:closed_economy}

\noindent
The altruism score update in Equations~\ref{eq:update} - \ref{eq:score_clamp} defines a
\textbf{closed economy}: points enter the system only when a
ride is given, and leave only when a ride is received.
No external source of altruism exists.
This section explains why that constraint is load-bearing,
not incidental.
Two natural extensions, (a) letting agents purchase points, and (b) granting premium members better service each collapse a
different pillar of the system.
We show this by tracing the failure through the existing
equations.

% ============================================================
\subsection{What the Closed Economy Currently
Guarantees}
% ============================================================

Recall the altruism update from the ARS model.
When driver $a_i$ gives a ride to rider $a_j$, the driver's
score increases by:
\begin{align}
  \Delta s_i^{t+1} \;=\; \alpha_s \, s_j^t
    \left(1 - \frac{d_i(\phi_i^t \| a_j) - d_i(\phi_i^t)}
                   {\text{max\_detour}}\right)
\label{eq:update}
\end{align}
and the rider's score decreases by:
\begin{align}
  \Delta s_j^{t+1} \;=\; -\beta_s \cdot \Delta s_i^{t+1}
\end{align}
Scores are clamped after updates to remain within valid bounds:
\begin{align}
s_i^{t+1} = \max\left(0,\min\left(1, s_i^t + \Delta s_i^{t+1}\right)\right).
\label{eq:score_clamp}
\end{align}

Three properties follow directly from this structure.

\textbf{(P1) Score reflects contribution.}
$s_i^t$ is a sufficient statistic for agent $a_i$'s net
contribution history.
A high score means the agent has driven more than they have
ridden, net of detour costs.
There is no other way to accumulate a high score.

\textbf{(P2) Role assignment is incorruptible.}
The role-switching rule in Section~3.5 assigns driver
obligations to agents with $s_i^t \leq 0.2$.
Because score is contribution-derived, this rule assigns
driving to agents who have not contributed enough --- exactly
the right agents.

\textbf{(P3) The equilibrium constraint holds.}
Equation~\ref{eq:update} requires $|D_t|/|R_t| \in [L, H]$ with $L > 1$.
This is sustained because low-scoring agents are pushed into
the driver role, maintaining a sufficient driver supply.

The two extensions below each destroy one or more of these
properties.

% ============================================================
\subsection{Extension 1: Purchasable Altruism
Points}
% ============================================================

Suppose agents may purchase additional altruism points at
price $p > 0$ per unit.
The score update in Equation~\ref{eq:score_clamp} becomes:
\begin{equation}
  s_i^{t+1} \;=\; \text{clip}\!\left(\,
    s_i^t \;+\; \Delta s_i^{t+1} \;+\; \lambda_i^t,
    \; 0,\; 1
  \right)
\end{equation}
where $\lambda_i^t \geq 0$ is the quantity purchased by
agent $a_i$ on day $t$. Compare (A) with Eq.~\ref{eq:score_clamp}, whereas (A) can be written alternatively as $s_i^{t+1}=max(0,min(1,s_i^t+\Delta s_i^{t+1}+\lambda_i^t))$. The only addition is the $\lambda_i^t$ term inside i.e. the purchased points. The $0,1$ are just the lower and upper bounds of the clamp.

\paragraph{\textbf{P1 breaks immediately.}}
With $\lambda_i^t > 0$, a high score no longer implies
contribution.
It could reflect driving history, or it could reflect
financial expenditure.
The system cannot tell them apart.

\paragraph{\textbf{P2 breaks as a consequence.}}
The role-switching rule assigns driver obligations based on
$s_i^t \leq 0.2$.
Under Equation~(A), any agent willing to pay can keep
$s_i^t > 0.2$ indefinitely, regardless of their actual
contribution.
The rule now assigns driving only to agents who are both
low-contributors \emph{and} unwilling to pay --- a much
smaller set than intended.

\paragraph{\textbf{P3 breaks as a consequence.}}
Let $\mathcal{B}^t \subseteq A$ denote the set of agents
who purchase points on day $t$.
As rational self-interest predicts $\mathcal{B}^t$ to grow
(purchasing is cheaper than driving for sufficiently
wealthy agents), the driver-rider ratio shifts toward:
\begin{equation}
  \frac{|D_t|}{|R_t|}
  \;\longrightarrow\;
  \frac{N - |\mathcal{B}^t|}{|\mathcal{B}^t|}
\end{equation}
As $|\mathcal{B}^t|$ grows, the driver supply shrinks.
A small pool of genuinely altruistic drivers ends up
serving a large pool of paying riders.
This is precisely the extractive dynamic that commercial
platforms already exhibit.
The closed economy was designed to prevent it.

\paragraph{\textbf{Summary.}}
Purchasable points corrupt the signal value of
$s_i^t$, disable the free-rider detection mechanism in
Section 3.5 of the main paper, and undermine the equilibrium constraint in
Equation~\ref{eq:equilibrium}.
All three failures follow from the single change in
Equation~(A).

\noindent \textbf{Remark 1.} $\lambda_i^t \geq 0$ is the quantity purchased by agent $a_i$
on day $t$ at price $p > 0$ per unit, determined entirely by the agent's rational self-interest rather than any system parameter.
The score update under a purchasable regime then becomes:
\begin{equation}
  s_i^{t+1} = \max\!\left(0,\;\min\!\left(1,\;
    s_i^t + \Delta s_i^{t+1} + \lambda_i^t\right)\right)
  \label{eq:purchased}
\end{equation}
Note that $\lambda_i^t$ is not a hyperparameter set by the system
designer. It is a \textit{decision variable} controlled entirely by the agent.
Given a purchase price $p > 0$, a rational agent will always buy exactly enough points to stay above the rider eligibility threshold $s = 0.2$, and no more.
Equation~\ref{eq:purchased} makes this precise: if an agent's score after the day's natural update
$s_i^t + \Delta s_i^{t+1}$ already exceeds $0.2$, they
purchase nothing ($\lambda_i^t = 0$); if it falls below, they purchase precisely the shortfall. 
This purchasing rule requires no assumption about agent wealth or utility functions beyond rationality. Any agent who prefers riding to driving will adopt it whenever $p$ is
finite. The collapse of the driver pool follows directly: since
$\lambda_i^t$ is always sufficient to maintain rider eligibility,
no rational agent ever needs to drive unless they choose to,
and the equilibrium constraint $|D_t|/|R_t| \geq L > 1$
can no longer be guaranteed by the altruism mechanism alone. \textit{The driver pool collapse is now a logical consequence of the purchasing rule, not just an empirical claim.}
% ============================================================
\subsection{Extension 2: Premium Membership}
% ============================================================

Suppose a tiered membership model grants premium subscribers
priority in rider matching.
Let $\mathcal{P}^t \subseteq A$ denote premium agents,
and let $\mu > 1$ be the priority multiplier applied to
their selection probability.
The MARL reward in Equation~\ref{eq:reward} is unchanged, but the
effective selection probability of a premium rider
$a_j \in \mathcal{P}^t$ relative to a standard rider
$a_k \notin \mathcal{P}^t$ with equal altruism score becomes:
\begin{equation}
  \frac{P(\text{select}\; a_j)}{P(\text{select}\; a_k)}
  \;=\;
  \mu \cdot
  \frac{s_j^t}{s_k^t}
  \;=\; \mu
  \qquad \text{when } s_j^t = s_k^t
\end{equation}
The multiplier $\mu$ is independent of contribution history.
A premium agent with $s_j^t = 0.21$ receives $\mu$ times the
matching probability of a high-contributing standard agent
with $s_k^t = 0.90$.
Payment overrides contribution.

\paragraph{\textbf{Fairness breaks.}}
The Benefit Distribution Analysis measures
inequality using Gini coefficients over the agent population.
In the closed economy, the Benefit Gini for distance savings
is already $\sim$0.35. This implies that moderate inequality arises naturally
from variation in contribution levels.
Under premium membership, distance benefits concentrate
among subscribers by construction.
More critically, this inequality is \textbf{structural and
permanent}: a standard agent cannot recover their relative
position through contribution alone, because $\mu$ is not
earned.
In the closed economy, a low-scoring agent can always improve
by driving.
Under premium membership, they cannot.
The self-correcting property of the system is removed.

\paragraph{\textbf{Driver supply erodes.}}
Standard agents observe that high contribution does not
produce proportionally better outcomes --- premium agents
are always prioritized ahead of them regardless of score.
The marginal value of driving, already non-monetary, falls
further.
This erodes voluntary driver participation for the same
structural reason as purchasable points: the link between
contribution and benefit is severed.

To see this formally, consider the altruism-based reward in
Equation~\ref{eq:reward}.
The altruism gain term $\Delta s_i^{t+1}$ is the driver's
primary non-monetary incentive.
If the score $s_i^{t+1}$ it produces translates into worse
matching outcomes than a premium subscription would, the
rational agent stops driving and subscribes instead.
The volunteer driver pool collapses from the top down:
the most mobile agents, exactly those the system most
needs as drivers are the most likely to afford a premium
subscription and exit the driver pool.

% ============================================================
\subsection{The Closed Economy as an Invariant}
% ============================================================

Both failure modes share the same structure.
Each introduces an external resource (money) that substitutes
for the internal resource (contribution) used to allocate
rights and rewards.
Once substitution is possible, $s_i^t$ loses its signal
value, and the behavioral incentives (reciprocity, voluntary driving,
self-correcting fairness) the closed economy
was designed to produce unfold. 

This is a specific instance of Goodhart's Law:
\textit{when a measure becomes a target, it ceases to be a
good measure.}
The altruism score functions as a reliable measure of
community contribution precisely because it cannot be
targeted by financial means.

ORACLE therefore enforces the following invariant across
all components of the model:
\begin{equation}
  \lambda_i^t \;=\; 0
  \qquad \forall\, a_i \in A,\quad \forall\, t \in [1, T]
\end{equation}
No external injection of altruism points is permitted.
Membership tiers, promotional credits, and referral bonuses
are all excluded by Equation~(D), even when they might
appear beneficial for user recruitment.
The only path to a high altruism score is driving.
The only consequence of a low altruism score is being
assigned to drive.

This circularity is not a limitation.
It is the mechanism by which the community sustains itself.

% ============================================================
\subsection{Connection to System Stability}
% ============================================================

The equilibrium constraint in Equation~(6) states:
\begin{align}
  \frac{|D_t|}{|R_t|} \;\in\; [L,\, H]
  \quad \text{with } L > 1
  \label{eq:equilibrium}
\end{align}
This is verified empirically in Section~5.1 via 99\%
confidence intervals on the time series data.
The invariant in Equation~(D) is the \emph{mechanism} that
makes this empirical result possible.
If we remove it  through either purchasable points or premium membership, there is no longer any guarantee that
the driver supply is self-replenishing.
The equilibrium in Equation~\ref{eq:equilibrium} ceases to hold, and the
confidence intervals in \autoref{tab:ratio} become meaningless.

The closed economy is therefore not merely a design choice
about fairness.
It is a \textbf{precondition for the system's measurable
stability properties}.

\bigskip
\noindent\textbf{Remark 2.}
Empirically demonstrating these failure modes,  simulating
a purchasable-points regime and measuring the collapse of
the driver-rider ratio via Equation~(B), or simulating
premium membership and tracking the Gini coefficient
trajectory relative to the closed-economy baseline would strengthen this argument further.
We leave this as a direction for future work.
The theoretical argument is sufficient to motivate the
design invariant for the present system.

\begin{table}[t]
\centering
\caption{99\% confidence intervals for the driver--rider ratio under different agent distributions.}
\label{tab:ratio}

\begin{tabular}{clcccc}
\toprule
$\mathbf{N}$ & \textbf{Distribution} & \textbf{Mean $\pm$ SD} & \textbf{CV} & $\mathbf{CI_{low}}$ & $\mathbf{CI_{high}}$ \\
\midrule

\multirow{4}{*}{100}
 & Gaussian + B/D   & $1.14\pm0.25$ & 0.21 & \textbf{1.07} & 1.20 \\
 & Gaussian, fixed  & $1.21\pm0.21$ & 0.17 & \textbf{1.16} & 1.27 \\
 & Uniform + B/D    & $1.13\pm0.26$ & 0.23 & \textbf{1.06} & 1.20 \\
 & Uniform, fixed   & $1.18\pm0.20$ & 0.17 & \textbf{1.13} & 1.23 \\

\midrule

\multirow{4}{*}{150}
 & Gaussian + B/D   & $1.14\pm0.16$ & 0.14 & \textbf{1.10} & 1.18 \\
 & Gaussian, fixed  & $1.22\pm0.17$ & 0.14 & \textbf{1.17} & 1.26 \\
 & Uniform + B/D    & $1.13\pm0.17$ & 0.15 & \textbf{1.09} & 1.18 \\
 & Uniform, fixed   & $1.22\pm0.20$ & 0.16 & \textbf{1.16} & 1.27 \\

\midrule

\multirow{4}{*}{200}
 & Gaussian + B/D   & $1.12\pm0.14$ & 0.12 & \textbf{1.08} & 1.16 \\
 & Gaussian, fixed  & $1.15\pm0.14$ & 0.12 & \textbf{1.11} & 1.19 \\
 & Uniform + B/D    & $1.12\pm0.16$ & 0.14 & \textbf{1.07} & 1.16 \\
 & Uniform, fixed   & $1.15\pm0.13$ & 0.11 & \textbf{1.11} & 1.18 \\

\bottomrule
\end{tabular}

\vspace{3pt}
{\scriptsize
SD: Standard deviation;\;
CV: Coefficient of variation;\;
CI: Confidence interval;\;
B/D: with birth–death dynamics.
}

\end{table}

%% ===================================================================
\section{ORACLE: Algorithm}
\label{sec:oracle}
%% ===================================================================

This section presents the full pseudocode for ORACLE
(One-network Actor--Critic for Learning in Cooperative Environments).
Algorithm~\ref{alg:oracle} describes the core training step;
Algorithm~\ref{alg:role_assignment} describes the daily role assignment
procedure; Algorithm~\ref{alg:ride_sharing} unifies the full multi-day
loop across both operational modes.

\subsection{Ego-Conditioned Attention Critic}

The critic $Q_\phi$ evaluates the joint observation--action pair
$(\mathbf{s}, \mathbf{u})$ from the perspective of a specified ego
agent $i$ using three separate encoding paths:
\begin{itemize}[leftmargin=*]
    \item \textbf{Query} -- ego agent's observation only, via a
          state encoder:
          $e_i^{\text{state}} = f_{\text{state}}(s_i;\,\phi)$.
    \item \textbf{Keys / Values} -- all other agents' joint
          (observation, action) pairs via a shared encoder:
          $e_j^{\text{sa}} = f_{\text{sa}}([s_j;\,u_j];\,\phi)$,
          $j \neq i$.
    \item \textbf{Ego action} -- embedded separately and concatenated
          with the state embedding and attended features before the
          final MLP:
          $e_i^{\text{act}} = f_{\text{act}}(u_i;\,\phi)$.
\end{itemize}
Scaled dot-product multi-head attention with temperature $T$ aggregates
the other-agent embeddings relative to the ego query.  The scalar
Q-value is:
\begin{equation}
    Q_\phi(\mathbf{s}, \mathbf{u} \mid \text{ego}{=}i)
    = \text{MLP}_\phi\!\Big(
        \big[\,
          e_i^{\text{state}};\;
          \text{MHA}_\phi\!\big(e_i^{\text{state}},\,
            \{e_j^{\text{sa}}\}_{j \neq i};\,T\big);\;
          e_i^{\text{act}}
        \,\big]
      \Big)
\end{equation}
A \emph{single} weight set $\phi$ serves all $N$ agent perspectives.
Agent identity is conveyed structurally via the ego index, not through
learned embeddings, making the parameter count $O(1)$ in $N$ and
enabling zero-shot generalisation to unseen population sizes.

\subsection{Deployment and Continual Adaptation}

In deployment, $\pi_\theta$ acts as a \emph{recommendation policy}:
it proposes a greedy action to each driver, but the human driver
decides whether to follow the suggestion.  Drivers who deviate from
the recommendation---picking a different rider, delaying, or
ignoring the suggestion entirely---generate \emph{off-policy
transitions} that the trained policy would never produce on its own.
These transitions are stored in $\mathcal{D}$ alongside compliant
ones.  PER automatically up-weights high-TD-error transitions,
which predominantly arise from human deviations, focusing adaptation
on exactly those states where the policy is most surprised by real
human behavior.  This human-in-the-loop mechanism provides genuine
off-policy experience beyond the greedy frontier, enabling policy
improvement under real-world distribution shift without any
engineered exploration that could harm service quality.

\begin{algorithm}[ht!]
\renewcommand{\arraystretch}{1.3}
\caption{ORACLE -- Core Training Step}\label{alg:oracle}
\small
\begin{algorithmic}[1]
    \STATE \textbf{Require:} Learning rates $\alpha_\pi, \alpha_Q$,
           discount $\gamma$, soft-update rate $\tau$,
           regularisation coefficient $\lambda$,
           exploration strategy
           $\mathcal{E} \in \{
             \epsilon\text{-greedy},\;\text{TAGS},\;\text{GST}\}$
    \STATE \textbf{Initialise:} Shared actor $\pi_\theta$,
           shared ego-conditioned attention critic $Q_\phi$
    \STATE \phantom{\textbf{Initialise:}} Target networks
           $\pi_{\theta'} \gets \pi_\theta$,\;
           $Q_{\phi'} \gets Q_\phi$
    \STATE \phantom{\textbf{Initialise:}} Prioritised replay buffer
           $\mathcal{D}$ with parameters
           $(\alpha_{\mathcal{D}},\,\beta_{\mathcal{D}})$
    \REPEAT
        \STATE Observe joint state
               $\mathbf{s} = (s_1, \dots, s_N)$
               and masks
               $\mathbf{m} = (\mathbf{m}_1, \dots, \mathbf{m}_N)$
        \FOR{each agent $i = 1, \dots, N$}
            \STATE $\ell_i \gets \pi_\theta(s_i) \odot \mathbf{m}_i$
                   \COMMENT{Masked logits}
            \STATE $u_i \gets \mathcal{E}(\ell_i)$
                   \COMMENT{Action via exploration strategy}
        \ENDFOR
        \STATE Execute $\mathbf{u}$;\;
               observe $\mathbf{r},\,\mathbf{s}',\,\mathbf{m}'$
        \STATE Store $(\mathbf{s}, \mathbf{u}, \mathbf{r},
               \mathbf{s}', \mathbf{m}, \mathbf{m}')$ in $\mathcal{D}$
        \IF{$|\mathcal{D}| \ge B$ \textbf{and} end of episode}
            \STATE Sample prioritised mini-batch
                   $\mathcal{B} =
                     \{(\mathbf{s}^{(k)},\,\mathbf{u}^{(k)},\,
                        \mathbf{r}^{(k)},\,\mathbf{s}'^{(k)},\,
                        w^{(k)})\}$ from $\mathcal{D}$
            \STATE \COMMENT{\textbf{--- Critic update
                   (all ego perspectives) ---}}
            \STATE $\hat{u}_j' \gets
                    \text{softmax}(\pi_{\theta'}(s_j')
                    \odot \mathbf{m}_j')$,\; $\forall j$
            \FOR{ego $i = 1, \dots, N$}
                \STATE $y_i \gets r_i
                        + \gamma(1 - \text{done})\,
                          Q_{\phi'}(\mathbf{s}',
                          \hat{\mathbf{u}}' \mid \text{ego}{=}i)$
                \STATE $\delta_i \gets
                        Q_\phi(\mathbf{s}, \mathbf{u}
                        \mid \text{ego}{=}i) - y_i$
            \ENDFOR
            \STATE $\mathcal{L}_Q \gets
                    \tfrac{1}{N}\sum_{i=1}^{N}
                    \mathbb{E}_\mathcal{B}[w\cdot\delta_i^2]$
            \STATE $\phi \gets \phi
                    - \alpha_Q\,\nabla_\phi\mathcal{L}_Q$
            \STATE \COMMENT{\textbf{--- Actor update
                   (shared policy, all perspectives) ---}}
            \STATE $\tilde{u}_j \gets \mathcal{E}_\theta(s_j,
                    \mathbf{m}_j)$, $\forall j$
            \STATE $\mathcal{L}_\pi \gets
                    -\tfrac{1}{N}\sum_{i=1}^{N}
                    \mathbb{E}_\mathcal{B}\!\big[
                      w\cdot Q_\phi(\mathbf{s},\tilde{\mathbf{u}}
                      \mid \text{ego}{=}i)\big]
                    + \lambda\|\theta\|_2^2$
            \STATE $\theta \gets \theta
                    - \alpha_\pi\,\nabla_\theta\mathcal{L}_\pi$
            \STATE \COMMENT{\textbf{--- Target network soft update ---}}
            \STATE $\theta' \gets \tau\theta + (1-\tau)\theta'$;\quad
                   $\phi' \gets \tau\phi + (1-\tau)\phi'$
            \STATE Update $\mathcal{D}$ priorities:
                   $\tfrac{1}{N}\sum_i|\delta_i|$
        \ENDIF
        \STATE Anneal $\mathcal{E}$,\;
               $\alpha_{\mathcal{D}}$,\;
               $\beta_{\mathcal{D}}$,\; and learning rates
    \UNTIL{convergence}
\end{algorithmic}
\end{algorithm}

\begin{algorithm}[ht!]
\caption{Daily Role Assignment}\label{alg:role_assignment}
\small
\begin{algorithmic}[1]
    \STATE \textbf{Input:} Active agents $\mathcal{A}_{\text{active}}$,
           altruism scores
           $\{s_i^d\}_{i \in \mathcal{A}_{\text{active}}}$
    \STATE \textbf{Output:} Role map
           $\mathbf{M}_i \in \{\text{driver},\,\text{rider}\}$,
           $\forall i$
    \STATE $s_{\max}^d \gets \max_i\,s_i^d$
    \FOR{each agent $i \in \mathcal{A}_{\text{active}}$}
        \IF{$s_i^d \le 0.2$}
            \STATE $\mathbf{M}_i \gets \text{driver}$
                   \COMMENT{Low altruism $\Rightarrow$ driver}
        \ELSE
            \STATE Draw $v_1 \sim \mathcal{U}(0,1)$
            \IF{$v_1 < 0.1$}
                \STATE $\mathbf{M}_i \gets
                        \textbf{random}(\text{driver},\,\text{rider})$
                        \COMMENT{Role diversification}
            \ELSE
                \STATE $p_{\text{rider}} \gets s_i^d / s_{\max}^d$
                \STATE Draw $v_2 \sim \mathcal{U}(0,1)$
                \STATE $\mathbf{M}_i \gets \begin{cases}
                    \text{rider},  & v_2 < p_{\text{rider}} \\
                    \text{driver}, & \text{otherwise}
                \end{cases}$
            \ENDIF
        \ENDIF
    \ENDFOR
    \STATE \textbf{return} $\mathbf{M}$
\end{algorithmic}
\end{algorithm}

\begin{algorithm}[ht!]
\caption{Multi-Day Altruistic Ride-Sharing (ARS)}\label{alg:ride_sharing}
\small
\begin{algorithmic}[1]
    \STATE \textbf{Require:} Days $D$,
           ORACLE networks $(\pi_\theta, Q_\phi)$,
           mode $\in \{\textsc{train},\;\textsc{deploy{+}adapt}\}$,
           replay buffer $\mathcal{D}$,
           batch size $B$
    \STATE \textbf{If train:} exploration strategy $\mathcal{E}$,
           episodes per day $E$
    \STATE Initialise $\mathcal{A}$ with altruism
           $\{s_i^0\}_{i \in \mathcal{A}}$
    \FOR{each day $d = 1, \dots, D$}
        \STATE $\mathcal{A}_{\text{active}} \gets
               \textsc{UpdatePopulation}(\mathcal{A}, d)$
               \COMMENT{Birth / dropout
               (Eq.~\ref{eq:n_active}--\ref{eq:n_dropout})}
        \STATE $\mathbf{M} \gets
               \textsc{AssignRoles}(\mathcal{A}_{\text{active}},
               \{s_i^d\})$
               \COMMENT{Alg.~\ref{alg:role_assignment}}
        \STATE \textbf{Set} $E_d \gets E$ if mode $=$ \textsc{train},
               else $E_d \gets 1$
               \COMMENT{One real operational day in deployment}
        \IF{mode $=$ \textsc{train}}
            \STATE Reset $\mathcal{E}$ to initial
                   temperature / $\epsilon$
        \ENDIF
        \FOR{episode $e = 1, \dots, E_d$}
            \STATE $\mathbf{s} \gets
                   \textsc{ResetEnvironment}(
                   \mathcal{A}_{\text{active}}, \mathbf{M})$
            \WHILE{not terminated}
                \FOR{each active driver $i \in \mathcal{D}_d$}
                    \STATE Compute $\mathbf{m}_i$ from local
                           $5{\times}5$ perception grid
                    \IF{mode $=$ \textsc{train}}
                        \STATE $u_i \gets
                               \mathcal{E}(\pi_\theta(s_i)
                               \odot \mathbf{m}_i)$
                               \COMMENT{Exploration enabled}
                    \ELSE
                        \STATE $\hat{u}_i \gets
                               \arg\max\,\text{softmax}(
                               \pi_\theta(s_i) \odot \mathbf{m}_i)$
                               \COMMENT{Recommend to human driver}
                        \STATE $u_i \gets$ human decision
                               \COMMENT{May differ from $\hat{u}_i$;
                               human deviations provide off-policy
                               experience}
                    \ENDIF
                \ENDFOR
                \STATE Execute $\mathbf{u}$;\;
                       observe $(\mathbf{s}', \mathbf{r})$
                \STATE $s_i^d \gets
                        \textsc{UpdateAltruism}(s_i^d,
                        u_i, \mathbf{r})$,\; $\forall i$
                \STATE Store $(\mathbf{s}, \mathbf{u},
                       \mathbf{r}, \mathbf{s}',
                       \mathbf{m}, \mathbf{m}')$
                       in $\mathcal{D}$
                \STATE Accumulate metrics;\quad
                       $\mathbf{s} \gets \mathbf{s}'$
            \ENDWHILE
        \ENDFOR
        \IF{$|\mathcal{D}| \ge B$}
            \STATE Run Alg.~\ref{alg:oracle} (lines 12--28)
                   \COMMENT{\textsc{deploy+adapt}: off-policy update
                   from human-generated experience; PER
                   up-weights high-error deviation transitions;
                   $O(1)$ cost in $N$}
            \IF{mode $=$ \textsc{train}}
                \STATE Anneal $\mathcal{E}$
            \ENDIF
        \ENDIF
        \STATE $\{s_i^{d+1}\} \gets \{s_i^d\}$
               \COMMENT{Persist altruism across days}
    \ENDFOR
    \STATE \textbf{Output:} Performance metrics,
           evolved altruism $\{s_i^D\}$
\end{algorithmic}
\end{algorithm}

%% ===================================================================
\section{State, Action, and Reward Details}
\label{sec:env_details}
%% ===================================================================

\subsection{Observation Space}

Each agent $a_i$ observes a structured state vector
$s_i \in \mathbb{Z}^{28}$ composed of three components:

\begin{itemize}[leftmargin=*]

\item \textbf{Spatial coordinates:}
The $(x_i, y_i)$ position of the agent in the discrete grid
environment.

\item \textbf{Role identifier:}
A binary variable $\text{role}_i \in \{0,1\}$ indicating the
agent's current role (0 = rider, 1 = driver).

\item \textbf{Local perception field:}
A $5 \times 5$ observation grid centered on the agent that
encodes nearby rider locations and local environmental context.
Internally this grid is stored as a flattened vector
$\text{grid}_i \in \{-1,0,\dots,N{-}1\}^{25}$, where each cell
contains the ID of a rider occupying that location, or $-1$ if
the cell is empty.
\end{itemize}

Together these components form a 28-dimensional observation vector.

\subsection{Action Space}

The action space for each agent is discrete:
\[
u_i \in \{0,1,\dots,N\}
\]

where $u_i = j$ ($j < N$) corresponds to selecting rider $a_j$
for pickup and $u_i = N$ represents declining all available
requests (no-op).

To ensure feasibility, a dynamic action mask $\mathbf{m}_i$
restricts available actions to riders visible in the local
$5 \times 5$ perception grid, together with the no-op action.
This prevents invalid assignments while maintaining
computational efficiency.

\subsection{Reward Formulation}

The reward received by driver $a_i$ for selecting rider $a_j$
is defined as:

\begin{equation}
r_i =
\tanh\!\Big(
\alpha_r \cdot d_j(\emptyset)
-
(1-\alpha_r) \cdot \Delta d_i(a_j)
+
\Delta s_i^+
\Big)
\end{equation}

where $d_j(\emptyset)$ denotes the standalone trip distance of
rider $a_j$, $\Delta d_i(a_j)$ represents the additional detour
incurred by driver $a_i$ to serve that rider, and
$\Delta s_i^+$ corresponds to the altruism score gained by the
driver for completing the ride.

Driver detour distances $\Delta d_i(a_j)$ are computed using
Dijkstra’s shortest-path algorithm on the weighted road grid.

Drivers that observe nearby riders but choose the no-op action
receive a small baseline reward $\tanh(0.1)$ to discourage
persistent inactivity.

\subsection{Hyperparameter Optimization and Exploration Strategy}

Training hyperparameters were selected using Bayesian hyperparameter
optimization to efficiently explore the parameter space while limiting
the number of training runs required. The search procedure evaluated
configurations over learning rates, replay buffer parameters, actor and
critic network sizes, and exploration schedules. Candidate configurations
were sampled using a Bayesian optimization framework that iteratively
refined the search distribution based on observed validation performance.
The final configuration used in all experiments is summarized in
Table~\ref{tab:hyperparams}.

\paragraph{\textbf{Exploration Strategy.}}
Exploration during training is performed using \\
\emph{Temperature-Annealed Gumbel-Softmax (TAGS)},
which provides a differentiable relaxation of categorical
action sampling. Instead of directly sampling discrete
actions, the policy samples from a Gumbel-Softmax
distribution whose temperature is gradually annealed
during training, allowing the distribution to transition
from smooth exploratory behavior to near-deterministic
action selection.

We also evaluated the Gapped Straight-Through (GST) estimator, in which actions are selected via argmax in the forward pass, while gradients are computed through a deterministic perturbation of the logits in the backward pass. The perturbation enforces a minimum gap between the selected and unselected logits, reducing gradient variance compared to the standard Gumbel-Softmax approach.

While~\cite{tilbury2023revisitinggumbelsoftmaxmaddpg} found GST superior in their benchmarks, in our preliminary experiments TAGS provided more stable learning, possibly due to differences in environment complexity or network architecture. Therefore,
all reported experiments use TAGS with the temperature
annealed from $1.0$ to $0.01$ over the first $80\%$
of training episodes.

\begin{table}[ht!]
\centering
\caption{Simulation and ORACLE Hyperparameters}
\label{tab:hyperparams}
\renewcommand{\arraystretch}{1.25}
\begin{tabular}{l l l}
\hline
\textbf{Parameter} & \textbf{Symbol} & \textbf{Value} \\
\hline \hline
\multicolumn{3}{l}{\textit{Environment}} \\
\hline
Grid size & $W \times H$ & $15 \times 15$ cells ($3.6 \times 4.2$\,km) \\
Agent population & $N$ & $\{100, 150, 200\}$ \\
Perception radius & --- & $5 \times 5$ grid \\
Vehicle capacity & $l_i$ & $4$ passengers \\
Altruism score range & $s_i^d$ & $[0, 1]$ \\
Reward trade-off & $\alpha_r$ & $0.4$ \\
Altruism scaling (driver) & $\alpha_s$ & $0.5$ \\
Altruism scaling (rider) & $\beta_s$ & $0.7$ \\
No-op reward & --- & $0.1$ \\
\hline
\multicolumn{3}{l}{\textit{ORACLE architecture}} \\
\hline
Actor hidden layers & --- & $256 \to 512$ \\
Critic hidden layers & --- & $256 \to 512$ \\
Attention embedding dim & --- & $128$ \\
Number of attention heads & --- & $4$ \\
Attention temperature & --- & $1.0$ \\
\hline
\multicolumn{3}{l}{\textit{Training}} \\
\hline
Critic learning rate & $\alpha_Q$ & $1 \times 10^{-4}$ \\
Actor learning rate & $\alpha_\pi$ & $1 \times 10^{-5}$ \\
Discount factor & $\gamma$ & $0.95$ \\
Soft-update rate & $\tau$ & $0.01$ \\
Regularisation & $\lambda$ & $1 \times 10^{-4}$ \\
Batch size & $B$ & $256$ \\
Replay buffer size & $|\mathcal{D}|$ & $50{,}000$ \\
PER $\alpha_{\mathcal{D}}$ & --- & $0.6$ (constant) \\
PER $\beta_{\mathcal{D}}$ & --- & $0.4 \to 1.0$ (linear) \\
Optimizer & --- & AdamW (clip-norm $0.5$) \\
\hline
\multicolumn{3}{l}{\textit{Exploration (Temperature-Annealed Gumbel Softmax)}} \\
\hline
Temperature (start $\to$ end) & --- & $1.0 \to 0.01$ \\
Decay fraction & --- & $80\%$ of episodes \\
\hline
\end{tabular}
\end{table}

%% ===================================================================
\section{Population Dynamics Model}
\label{sec:pop_dynamics}
%% ===================================================================

The proposed dynamic population model allows agents to enter
(\emph{birth}) and exit (\emph{dropout}) the system over time.  Agents
occupy one of three states: \emph{active}, \emph{dropped-out}, or
\emph{unenrolled}.

\subsection{Dropout Process}

Agent dropout probability is governed by individual altruism scores
$s_i^d$ (Eq.~\ref{eq:death-probability}).  Low-altruism agents exhibit
exponentially higher dropout rates; high-altruism agents maintain stable
participation.

\begin{equation}
\label{eq:death-probability}
P_{\text{dropout}}(s_i^d) =
\begin{cases}
    \alpha_{bd} \cdot e^{-\beta_{bd}\, s_i^d}
        & \text{if } s_i^d < s_{\text{th}}^d \\[4pt]
    \gamma_{bd} - \delta_{bd} \cdot
        \dfrac{s_i^d - s_{\text{th}}^d}
              {s_{\max}^d - s_{\text{th}}^d}
        & \text{if } s_i^d \ge s_{\text{th}}^d
\end{cases}
\end{equation}

\subsection{Birth Process}

Unenrolled agents may join based on a multi-factor probability model
(Eq.~\ref{eq:birth-probability}), where
$\rho(d) = (N - N_{\text{never}}(d)) / N$ is the community adoption
rate.

\begin{equation}
\label{eq:birth-probability}
P_{\text{birth}}(d) =
    P_{\text{base}}(\rho)
    \cdot F_{\text{phase}}(\rho)
    \cdot F_{\text{urgency}}(d)
    \cdot F_{\text{network}}(N_{\text{active}})
    \cdot F_{\text{reputation}}(\bar{s})
\end{equation}

\begin{itemize}[leftmargin=*]
    \item $F_{\text{phase}}$: adoption phase multiplier based on Rogers'
          diffusion theory (innovators $\to$ laggards).
    \item $F_{\text{urgency}}$: time-sensitive pressure to join
          ($F_{\text{urgency}} \in [1.0, 3.0]$).
    \item $F_{\text{network}}$: positive feedback after $30\%$ adoption;
          congestion effects beyond $85\%$.
    \item $F_{\text{reputation}} = 1 + 0.4\,(\bar{s} - 0.5)$: influence
          of mean altruism $\bar{s}$ on newcomer interest.
\end{itemize}

New agents receive initial altruism reflecting behavioural trends across
adoption stages: (1)~early adopters receive higher scores;
(2)~network effects boost scores under high participation;
(3)~scarcity incentives add FOMO bonuses;
(4)~reputation adjustments reflect recent dropout trends.

\subsection{Population Dynamics}

Population counts evolve as:
\begin{align}
    N_{\text{active}}(d{+}1)  &= N_{\text{active}}(d)
        + B(d) - D(d) + R(d) \label{eq:n_active} \\
    N_{\text{never}}(d{+}1)   &= N_{\text{never}}(d)
        - B(d) \label{eq:n_never} \\
    N_{\text{dropout}}(d{+}1) &= N_{\text{dropout}}(d)
        + D(d) - R(d) \label{eq:n_dropout}
\end{align}
where $B(d)$, $D(d)$, and $R(d)$ denote the number of births, dropouts,
and reintegrations on day $d$.  Dropped-out agents retain their altruism
scores and may rejoin with probability $1 - P_{\text{dropout}}(s_i^d)$.

%% ===================================================================
\section{Baseline Implementation Details}
\label{sec:baselines}
%% ===================================================================

\subsection{Particle Swarm Optimisation (PSO)}

A swarm of particles is used, where each particle $\mathbf{x}$
represents a feasible rider-to-driver assignment.  The fitness function
balances rider benefit, driver detour, and altruism:
\begin{equation}
    f(\mathbf{x}) =
        \alpha_r \sum_{j \in \mathcal{R}} d_j(\emptyset)
        - (1 - \alpha_r) \sum_{i \in \mathcal{D}}
          \big(d_i(\phi_i^d \| a_j) - d_i(\phi_i^d)\big)
        + \beta_{\text{pso}} \sum_{(i,j) \in \mathbf{x}} s_{i,j}
\end{equation}

The swarm is initialised using a rider-centric geographic strategy.
Velocity and position updates follow the standard PSO equations:
\begin{align}
    \mathbf{v}_i^{t+1} &= w\,\mathbf{v}_i^t
        + c_1 r_1\,(\mathbf{p}_i - \mathbf{x}_i^t)
        + c_2 r_2\,(\mathbf{g} - \mathbf{x}_i^t) \\
    \mathbf{x}_i^{t+1} &= \mathbf{x}_i^t + \mathbf{v}_i^{t+1}
\end{align}
where $\mathbf{p}_i$ and $\mathbf{g}$ are the personal and global
best-known solutions.  A repair mechanism enforces capacity constraints
after each update.

%% ===================================================================
\section{Performance Metrics}
\label{sec:metrics}
%% ===================================================================

We define the following metrics to evaluate ARS.

\begin{enumerate}[leftmargin=*]

    \item \textbf{Total Distance and Carbon Emissions.}
    \[
        \text{DIST}_{\text{tot}} =
            \sum_{d=1}^{D} \sum_{i=1}^{N}
            \text{travel}(a_i, d),
        \qquad
        \text{CO}_{2} =
            \text{DIST}_{\text{tot}} \times \epsilon_{\text{avg}}
    \]
    where $\epsilon_{\text{avg}}$ is the average emission factor per
    unit distance.

    \item \textbf{Detour Factor.}
    \[
        \text{DET}_{\text{ratio}}(a_i) =
            \sum_{d=1}^{D}
            \frac{d_i(\phi_i^d)}{d_i(\emptyset)}
    \]

    \item \textbf{Average Trip Time.}
    \[
        \text{TIME}_{\text{avg}}(d) =
            \frac{1}{|\mathcal{A}_{\text{active}}|}
            \sum_{a \in \mathcal{A}_{\text{active}}}
            \text{time}(a, d)
    \]

    \item \textbf{Vehicle Utilisation.}
    \[
        \text{UTIL}_{\text{avg}}(d) =
            \frac{|\mathcal{D}_d| +
                  \sum_{a_j \in \mathcal{R}_d}
                  \mathbb{I}(a_j \text{ travelling at } d)}
                 {|\mathcal{D}_d|}
    \]

    \item \textbf{Road Traffic Density.}
    \begin{align*}
        \text{DENSE}(d) &=
            \sum_{c \in \text{Grid}}
            \mathbb{I}\!\big(\text{density}(c) >
            \rho_{\text{threshold}}\big) \\
        \text{density}(c) &=
            \mathbb{E}_{t}\!
            \sum_{i=1}^{N}
            \mathbb{I}\!\big(\text{loc}(a_i, t)
            = c\big)
    \end{align*}

    \item \textbf{Rider Acceptance Rate.}
    \[
        \text{ACCEPT}(d) =
            \frac{\sum_{a_j \in \mathcal{R}_d}
                  \mathbb{I}(a_j \text{ picked up})}
                 {|\mathcal{R}_d|}
    \]

    \item \textbf{Benefit Distribution Analysis.}
    \begin{enumerate}[leftmargin=*]
        \item \textit{3D Lorenz Surface}: cumulative joint distribution
              of personal benefit (distance saved) and community
              contribution (traffic reduction).
        \item \textit{Gini Coefficient}: calculated separately per
              benefit dimension from the 2D projections.
    \end{enumerate}

    \item \textbf{Reintegration Score.}
    \[
        \text{REINT} =
            \alpha\, R_{\text{basic}}
            + \beta\, R_{\text{time}}
            + \gamma\, R_{\text{quick}}
            + \delta\, R_{\text{stable}}
    \]
    \begin{equation}
    \begin{aligned}
        R_{\text{basic}} &= \frac{|\mathcal{R}_{\text{re}}|}
                                  {|\mathcal{D}_{\text{out}}|}
        &\quad
        R_{\text{time}} &= \frac{1}{|\mathcal{R}_{\text{re}}|}
            \sum_{r \in \mathcal{R}_{\text{re}}}
            e^{-\lambda_{\text{re}}(d_r - d_o)} \\
        R_{\text{quick}} &=
            \frac{|\{r \in \mathcal{R}_{\text{re}} :
                    d_r - d_o \le \tau_{\text{re}}\}|}
                 {|\mathcal{R}_{\text{re}}|}
        &\quad
        R_{\text{stable}} &= \frac{|\mathcal{S}|}{|\mathcal{U}|}
    \end{aligned}
    \end{equation}
    where $\mathcal{R}_{\text{re}}$ is the set of reintegration events
    $r = (a_i, d_o, d_r)$,
    $\mathcal{D}_{\text{out}}$ is the set of all dropout events,
    $\mathcal{S}$ is the set of agents who returned and remained stable,
    and $\mathcal{U}$ is the set of unique returning agents.

\end{enumerate}

%% ===================================================================
\section{Additional Experimental Results}

\begin{figure}[H]
\centering
\includegraphics[width=\textwidth]{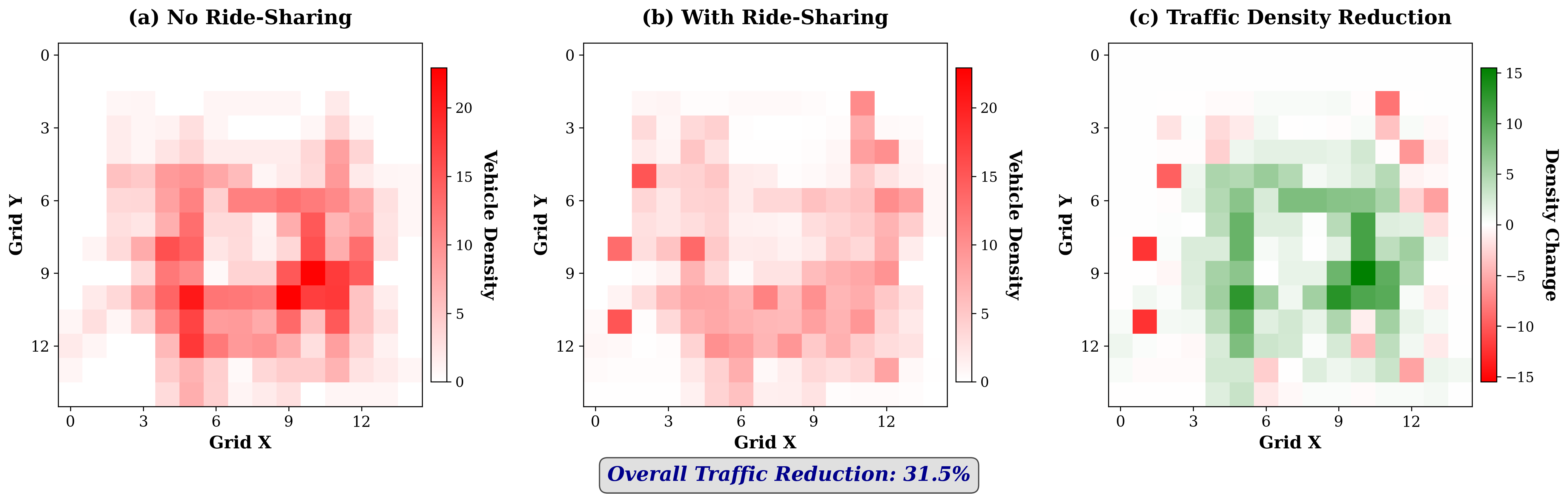}
\vspace{1mm}
\caption{Comparison of traffic density maps illustrating the reduction in urban congestion achieved by the ARS framework. The figure displays vehicle density for (a) the No Ride-Sharing baseline and (b) the With Ride-Sharing scenario. Subplot (c) visualizes the difference, quantifying a significant overall traffic reduction of 31.5\%.}
\label{fig:traffic_density_appendix}
\end{figure}

\begin{figure}[H]
\centering
\begin{subfigure}{0.48\textwidth}
\centering
\includegraphics[width=\linewidth]{{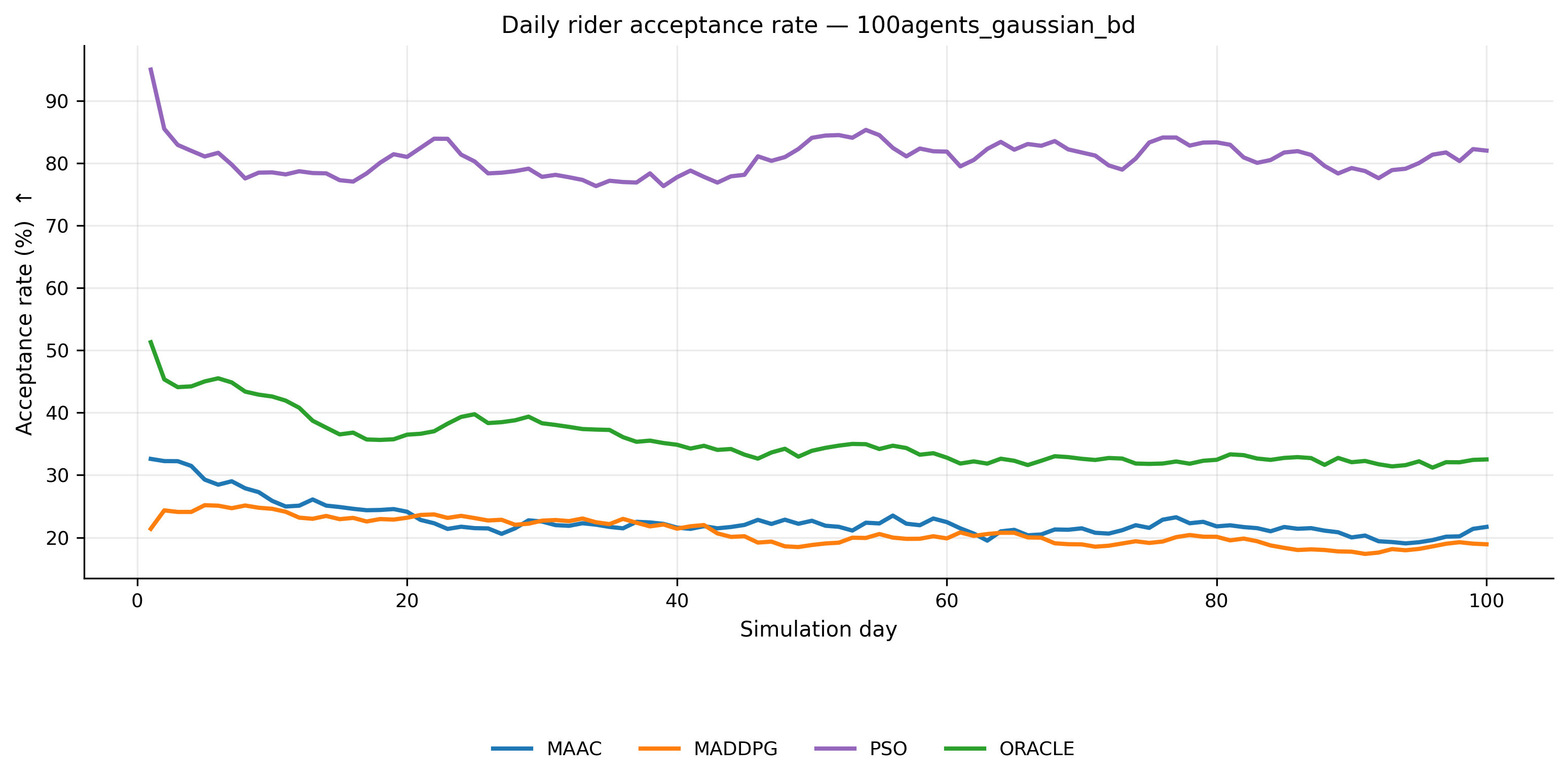}}
\caption{Gaussian allocation; agents may enter and exit (Birth–Death)}
\end{subfigure}
\hfill
\begin{subfigure}{0.48\textwidth}
\centering
\includegraphics[width=\linewidth]{{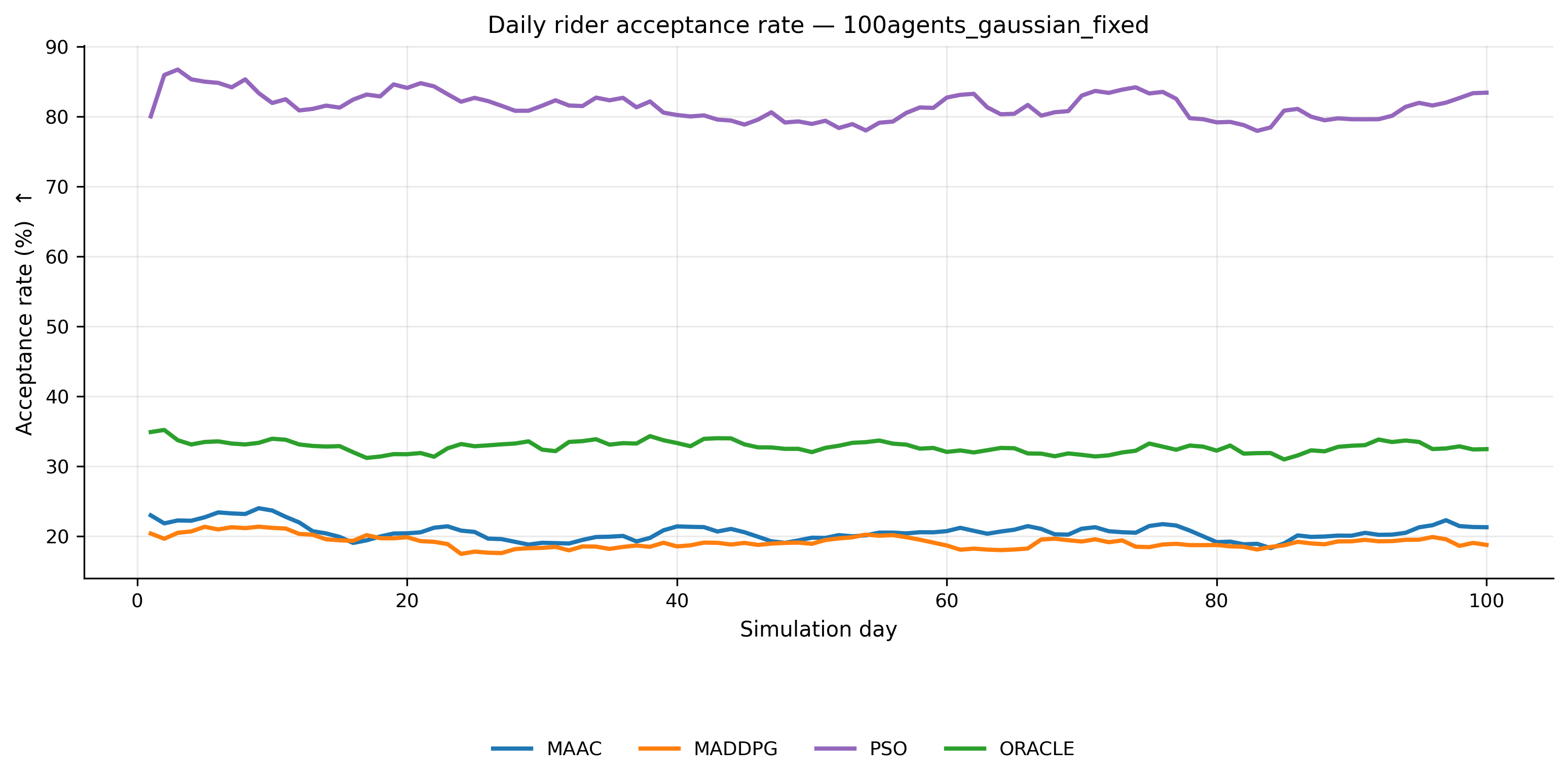}}
\caption{Gaussian allocation; no entry and exit of agents}
\end{subfigure}

\vspace{0.5em}

\begin{subfigure}{0.48\textwidth}
\centering
\includegraphics[width=\linewidth]{{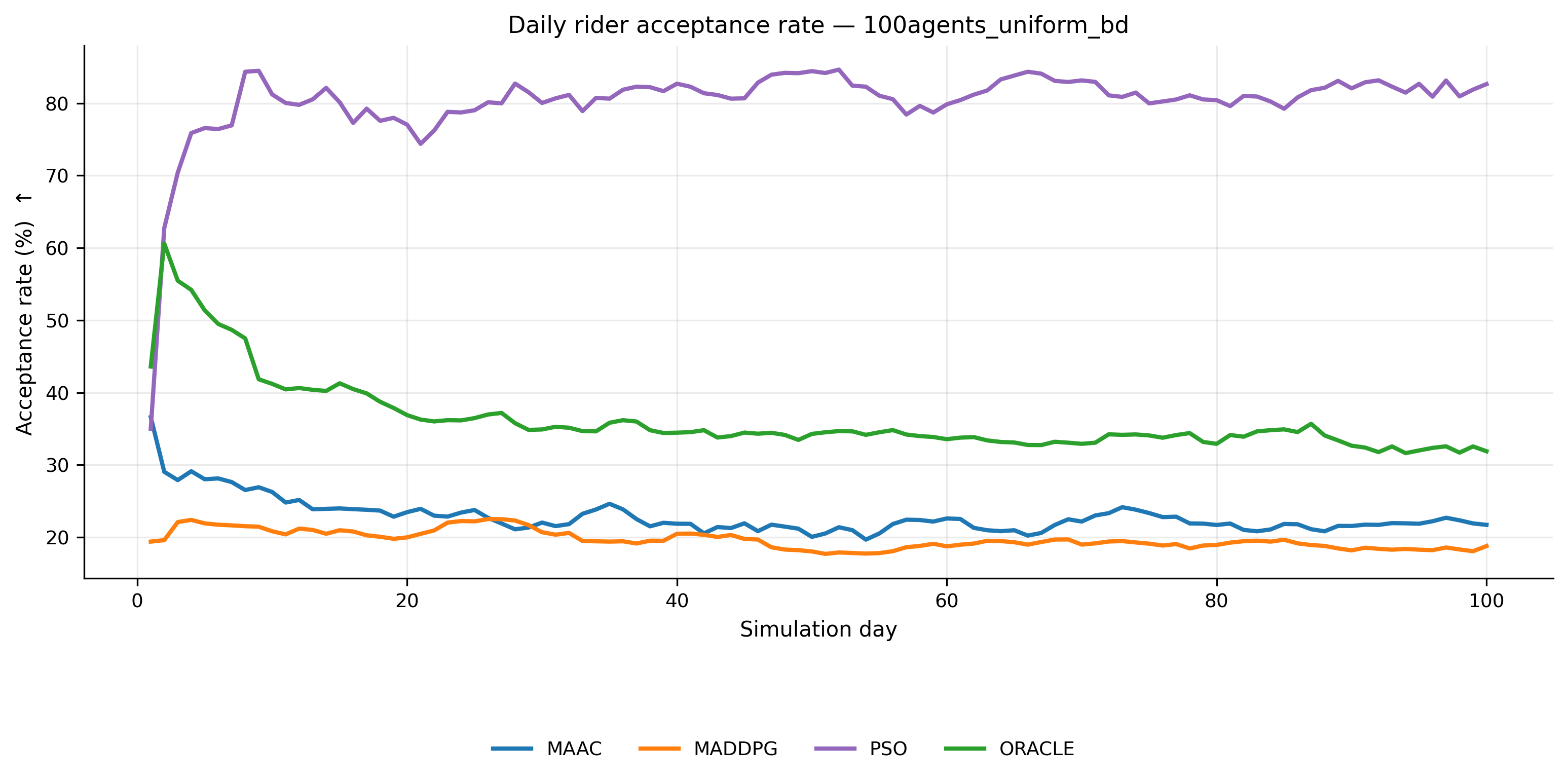}}
\caption{Uniform allocation; agents may enter and exit (Birth–Death)}
\end{subfigure}
\hfill
\begin{subfigure}{0.48\textwidth}
\centering
\includegraphics[width=\linewidth]{{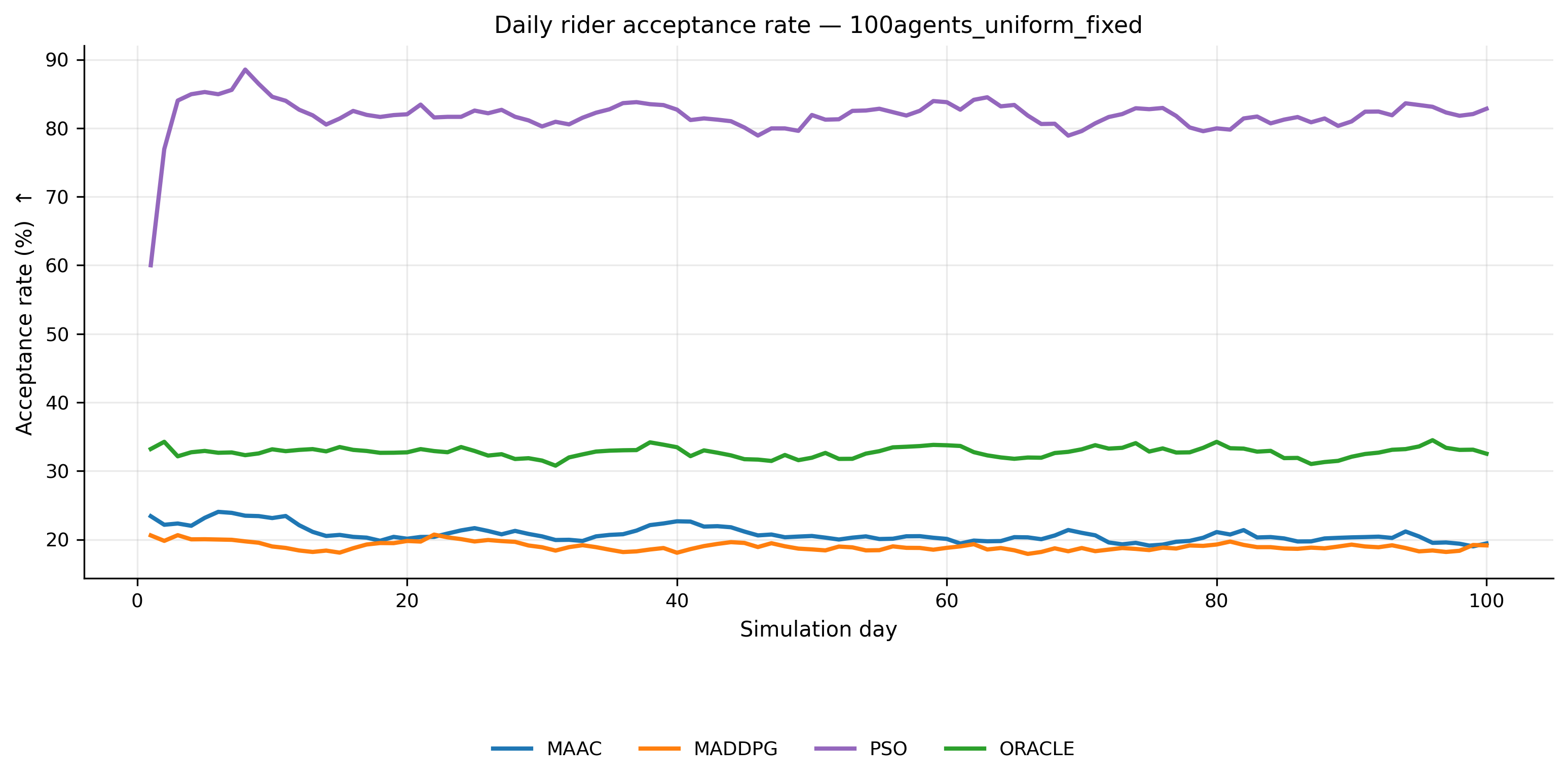}}
\caption{Uniform allocation; no entry and exit of agents}
\end{subfigure}

\caption{Daily acceptance rate with 100 agents.}
\end{figure}

\begin{figure}[H]
\centering
\begin{subfigure}{0.48\textwidth}
\centering
\includegraphics[width=\linewidth]{{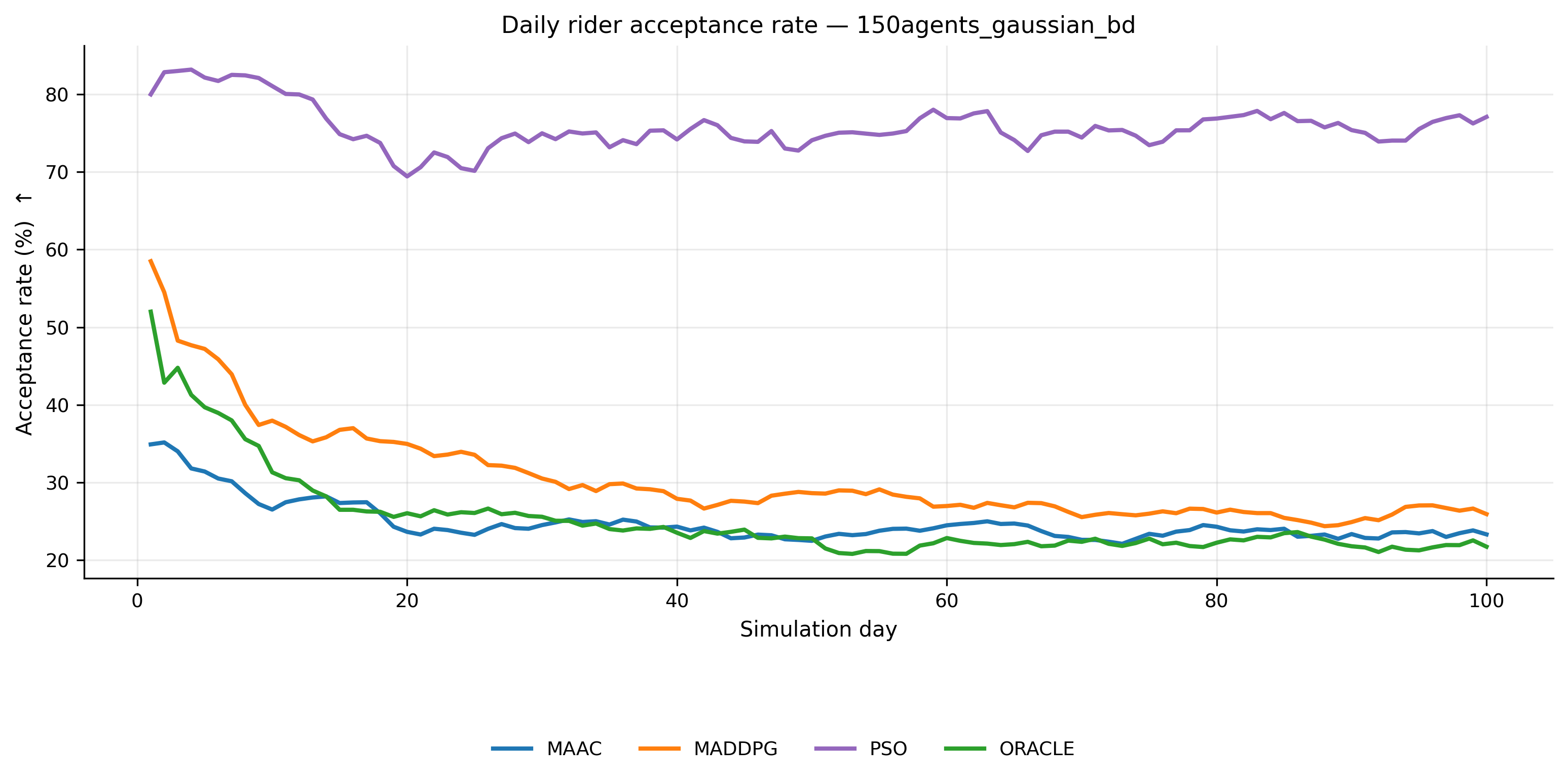}}
\caption{Gaussian allocation; agents may enter and exit (Birth–Death)}
\end{subfigure}
\hfill
\begin{subfigure}{0.48\textwidth}
\centering
\includegraphics[width=\linewidth]{{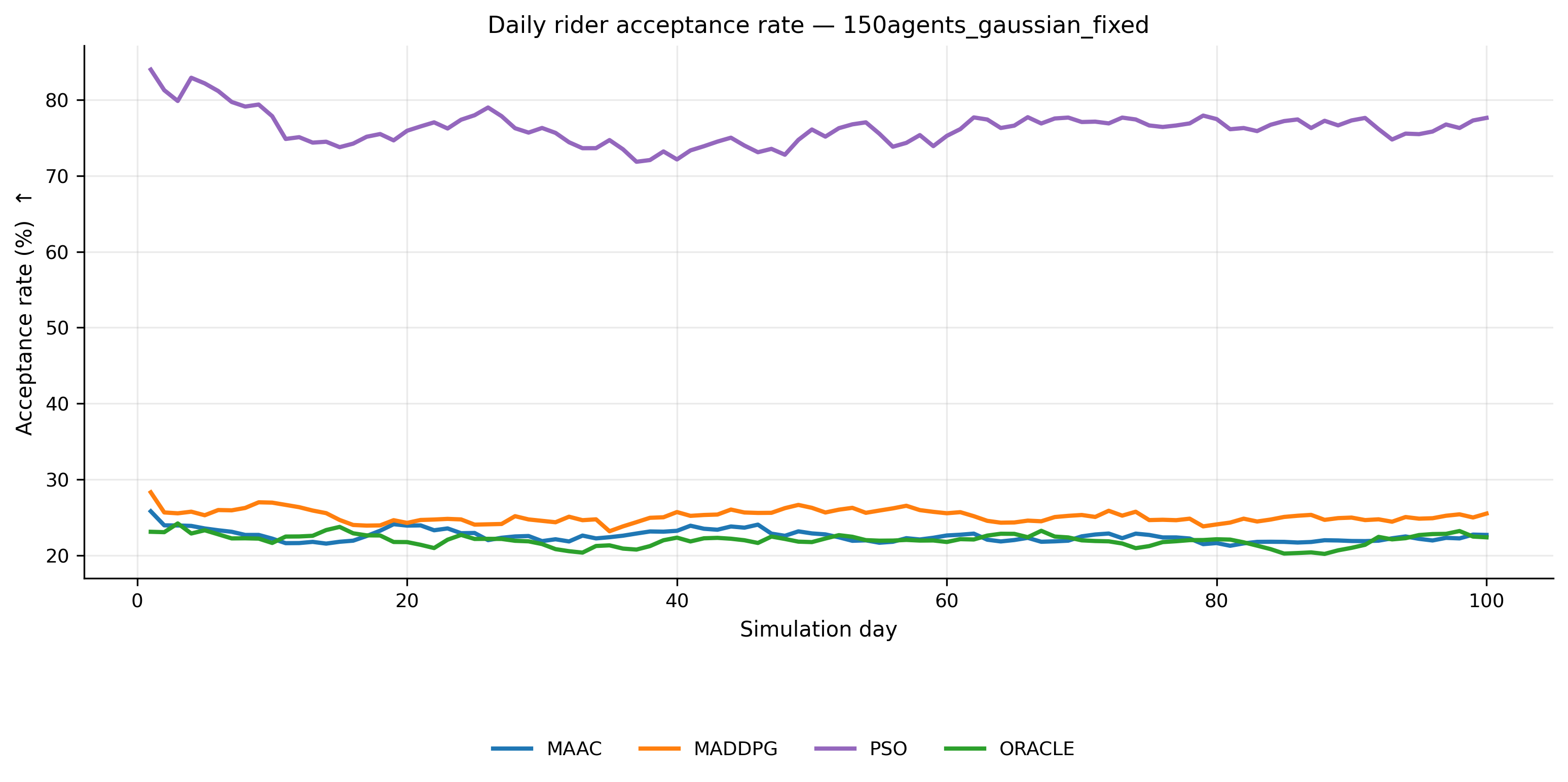}}
\caption{Gaussian allocation; no entry and exit of agents}
\end{subfigure}

\vspace{0.5em}

\begin{subfigure}{0.48\textwidth}
\centering
\includegraphics[width=\linewidth]{{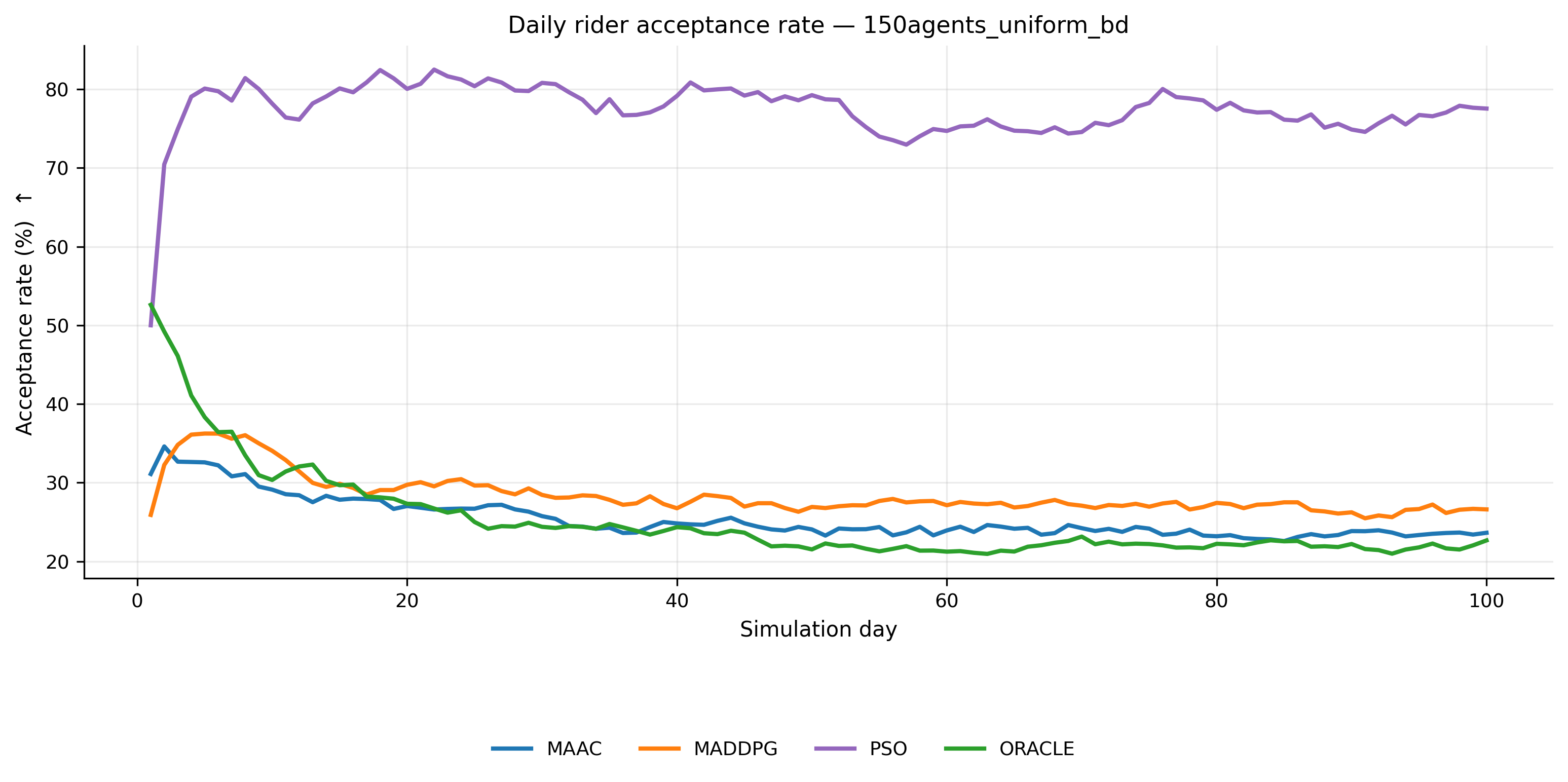}}
\caption{Uniform allocation; agents may enter and exit (Birth–Death)}
\end{subfigure}
\hfill
\begin{subfigure}{0.48\textwidth}
\centering
\includegraphics[width=\linewidth]{{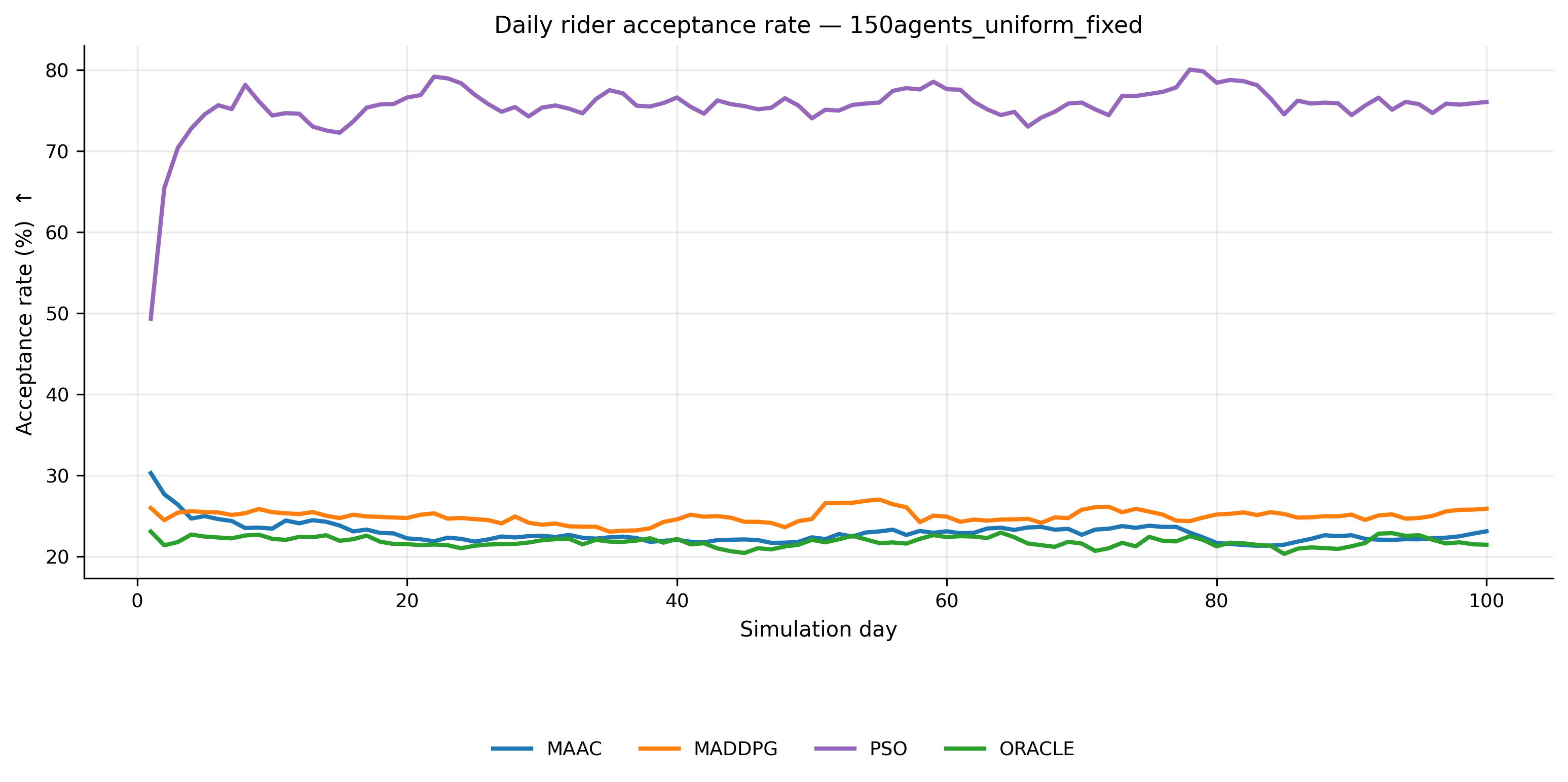}}
\caption{Uniform allocation; no entry and exit of agents}
\end{subfigure}

\caption{Daily acceptance rate with 150 agents.}
\end{figure}

\begin{figure}[H]
\centering
\begin{subfigure}{0.48\textwidth}
\centering
\includegraphics[width=\linewidth]{{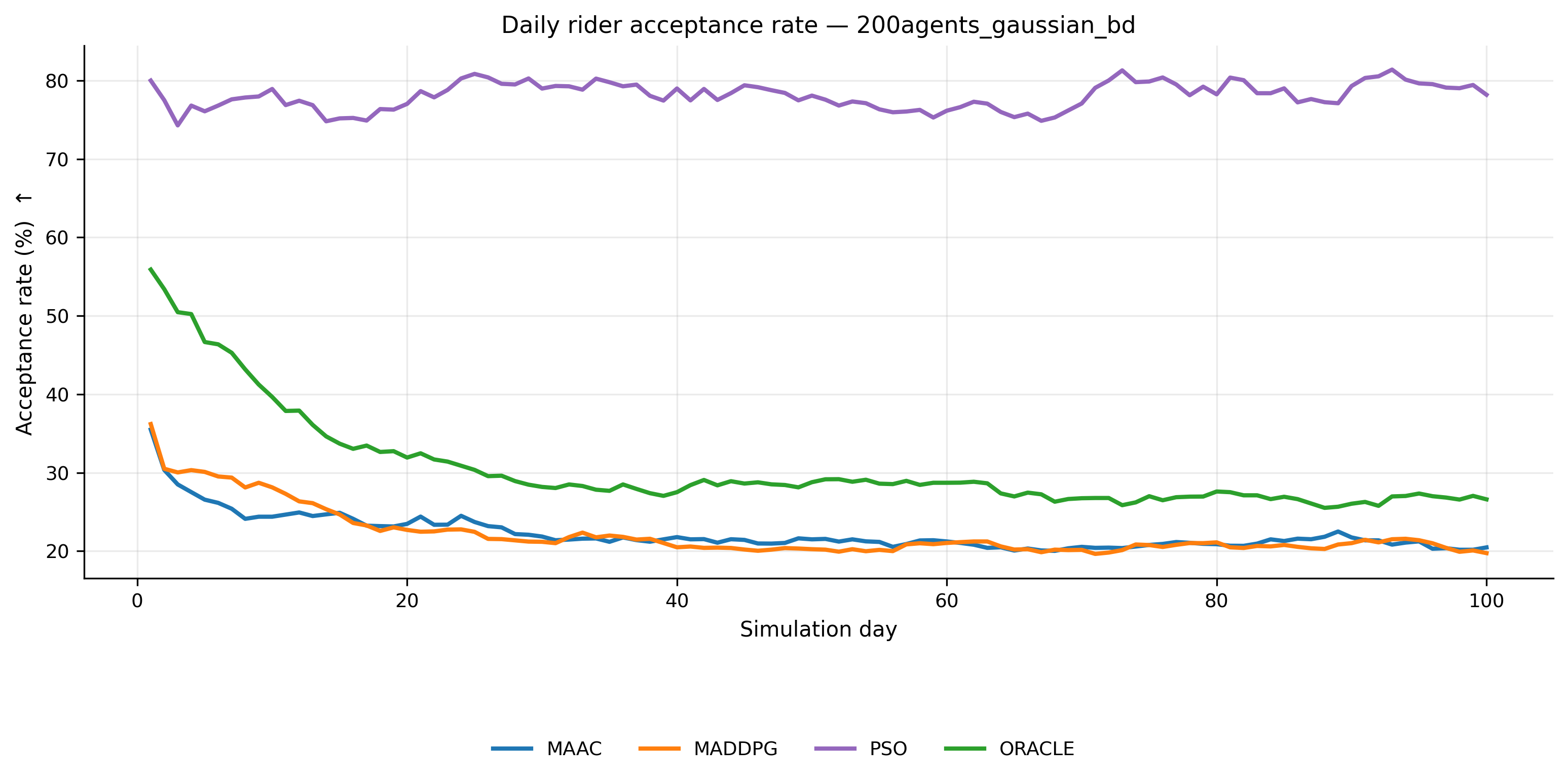}}
\caption{Gaussian allocation; agents may enter and exit (Birth–Death)}
\end{subfigure}
\hfill
\begin{subfigure}{0.48\textwidth}
\centering
\includegraphics[width=\linewidth]{{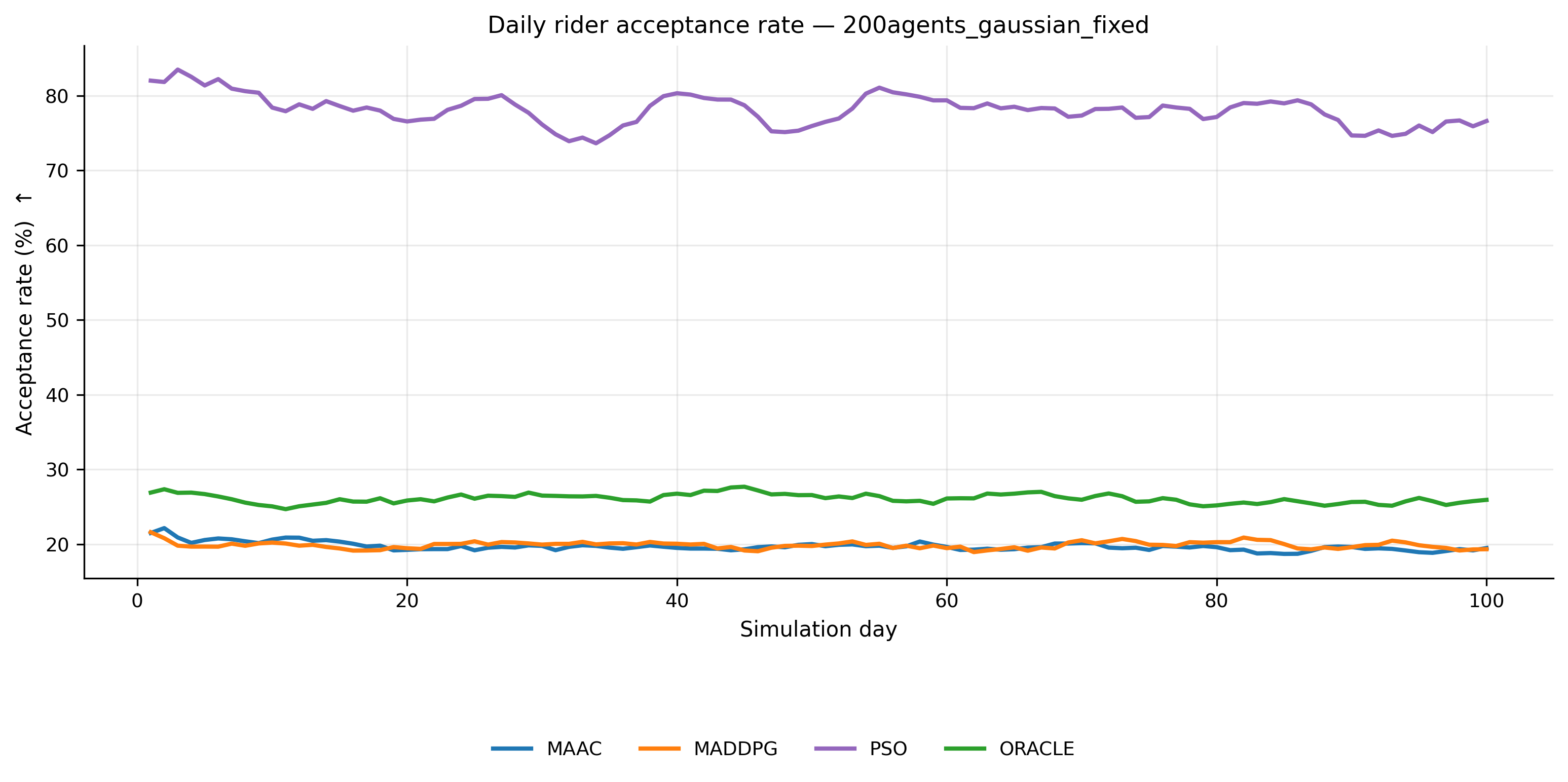}}
\caption{Gaussian allocation; no entry and exit of agents}
\end{subfigure}

\vspace{0.5em}

\begin{subfigure}{0.48\textwidth}
\centering
\includegraphics[width=\linewidth]{{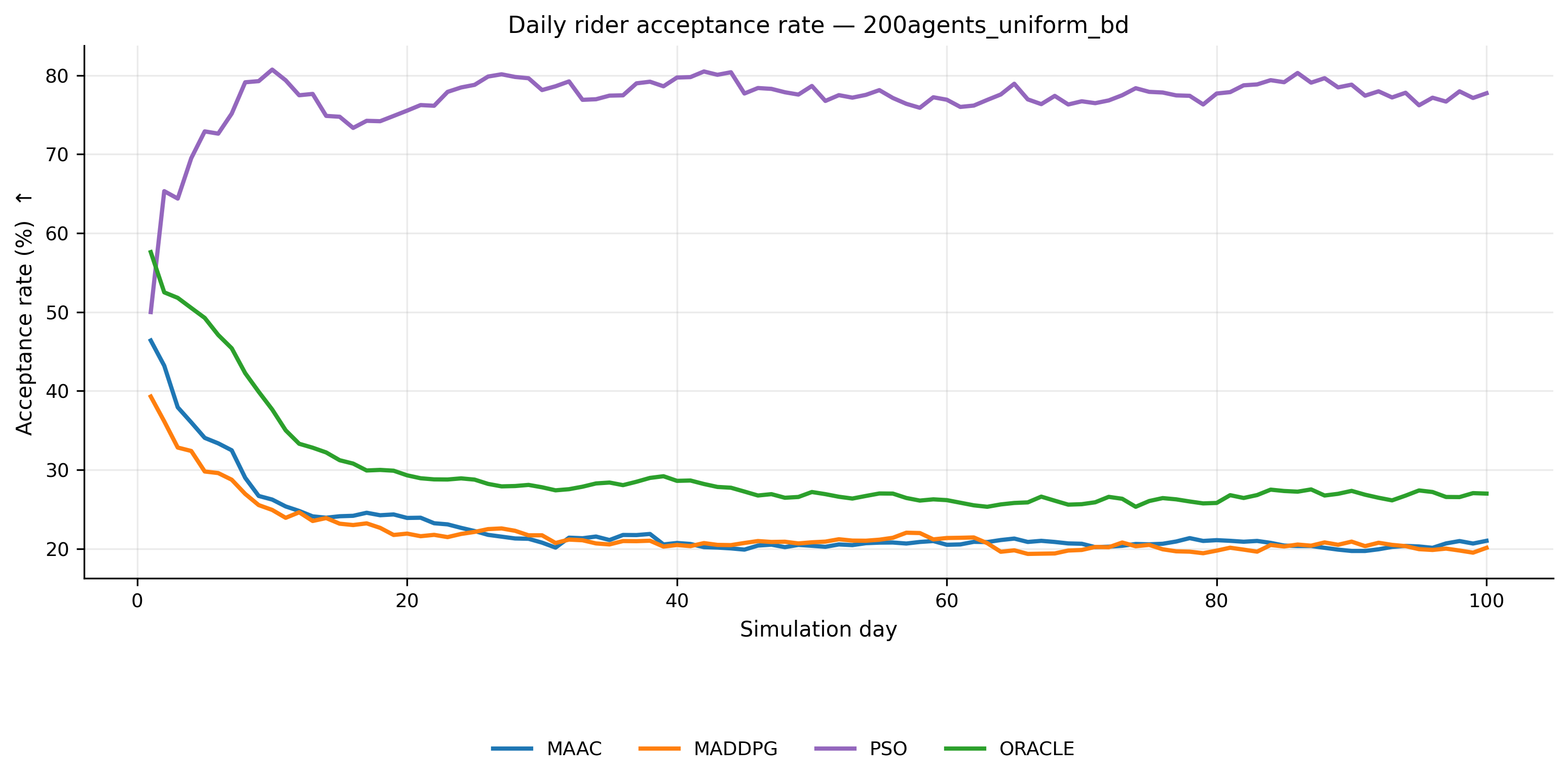}}
\caption{Uniform allocation; agents may enter and exit (Birth–Death)}
\end{subfigure}
\hfill
\begin{subfigure}{0.48\textwidth}
\centering
\includegraphics[width=\linewidth]{{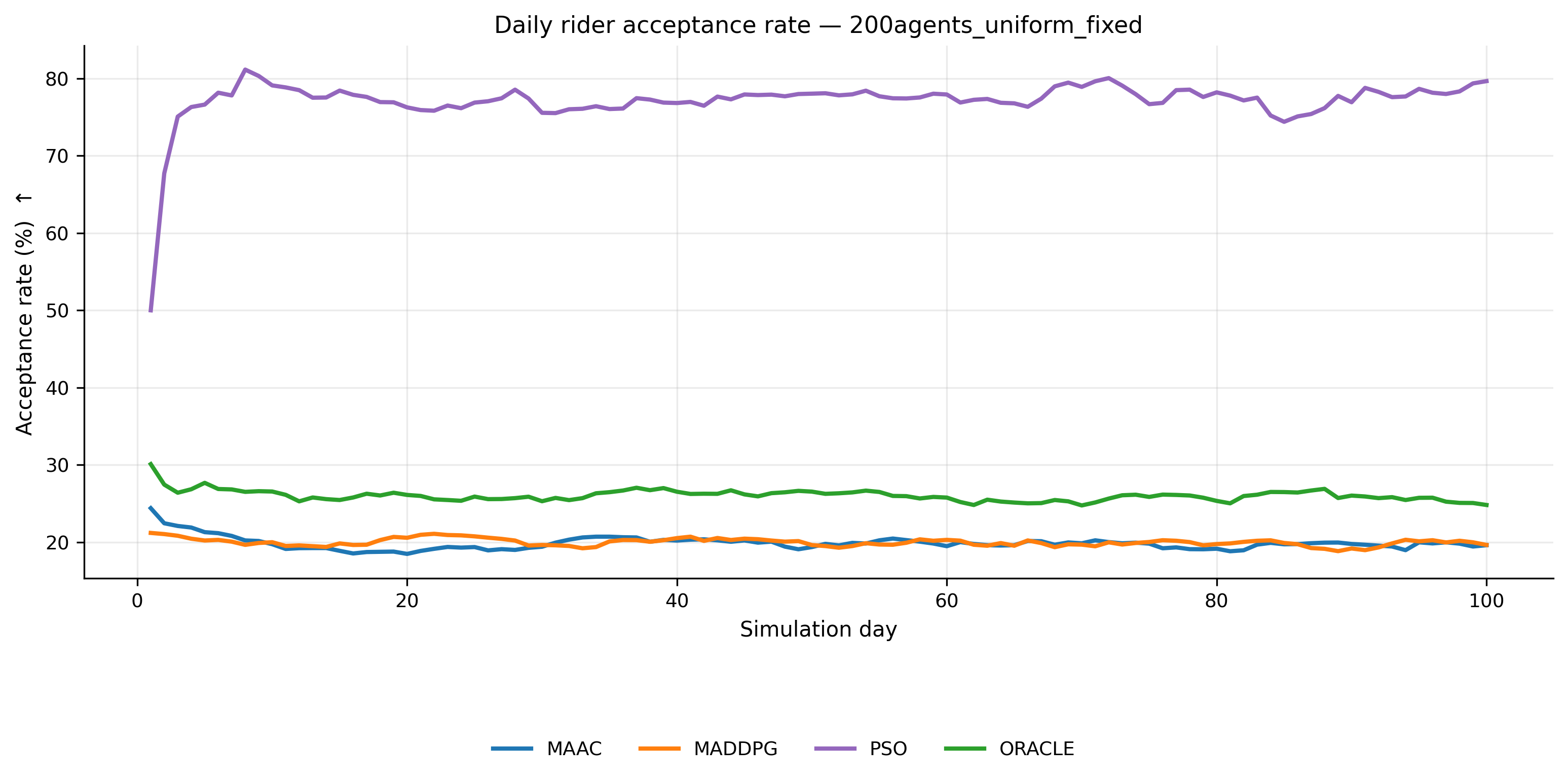}}
\caption{Uniform allocation; no entry and exit of agents}
\end{subfigure}

\caption{Daily acceptance rate with 200 agents.}
\end{figure}

\begin{figure}[H]
\centering
\begin{subfigure}{0.48\textwidth}
\centering
\includegraphics[width=\linewidth]{{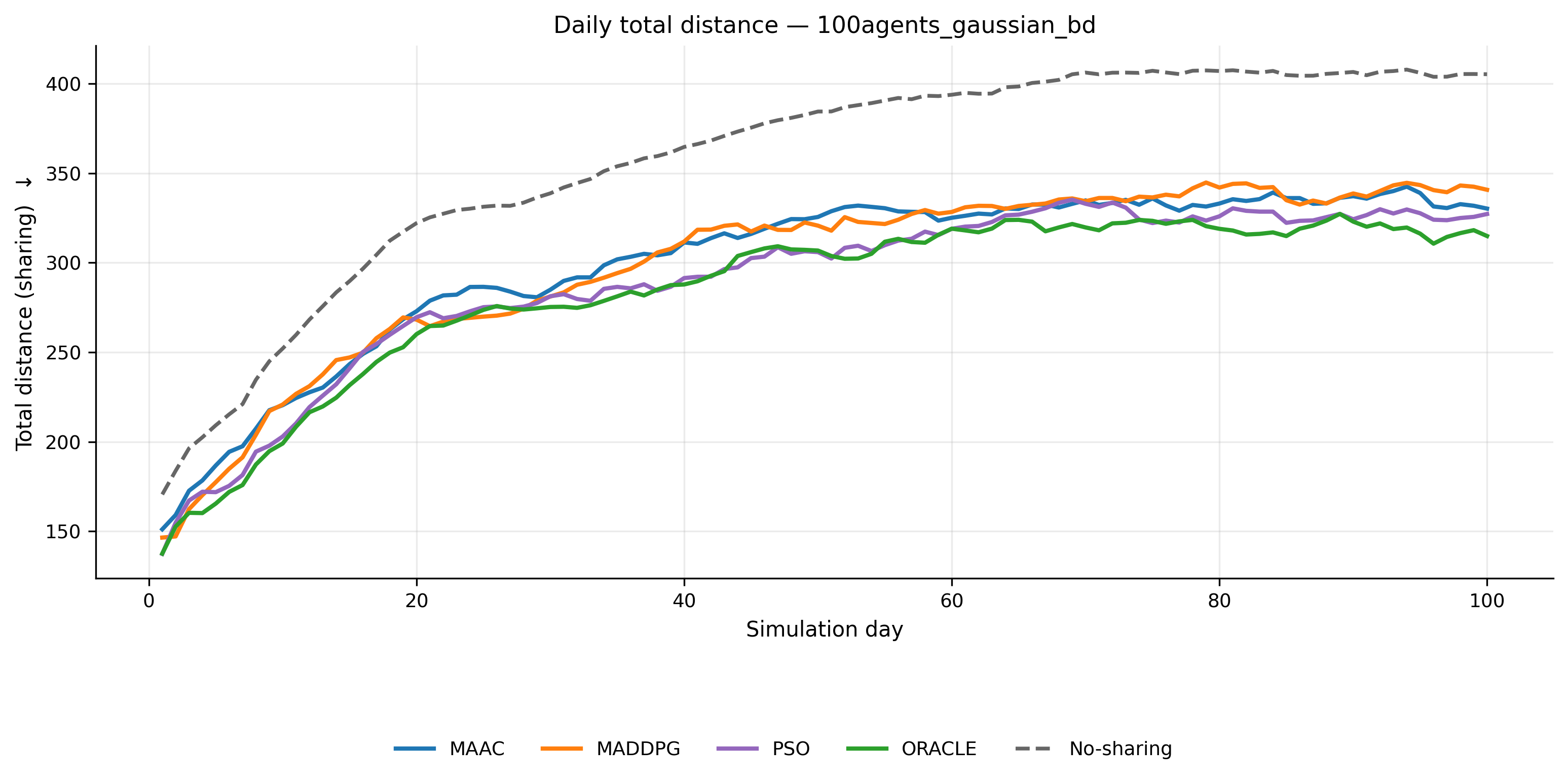}}
\caption{Gaussian allocation; agents may enter and exit (Birth–Death)}
\end{subfigure}
\hfill
\begin{subfigure}{0.48\textwidth}
\centering
\includegraphics[width=\linewidth]{{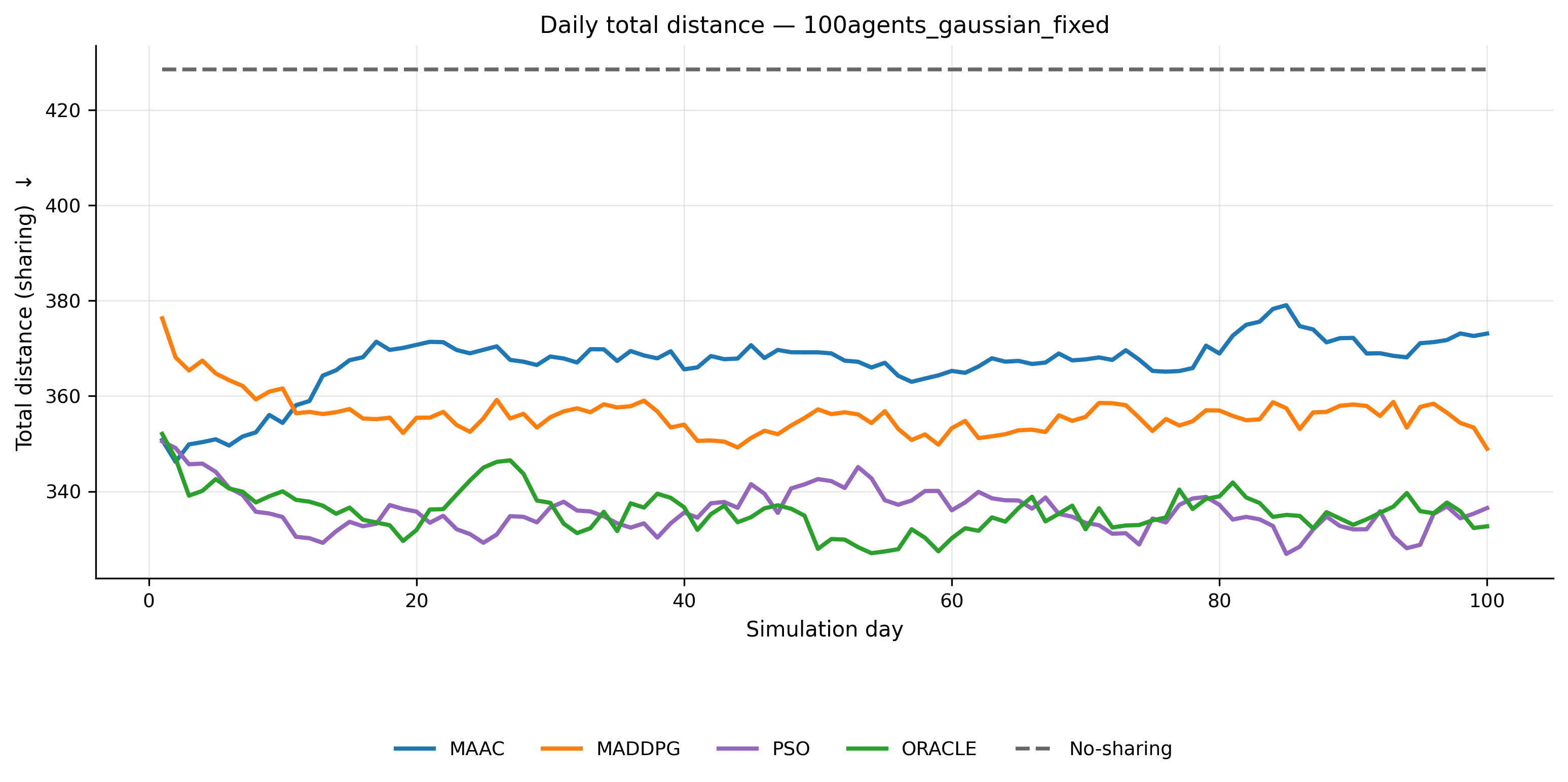}}
\caption{Gaussian allocation; no entry and exit of agents}
\end{subfigure}

\vspace{0.5em}

\begin{subfigure}{0.48\textwidth}
\centering
\includegraphics[width=\linewidth]{{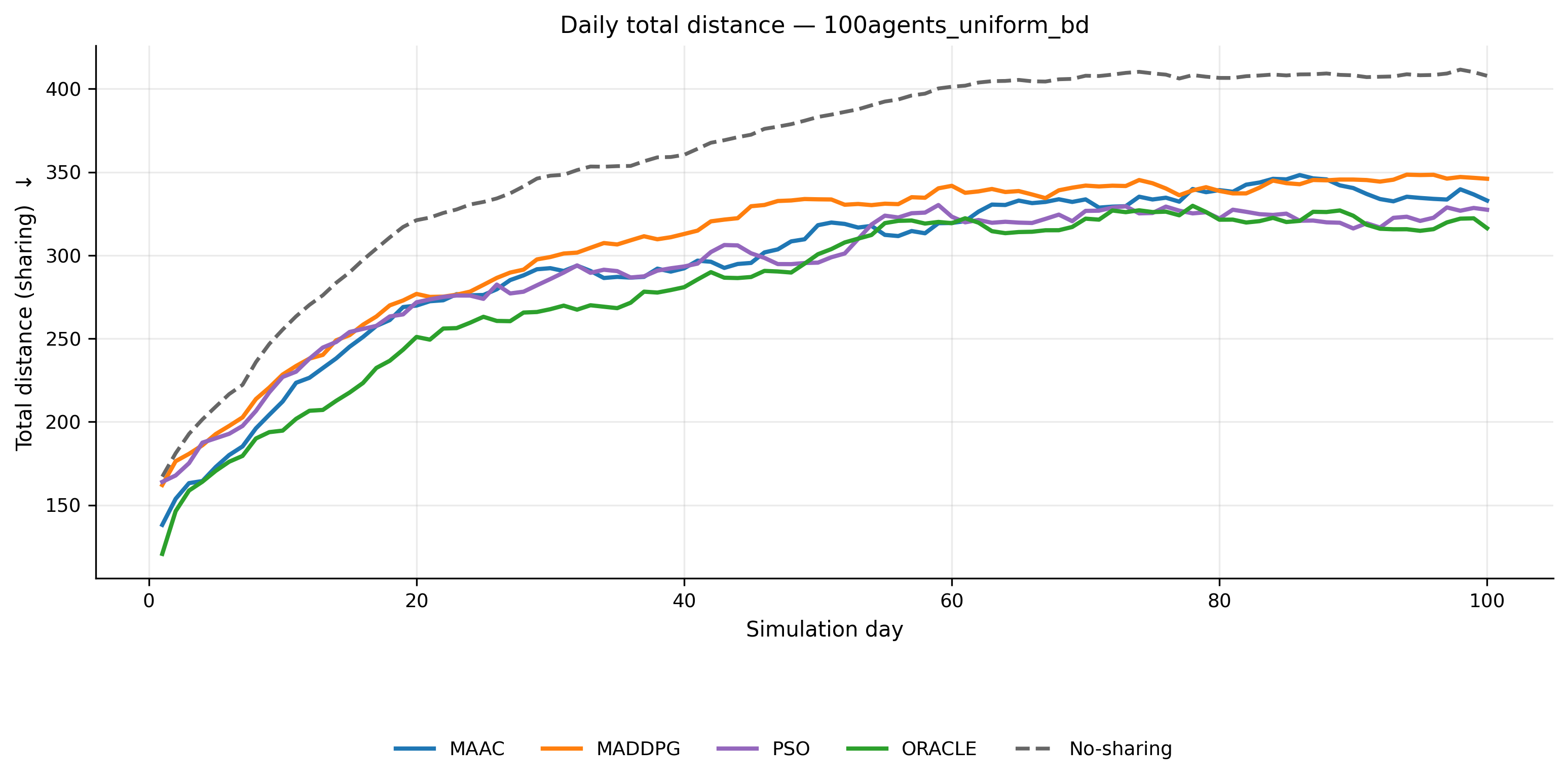}}
\caption{Uniform allocation; agents may enter and exit (Birth–Death)}
\end{subfigure}
\hfill
\begin{subfigure}{0.48\textwidth}
\centering
\includegraphics[width=\linewidth]{{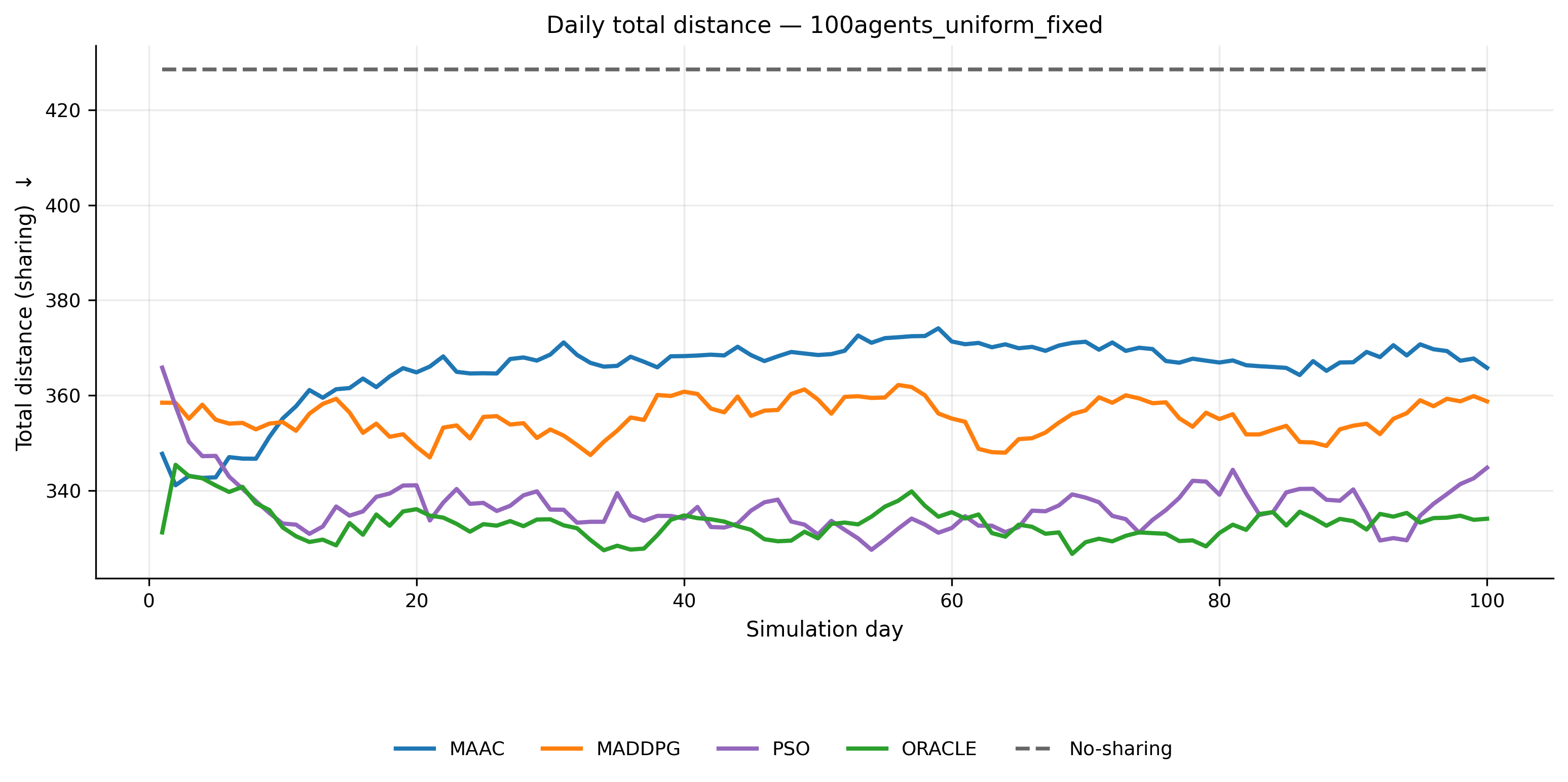}}
\caption{Uniform allocation; no entry and exit of agents}
\end{subfigure}

\caption{Daily travel distance with 100 agents.}
\end{figure}

\begin{figure}[H]
\centering
\begin{subfigure}{0.48\textwidth}
\centering
\includegraphics[width=\linewidth]{{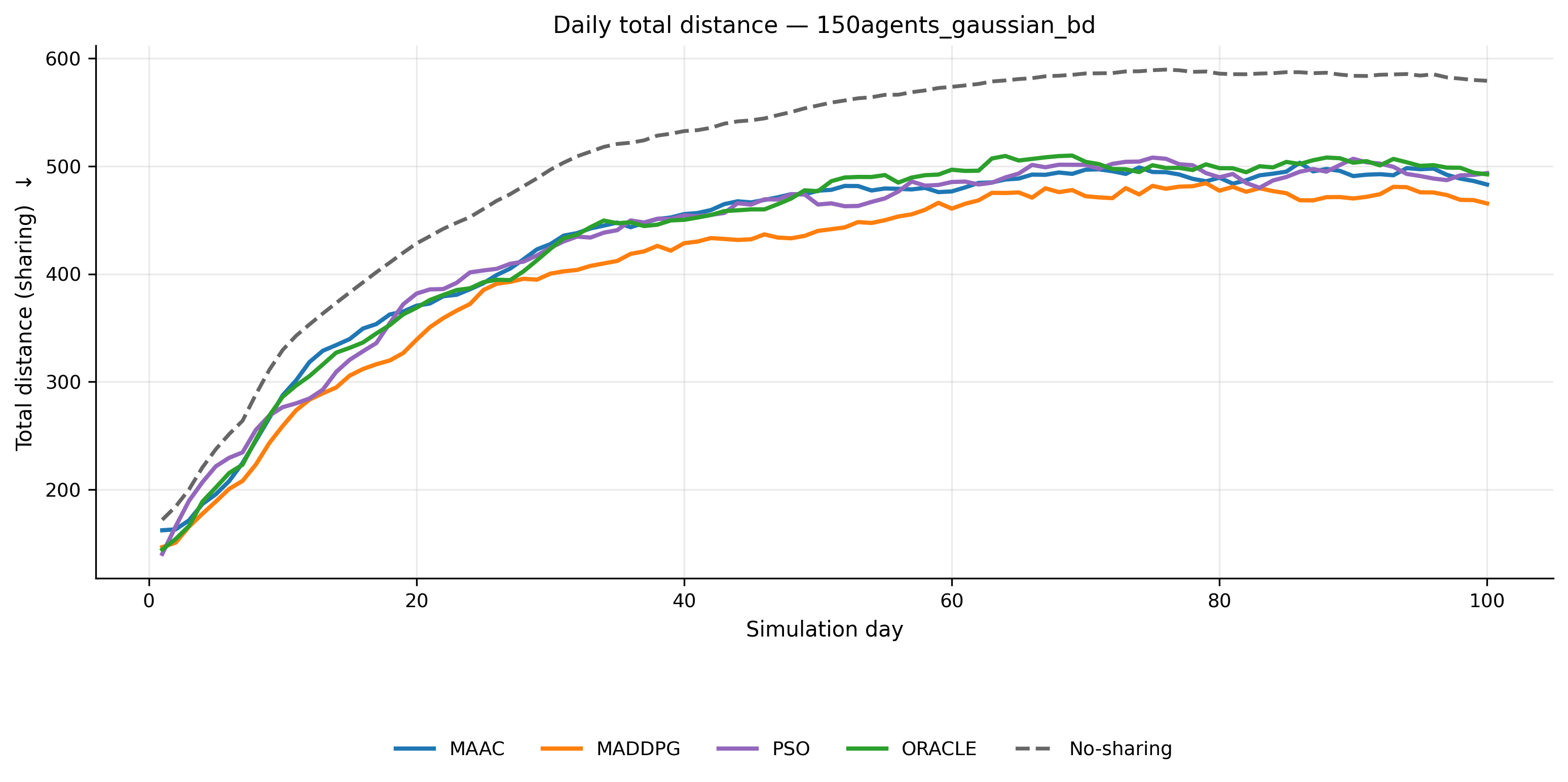}}
\caption{Gaussian allocation; agents may enter and exit (Birth–Death)}
\end{subfigure}
\hfill
\begin{subfigure}{0.48\textwidth}
\centering
\includegraphics[width=\linewidth]{{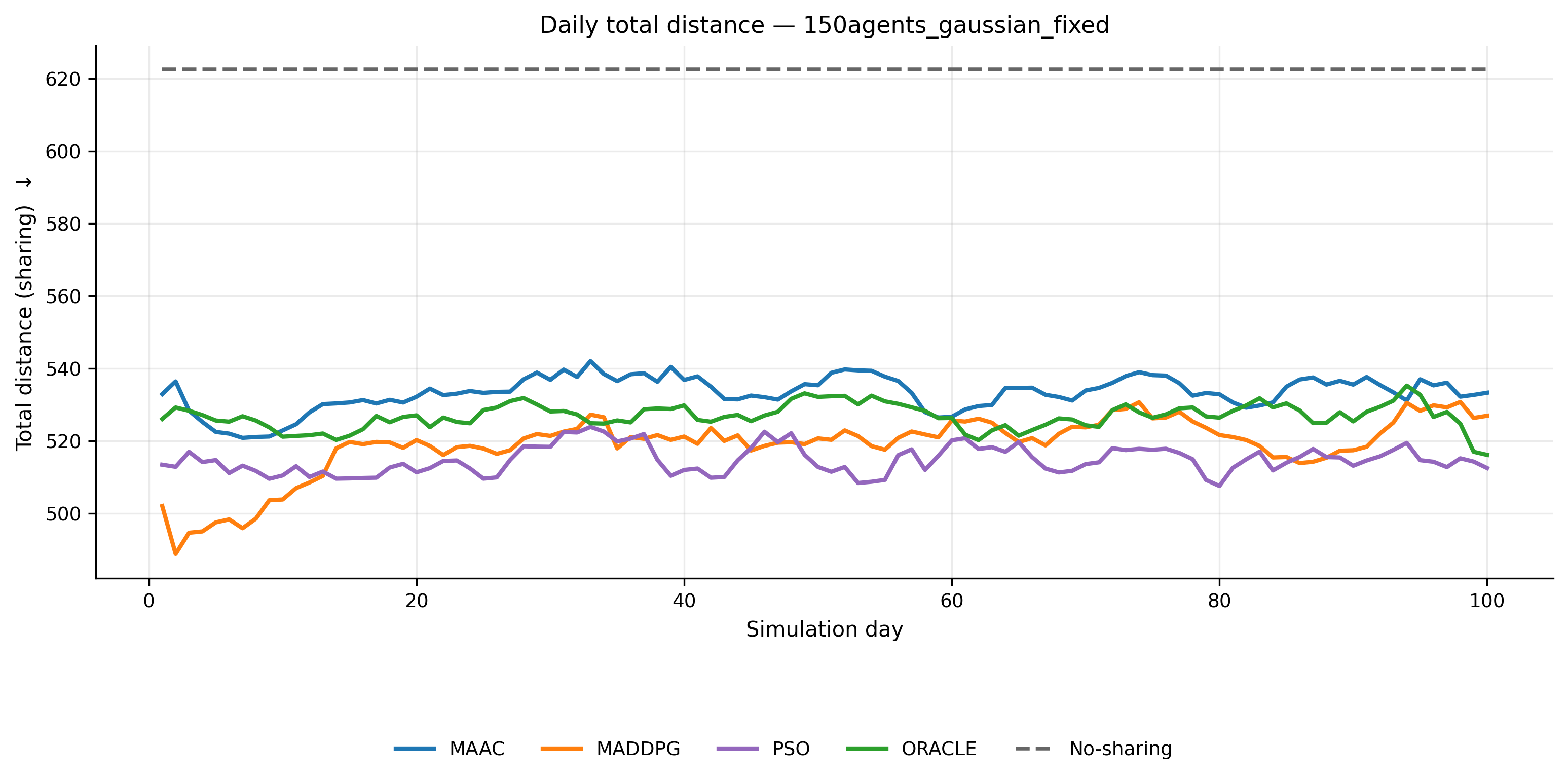}}
\caption{Gaussian allocation; no entry and exit of agents}
\end{subfigure}

\vspace{0.5em}

\begin{subfigure}{0.48\textwidth}
\centering
\includegraphics[width=\linewidth]{{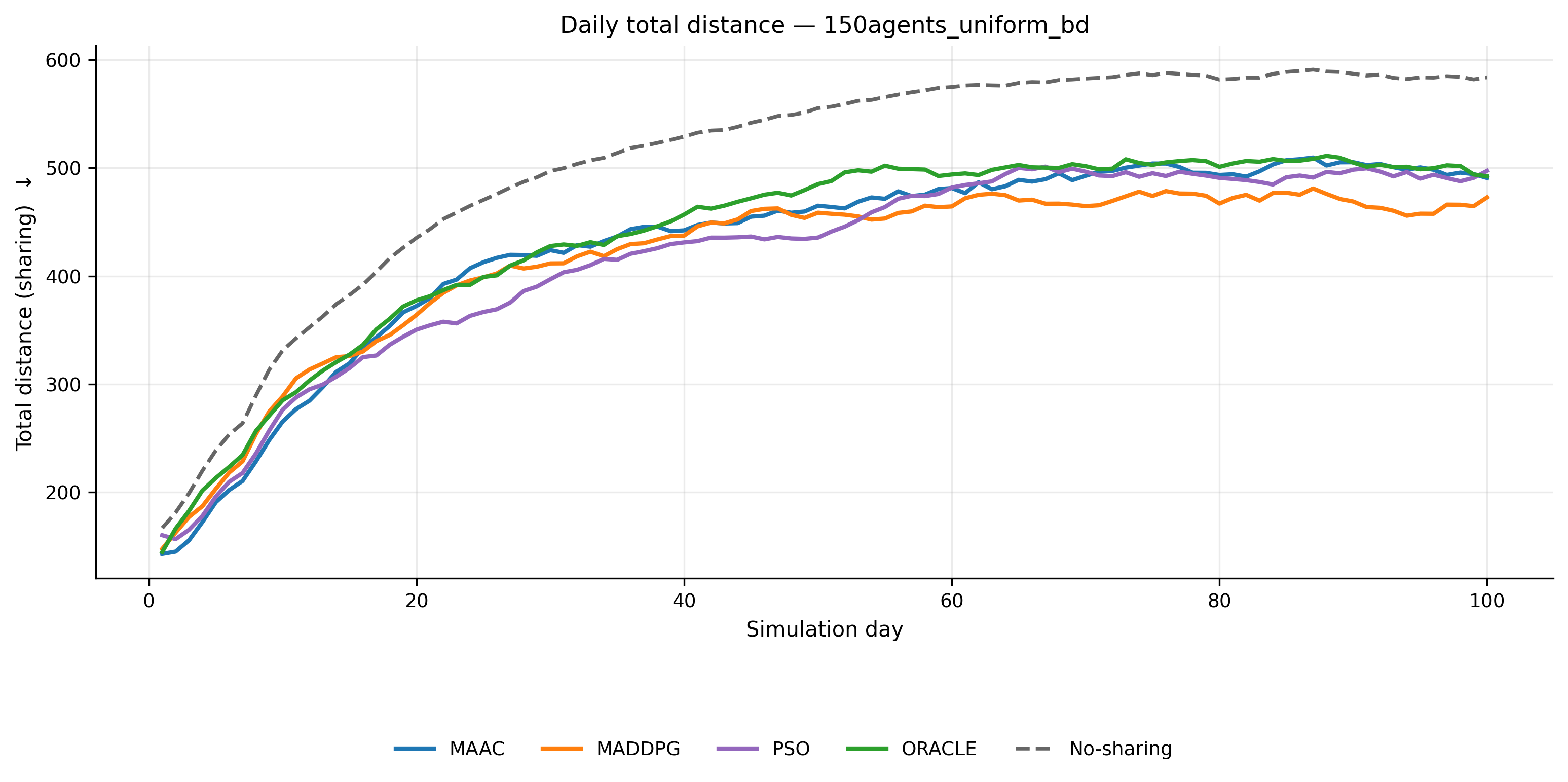}}
\caption{Uniform allocation; agents may enter and exit (Birth–Death)}
\end{subfigure}
\hfill
\begin{subfigure}{0.48\textwidth}
\centering
\includegraphics[width=\linewidth]{{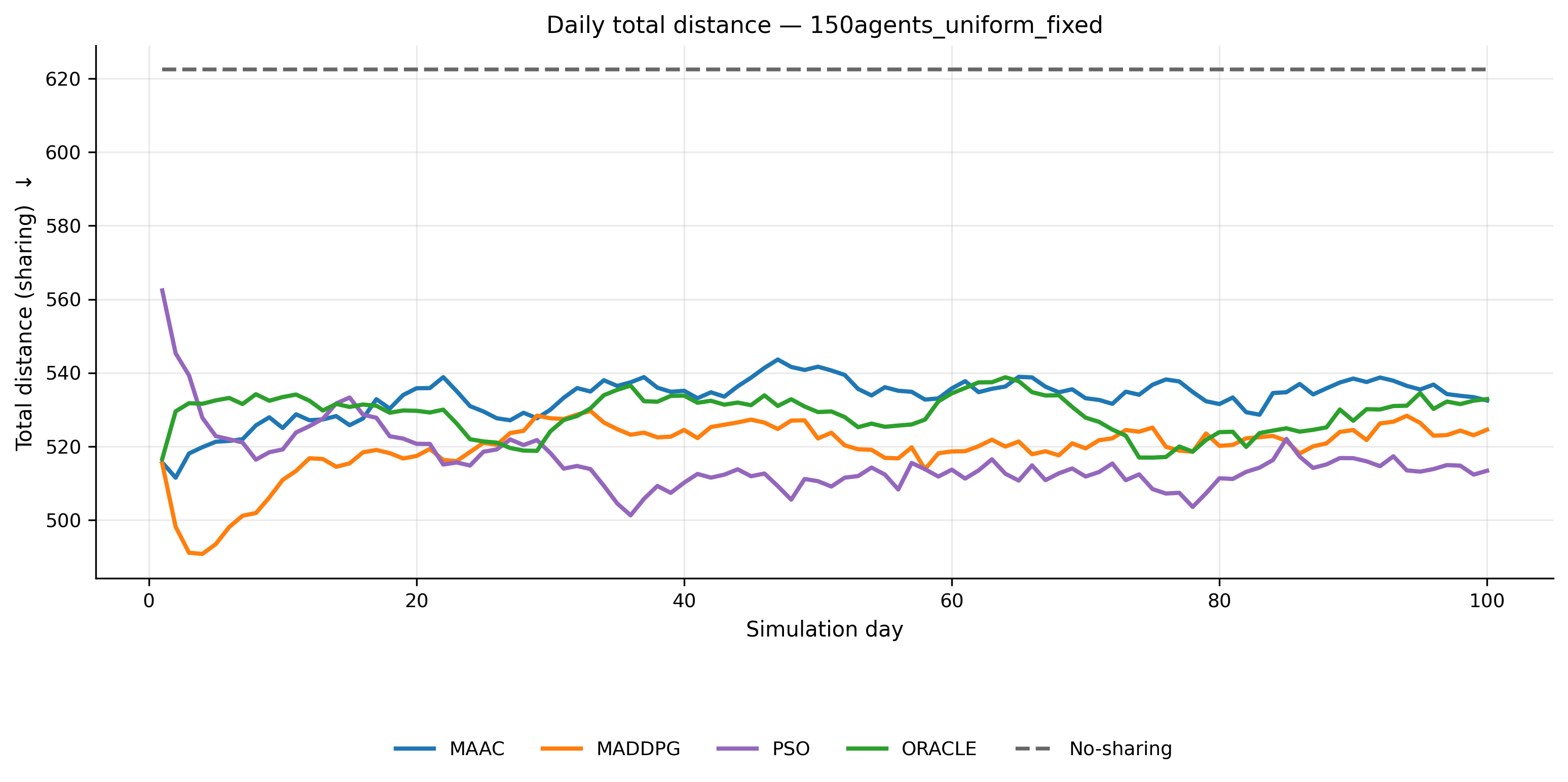}}
\caption{Uniform allocation; no entry and exit of agents}
\end{subfigure}

\caption{Daily travel distance with 150 agents.}
\end{figure}

\begin{figure}[H]
\centering
\begin{subfigure}{0.48\textwidth}
\centering
\includegraphics[width=\linewidth]{{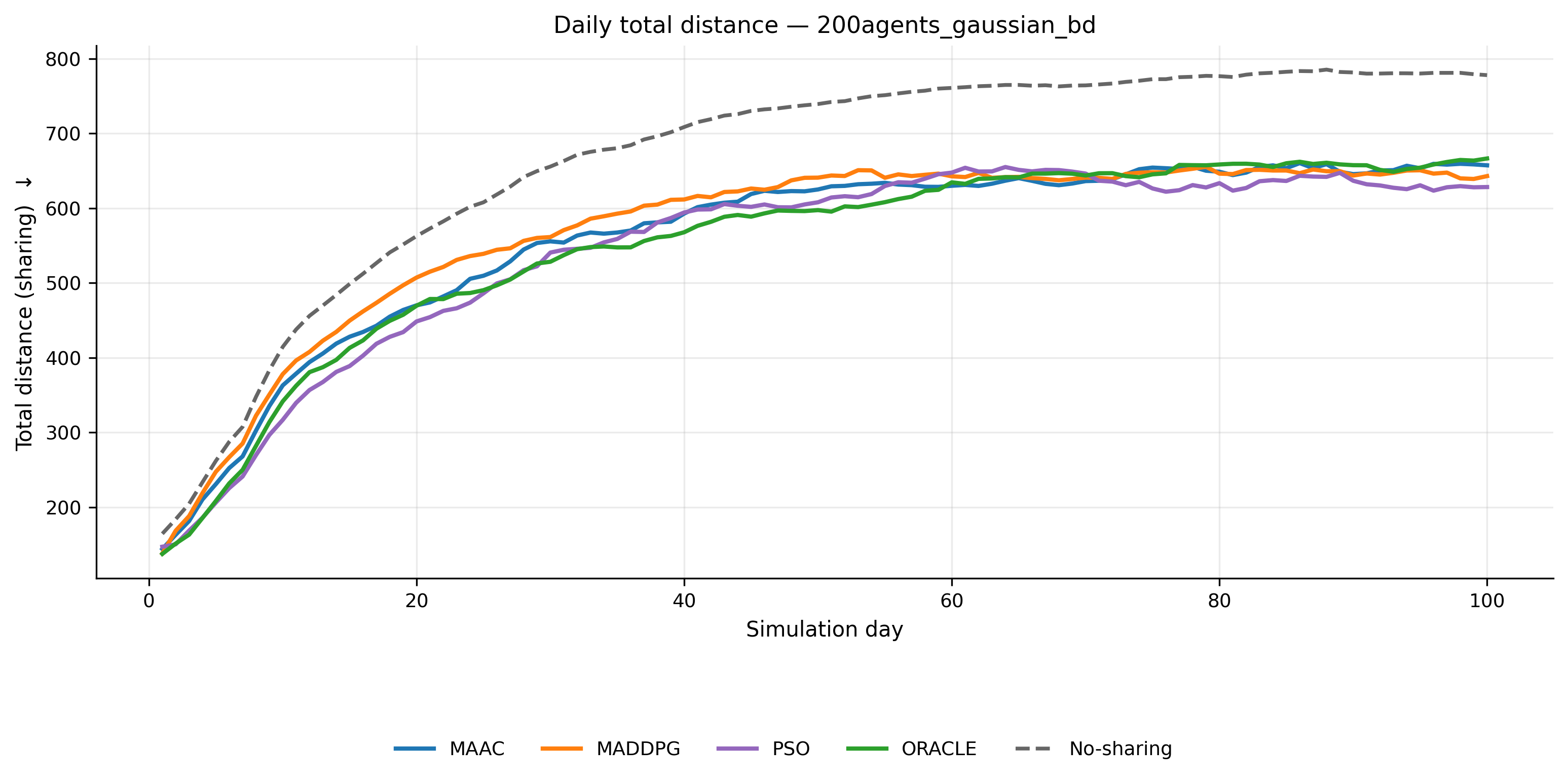}}
\caption{Gaussian allocation; agents may enter and exit (Birth–Death)}
\end{subfigure}
\hfill
\begin{subfigure}{0.48\textwidth}
\centering
\includegraphics[width=\linewidth]{{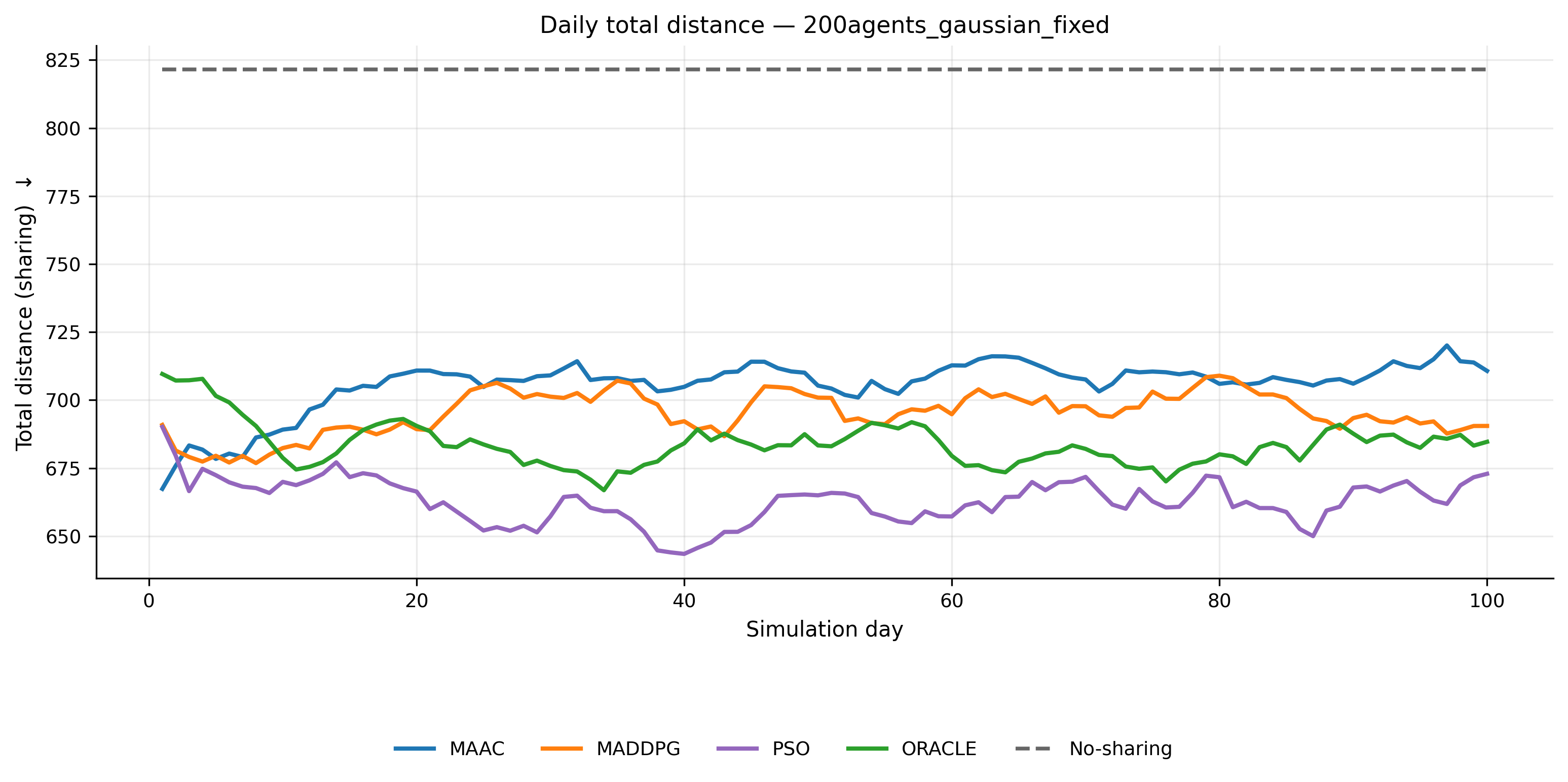}}
\caption{Gaussian allocation; no entry and exit of agents}
\end{subfigure}

\vspace{0.5em}

\begin{subfigure}{0.48\textwidth}
\centering
\includegraphics[width=\linewidth]{{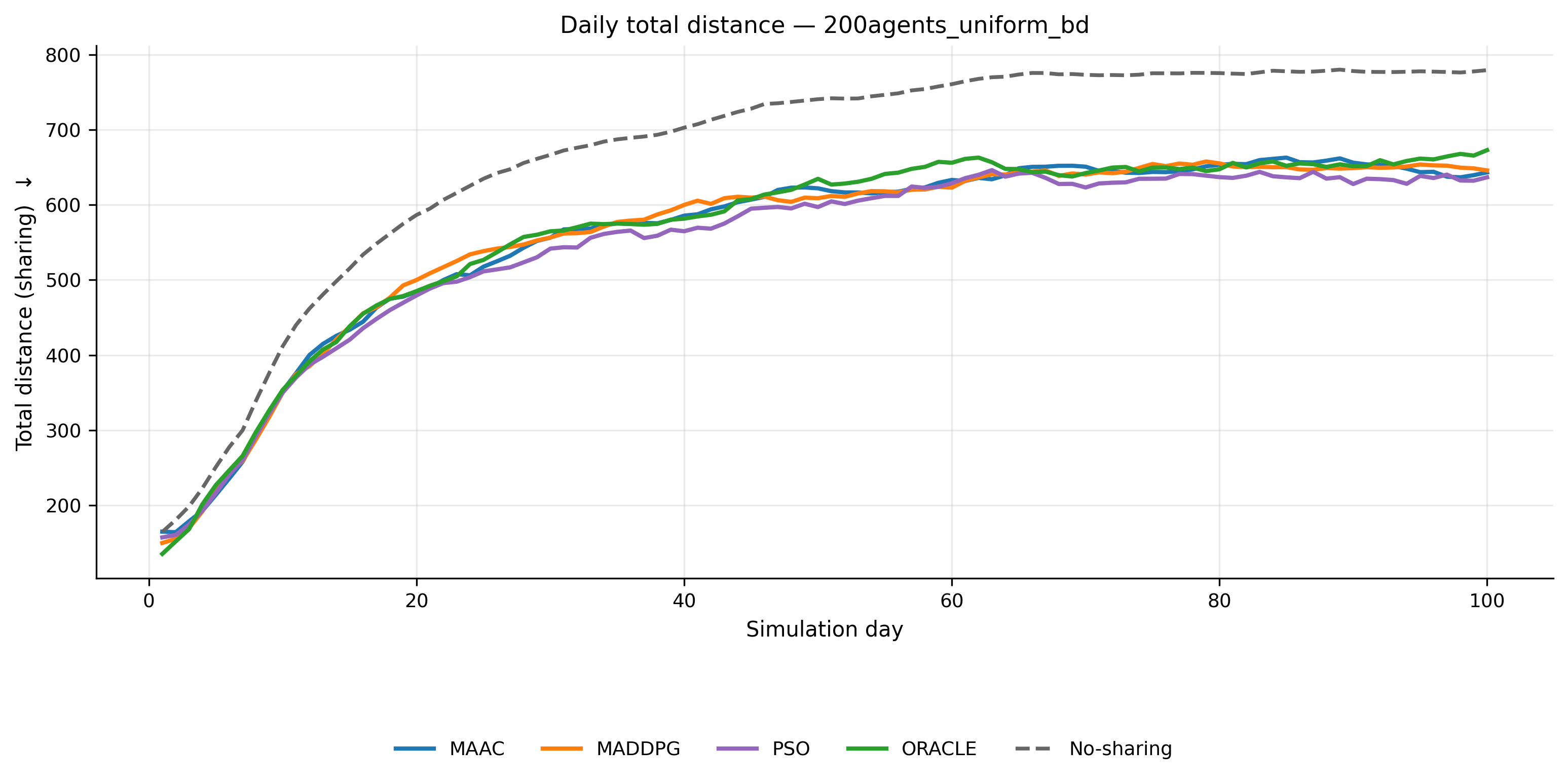}}
\caption{Uniform allocation; agents may enter and exit (Birth–Death)}
\end{subfigure}
\hfill
\begin{subfigure}{0.48\textwidth}
\centering
\includegraphics[width=\linewidth]{{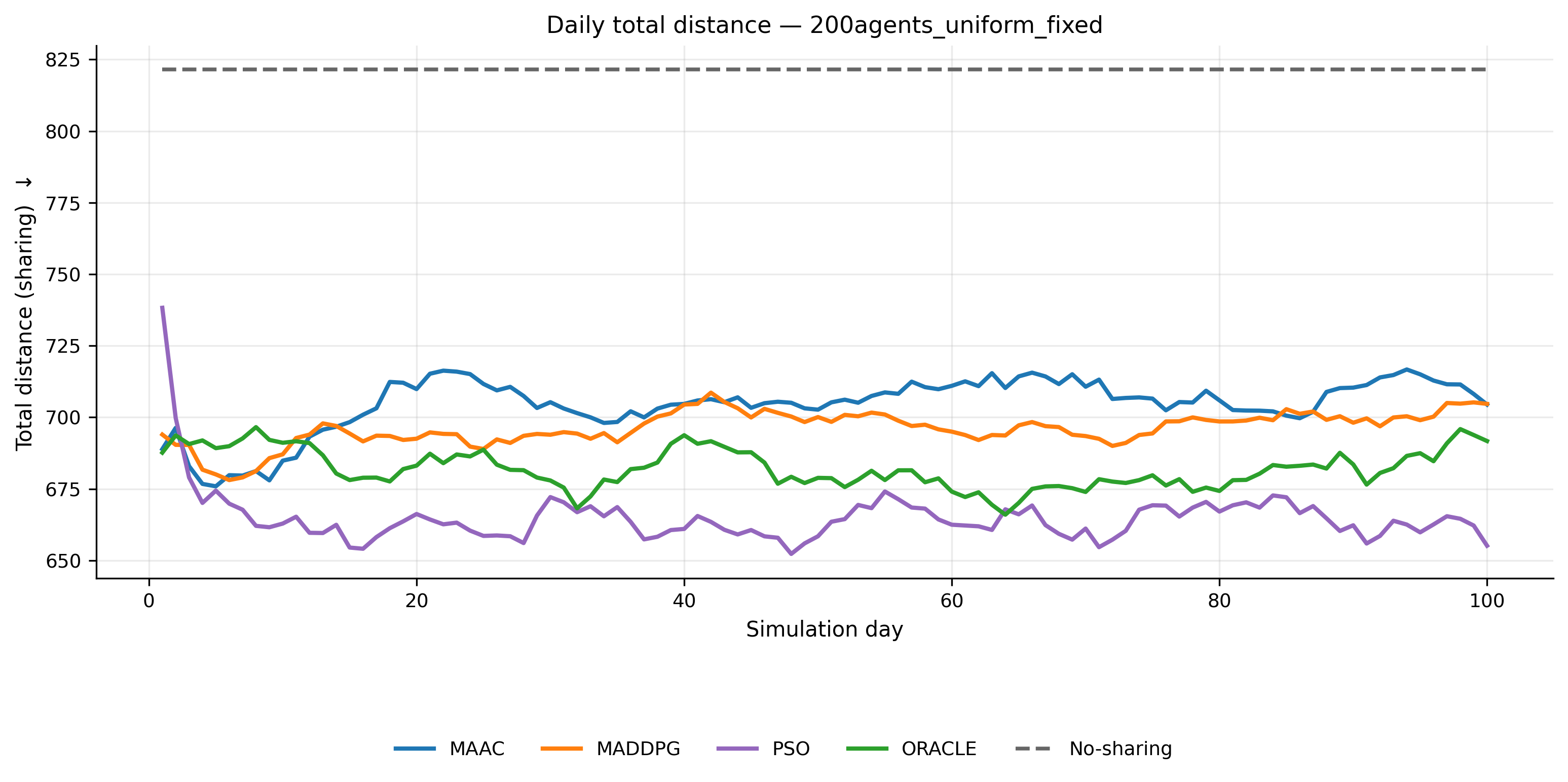}}
\caption{Uniform allocation; no entry and exit of agents}
\end{subfigure}

\caption{Daily travel distance with 200 agents.}
\end{figure}

\begin{figure}[H]
\centering
\begin{subfigure}{0.48\textwidth}
\centering
\includegraphics[width=\linewidth]{{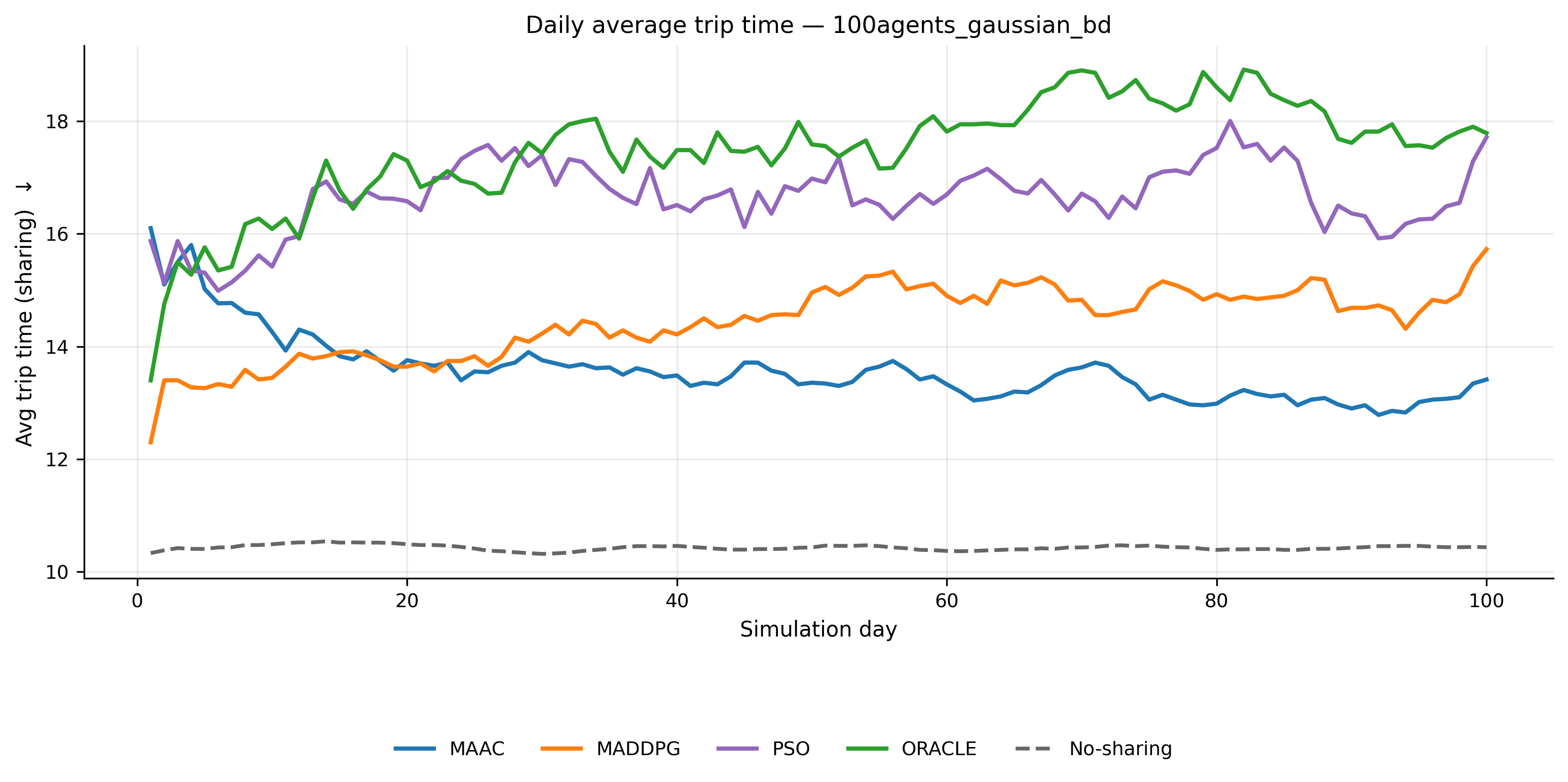}}
\caption{Gaussian allocation; agents may enter and exit (Birth–Death)}
\end{subfigure}
\hfill
\begin{subfigure}{0.48\textwidth}
\centering
\includegraphics[width=\linewidth]{{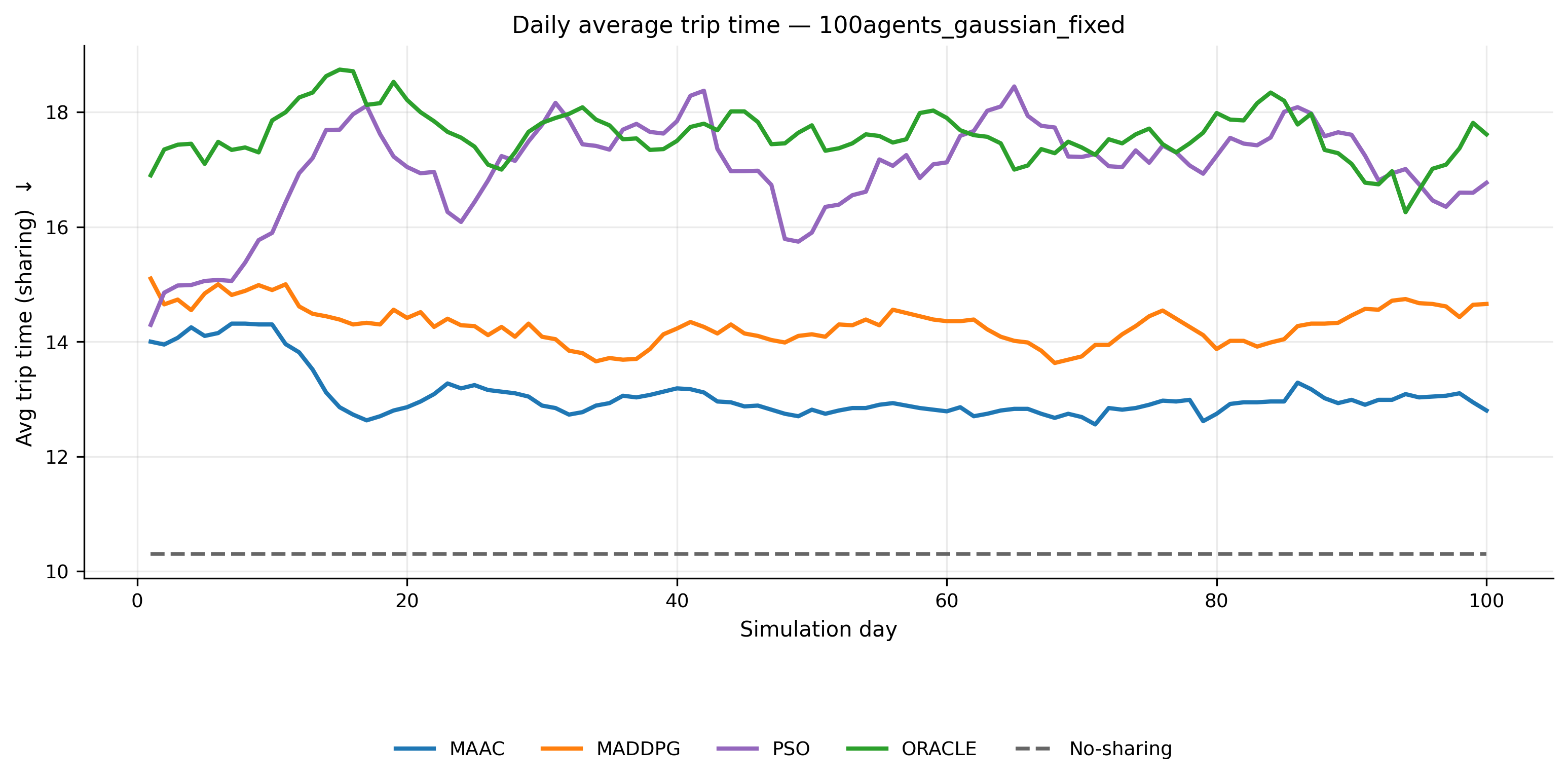}}
\caption{Gaussian allocation; no entry and exit of agents}
\end{subfigure}

\vspace{0.5em}

\begin{subfigure}{0.48\textwidth}
\centering
\includegraphics[width=\linewidth]{{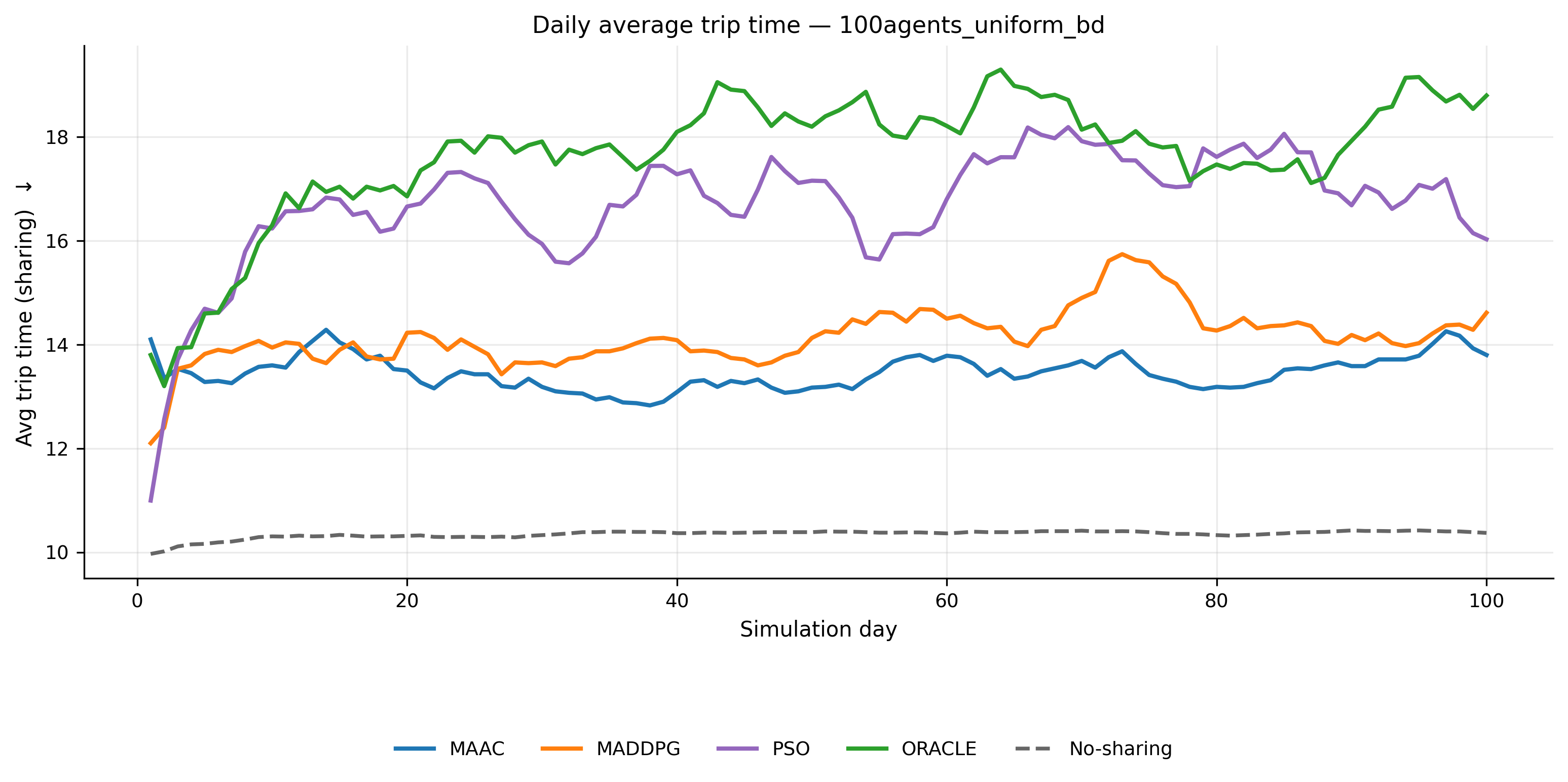}}
\caption{Uniform allocation; agents may enter and exit (Birth–Death)}
\end{subfigure}
\hfill
\begin{subfigure}{0.48\textwidth}
\centering
\includegraphics[width=\linewidth]{{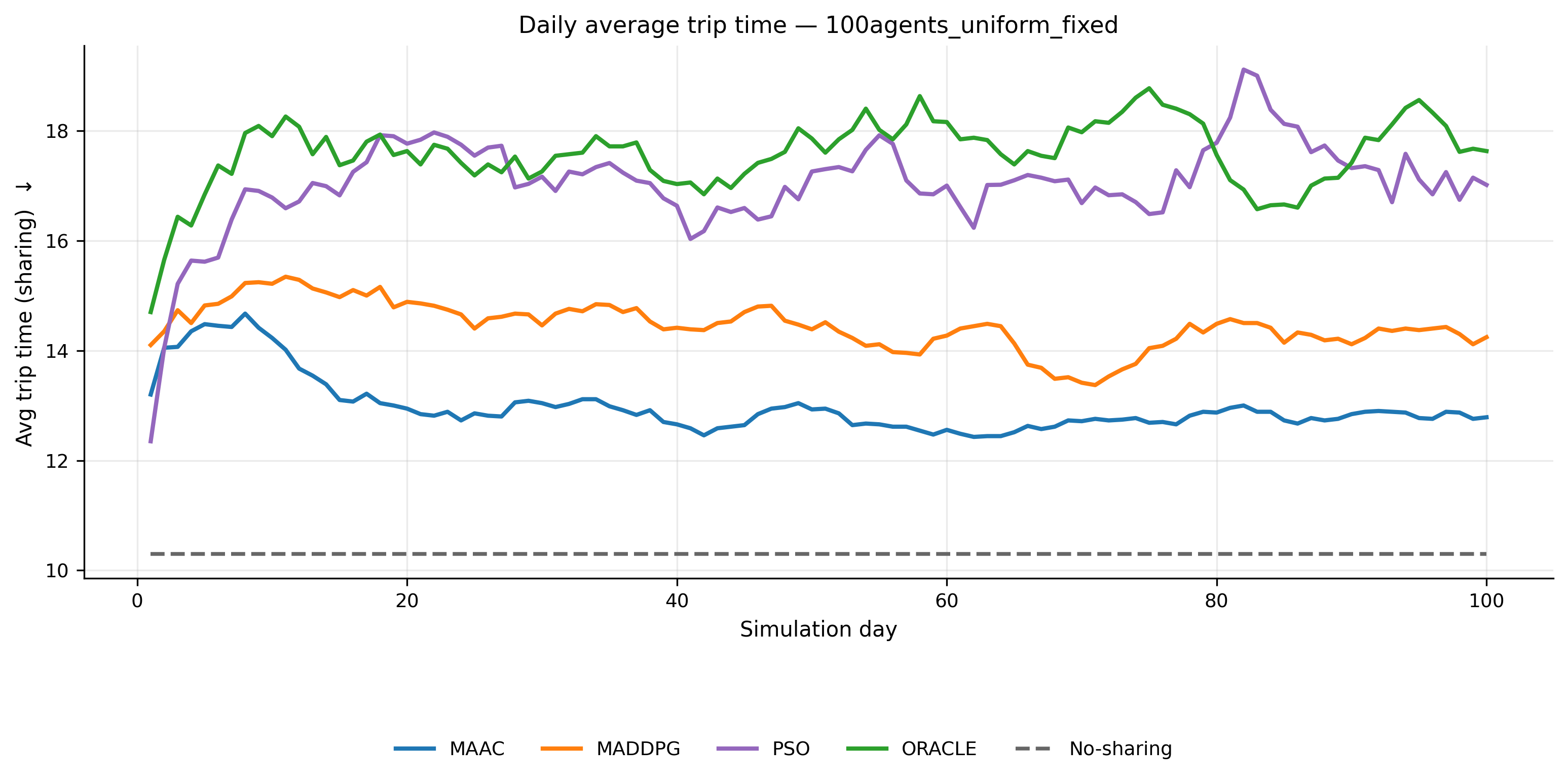}}
\caption{Uniform allocation; no entry and exit of agents}
\end{subfigure}

\caption{Daily trip time with 100 agents.}
\end{figure}

\begin{figure}[H]
\centering
\begin{subfigure}{0.48\textwidth}
\centering
\includegraphics[width=\linewidth]{{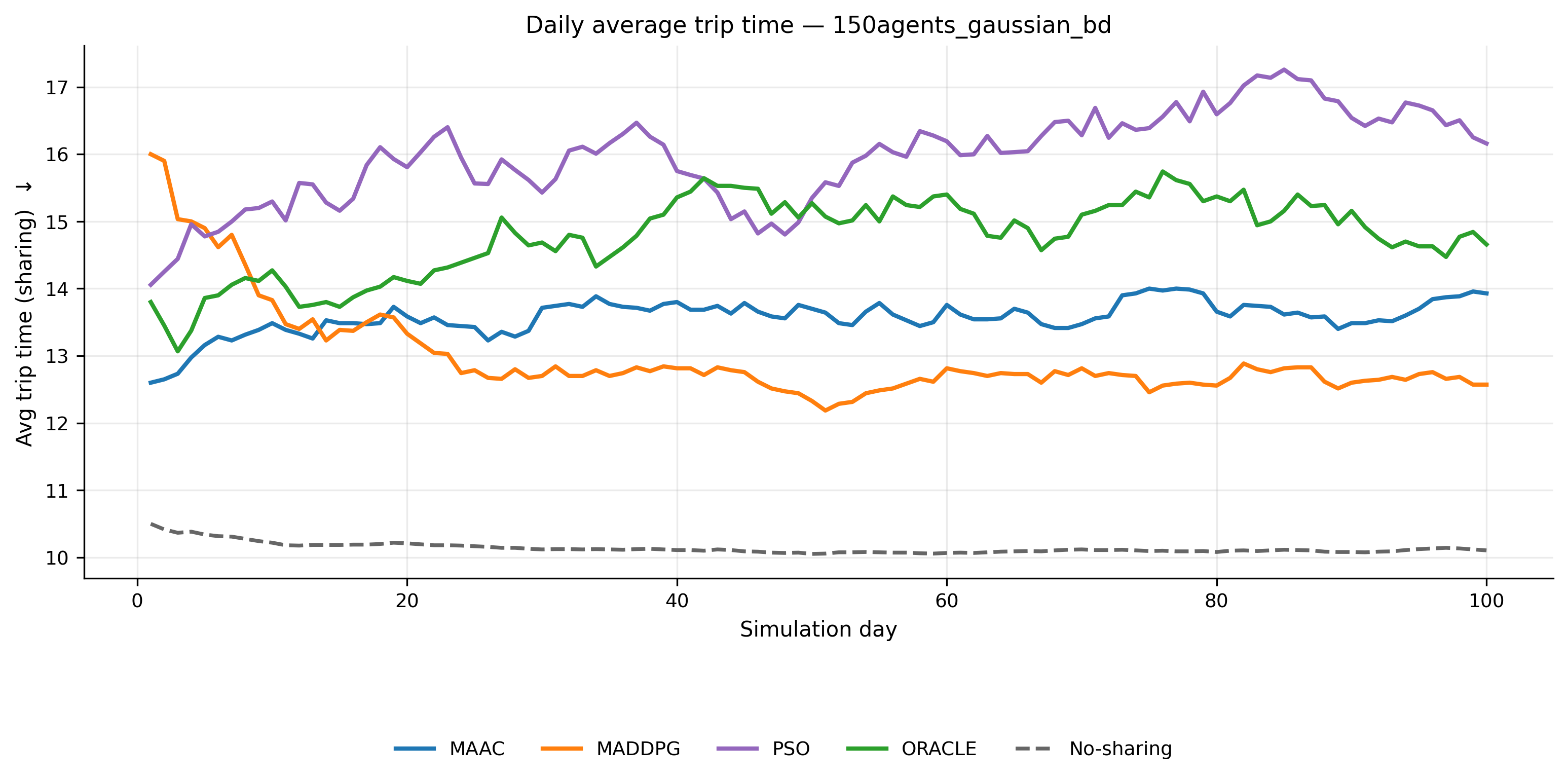}}
\caption{Gaussian allocation; agents may enter and exit (Birth–Death)}
\end{subfigure}
\hfill
\begin{subfigure}{0.48\textwidth}
\centering
\includegraphics[width=\linewidth]{{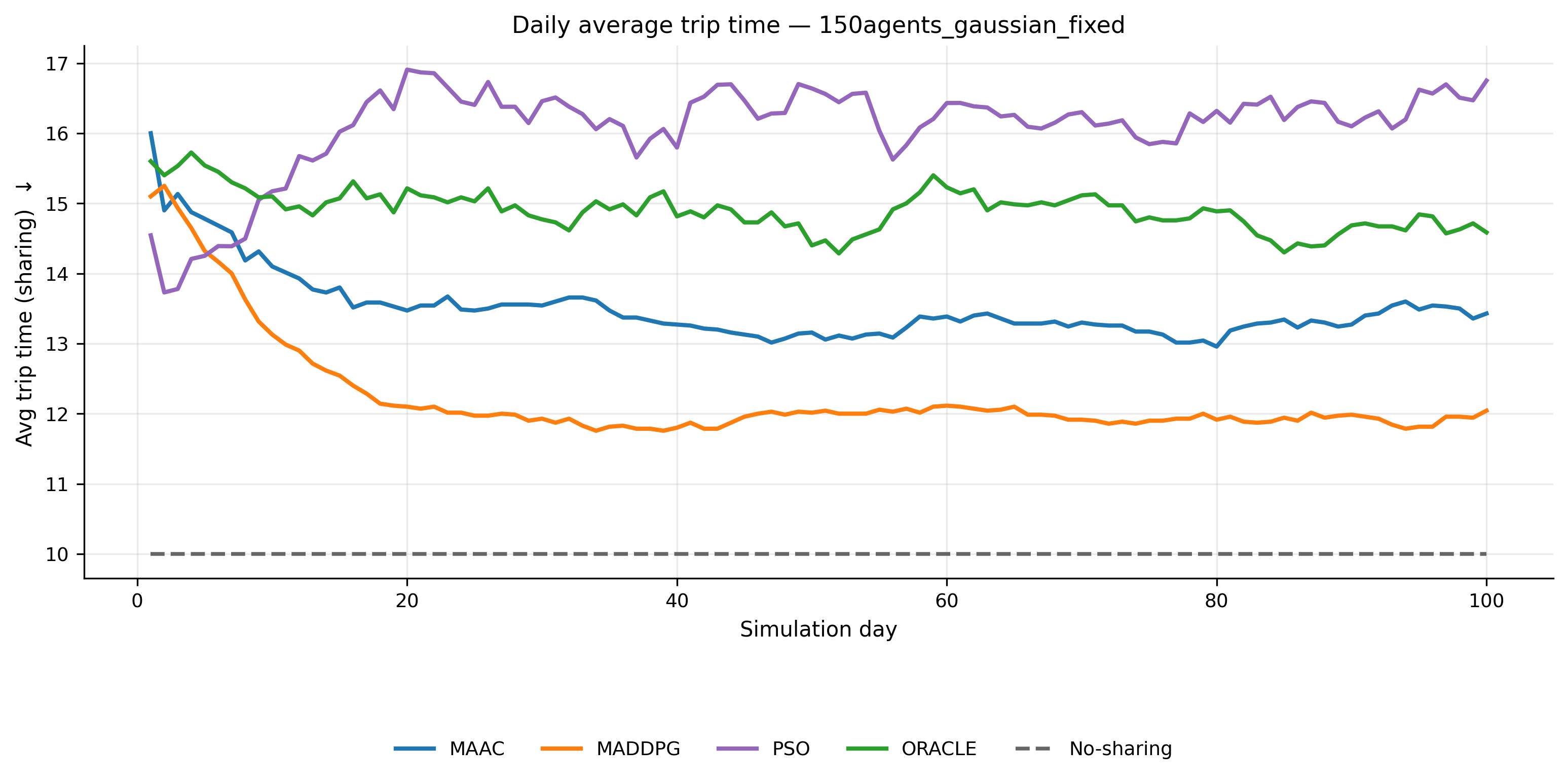}}
\caption{Gaussian allocation; no entry and exit of agents}
\end{subfigure}

\vspace{0.5em}

\begin{subfigure}{0.48\textwidth}
\centering
\includegraphics[width=\linewidth]{{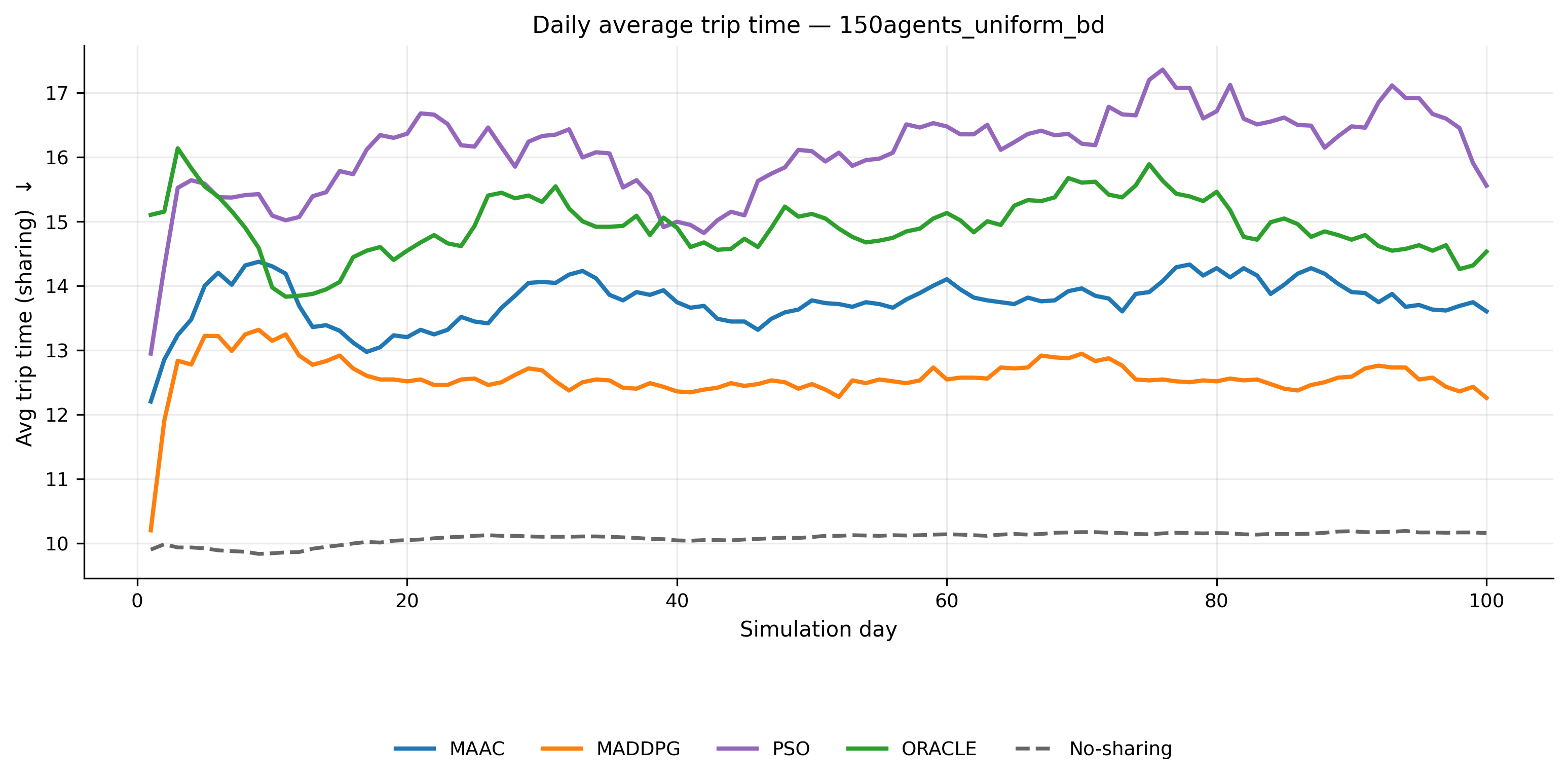}}
\caption{Uniform allocation; agents may enter and exit (Birth–Death)}
\end{subfigure}
\hfill
\begin{subfigure}{0.48\textwidth}
\centering
\includegraphics[width=\linewidth]{{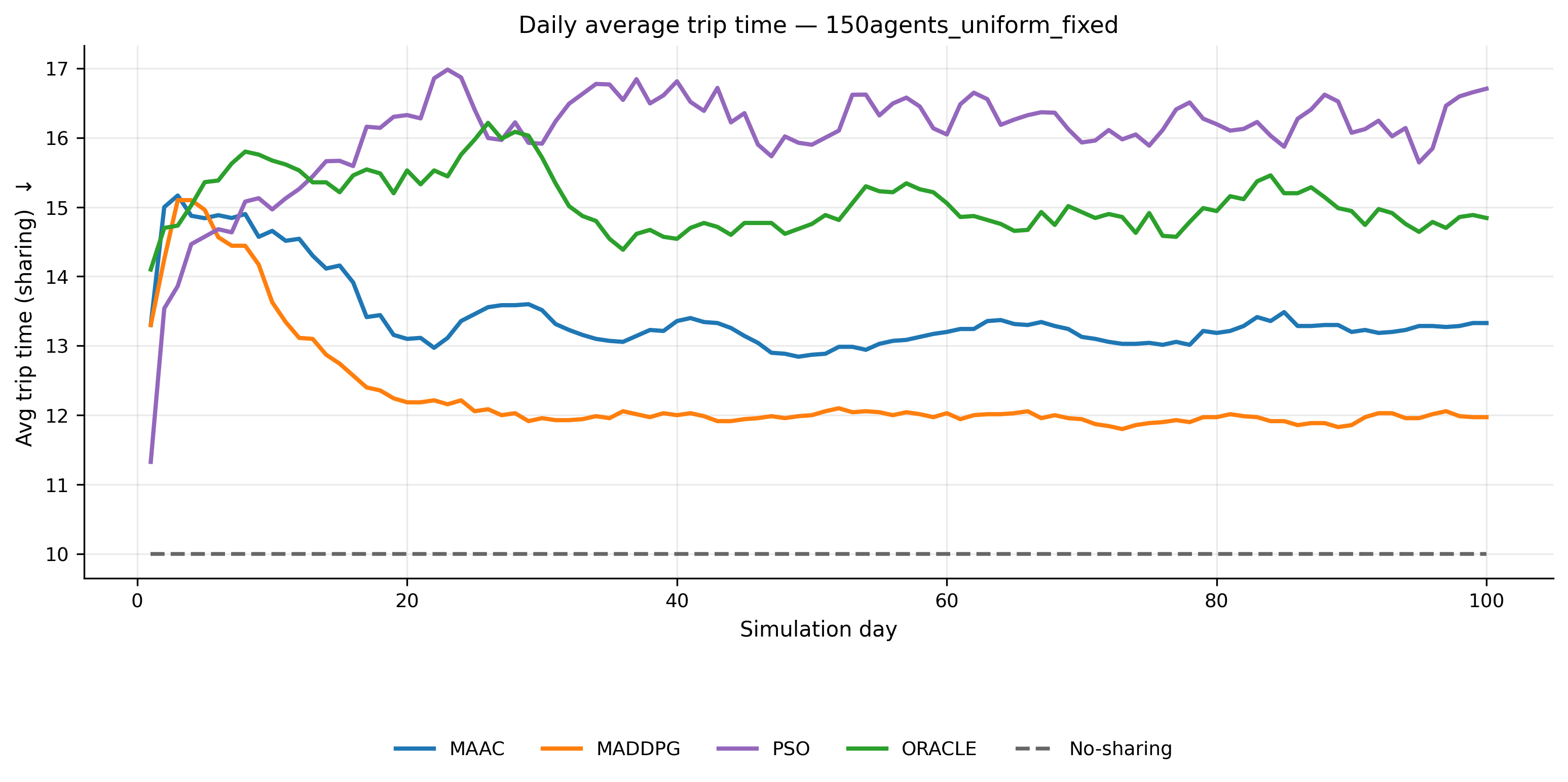}}
\caption{Uniform allocation; no entry and exit of agents}
\end{subfigure}

\caption{Daily trip time with 150 agents.}
\end{figure}

\begin{figure}[H]
\centering
\begin{subfigure}{0.48\textwidth}
\centering
\includegraphics[width=\linewidth]{{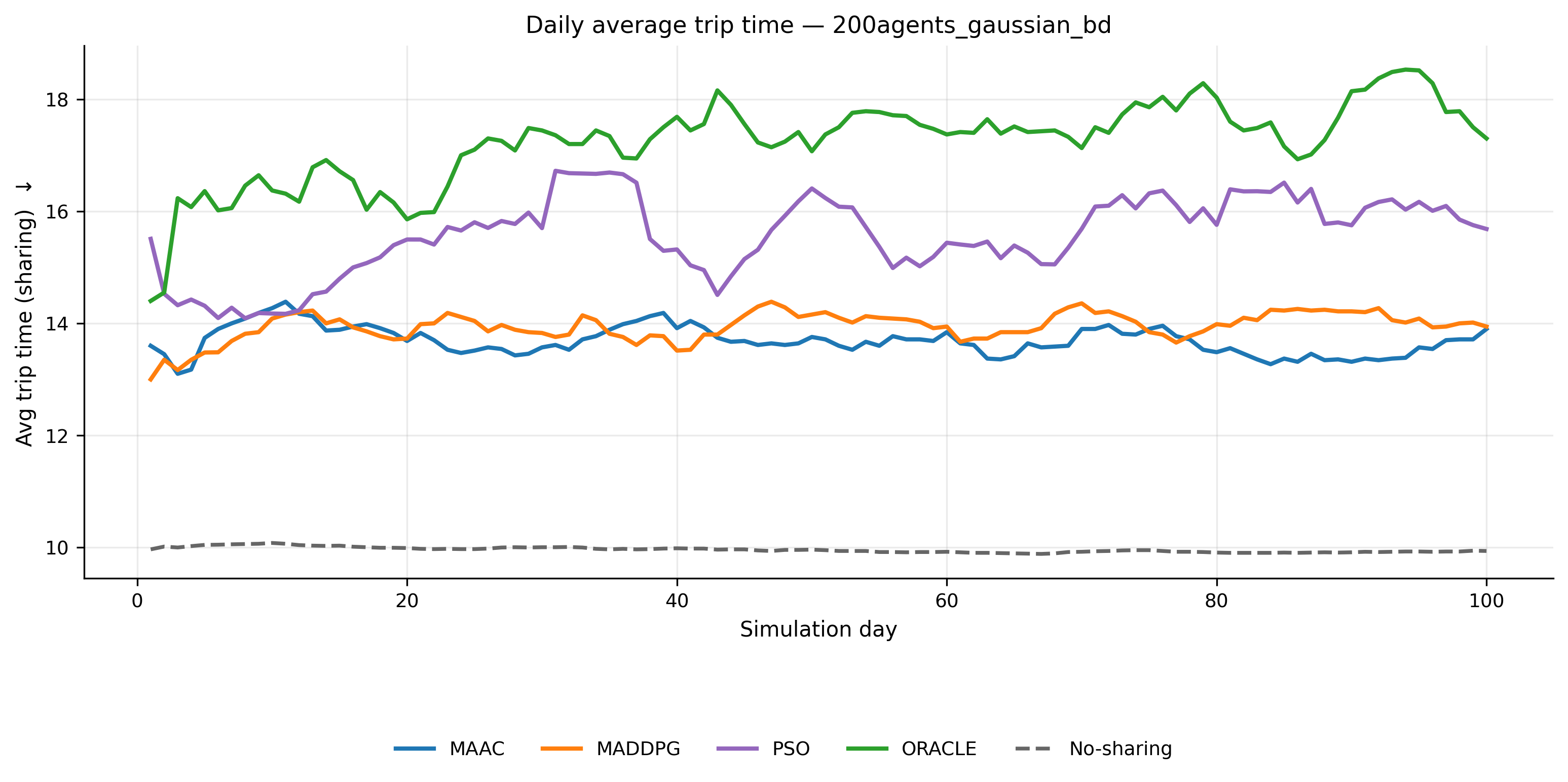}}
\caption{Gaussian allocation; agents may enter and exit (Birth–Death)}
\end{subfigure}
\hfill
\begin{subfigure}{0.48\textwidth}
\centering
\includegraphics[width=\linewidth]{{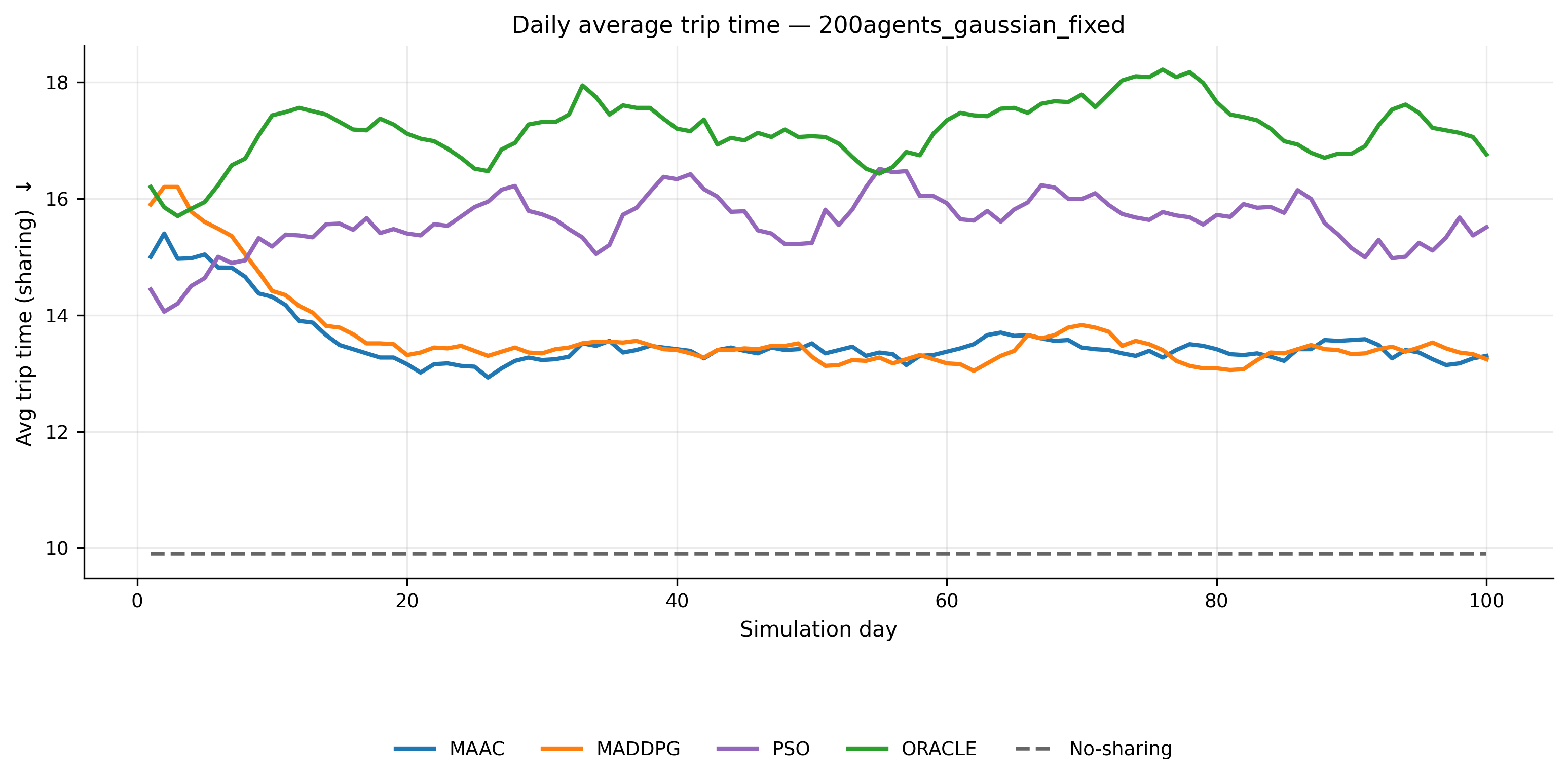}}
\caption{Gaussian allocation; no entry and exit of agents}
\end{subfigure}

\vspace{0.5em}

\begin{subfigure}{0.48\textwidth}
\centering
\includegraphics[width=\linewidth]{{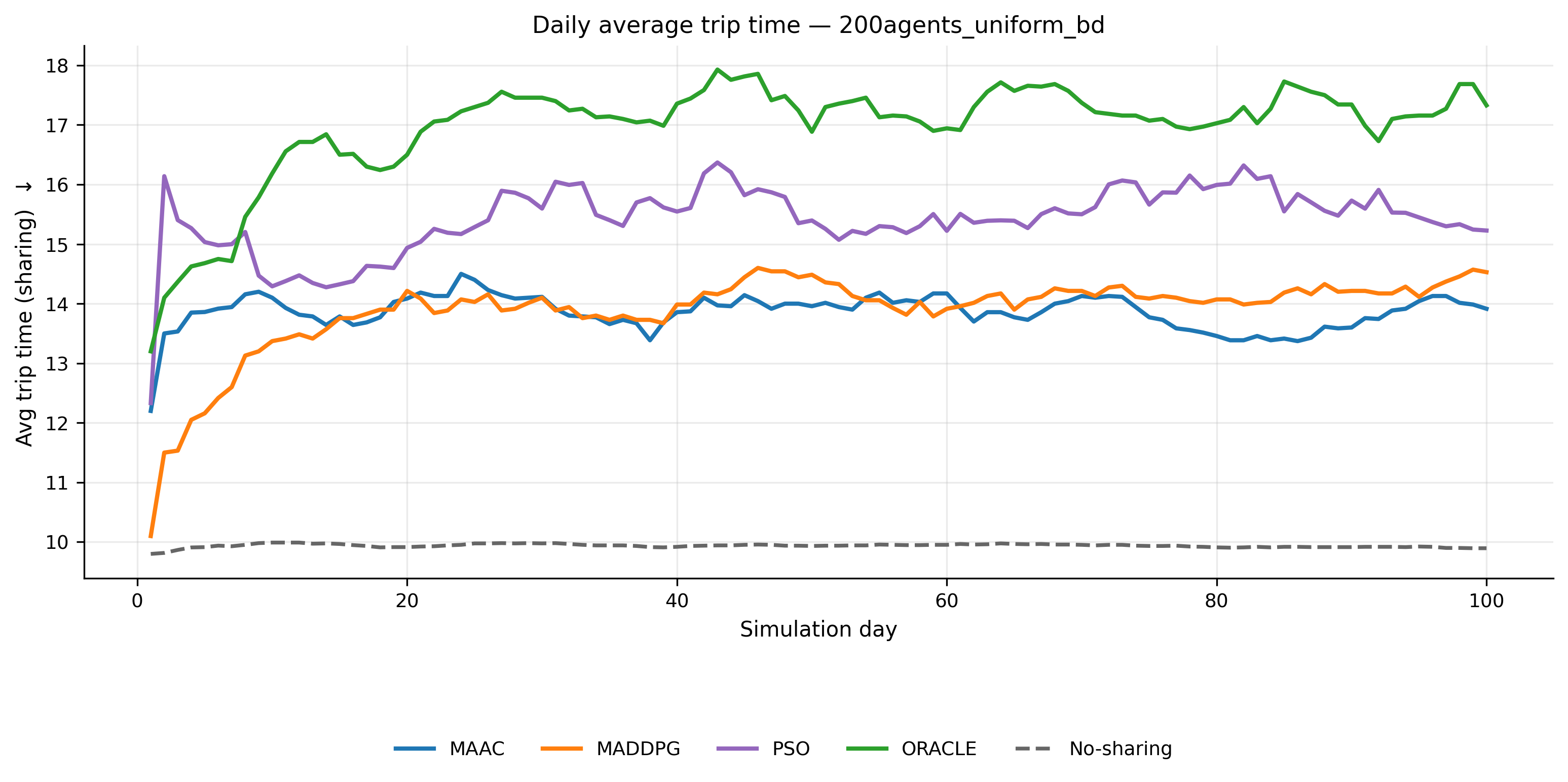}}
\caption{Uniform allocation; agents may enter and exit (Birth–Death)}
\end{subfigure}
\hfill
\begin{subfigure}{0.48\textwidth}
\centering
\includegraphics[width=\linewidth]{{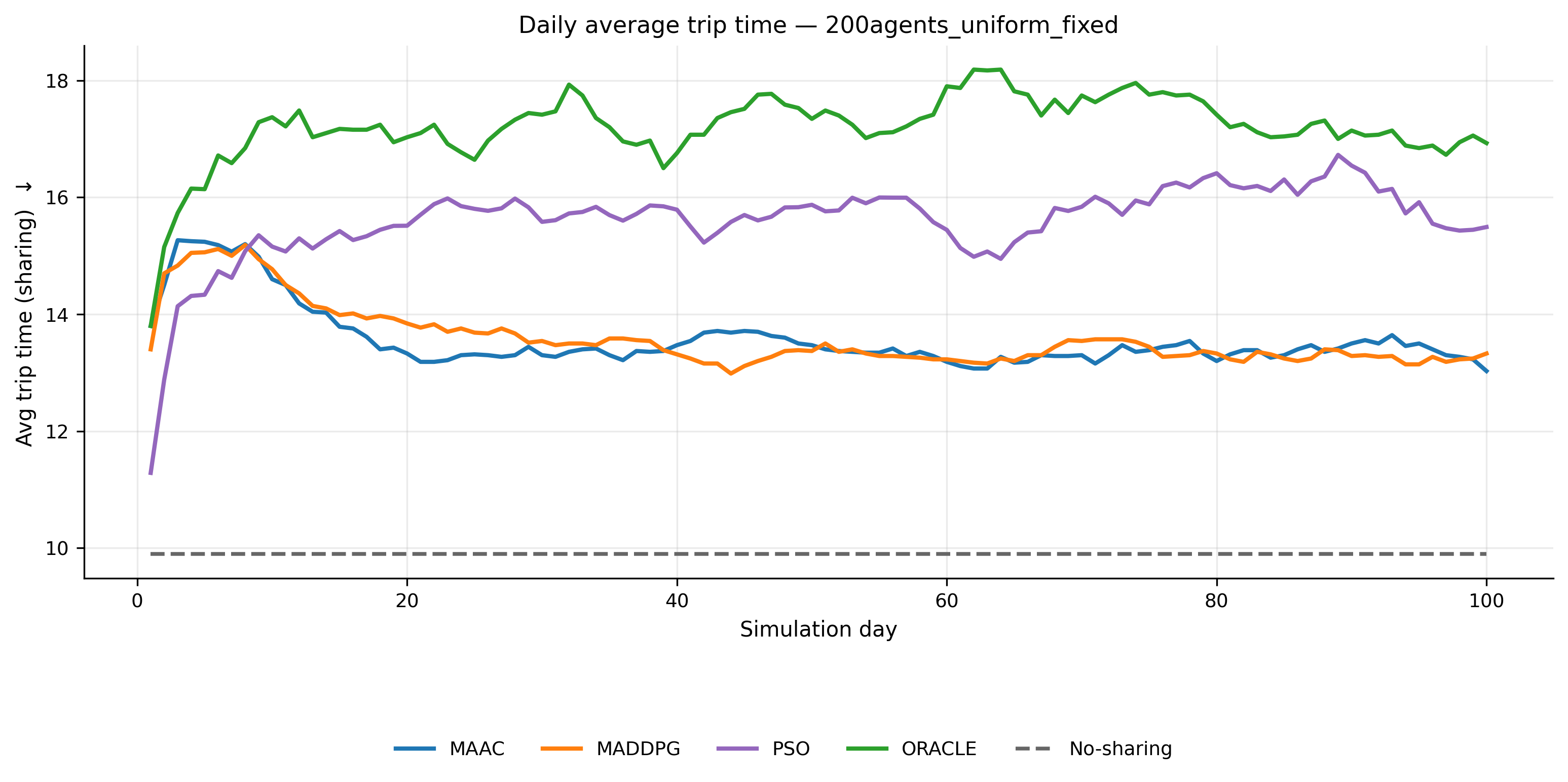}}
\caption{Uniform allocation; no entry and exit of agents}
\end{subfigure}

\caption{Daily trip time with 200 agents.}
\end{figure}

\begin{figure}[H]
\centering
\begin{subfigure}{0.48\textwidth}
\centering
\includegraphics[width=\linewidth]{{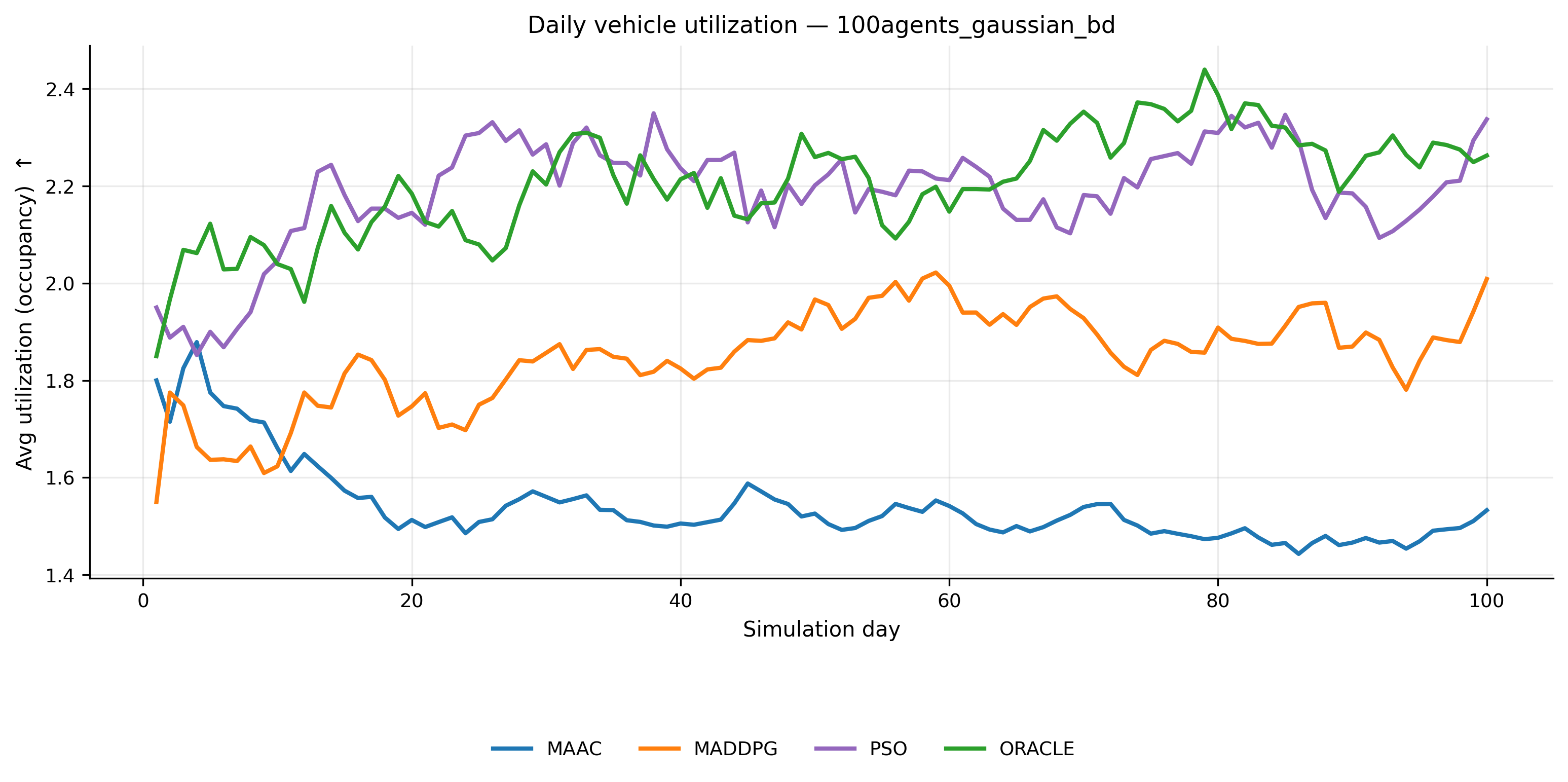}}
\caption{Gaussian allocation; agents may enter and exit (Birth–Death)}
\end{subfigure}
\hfill
\begin{subfigure}{0.48\textwidth}
\centering
\includegraphics[width=\linewidth]{{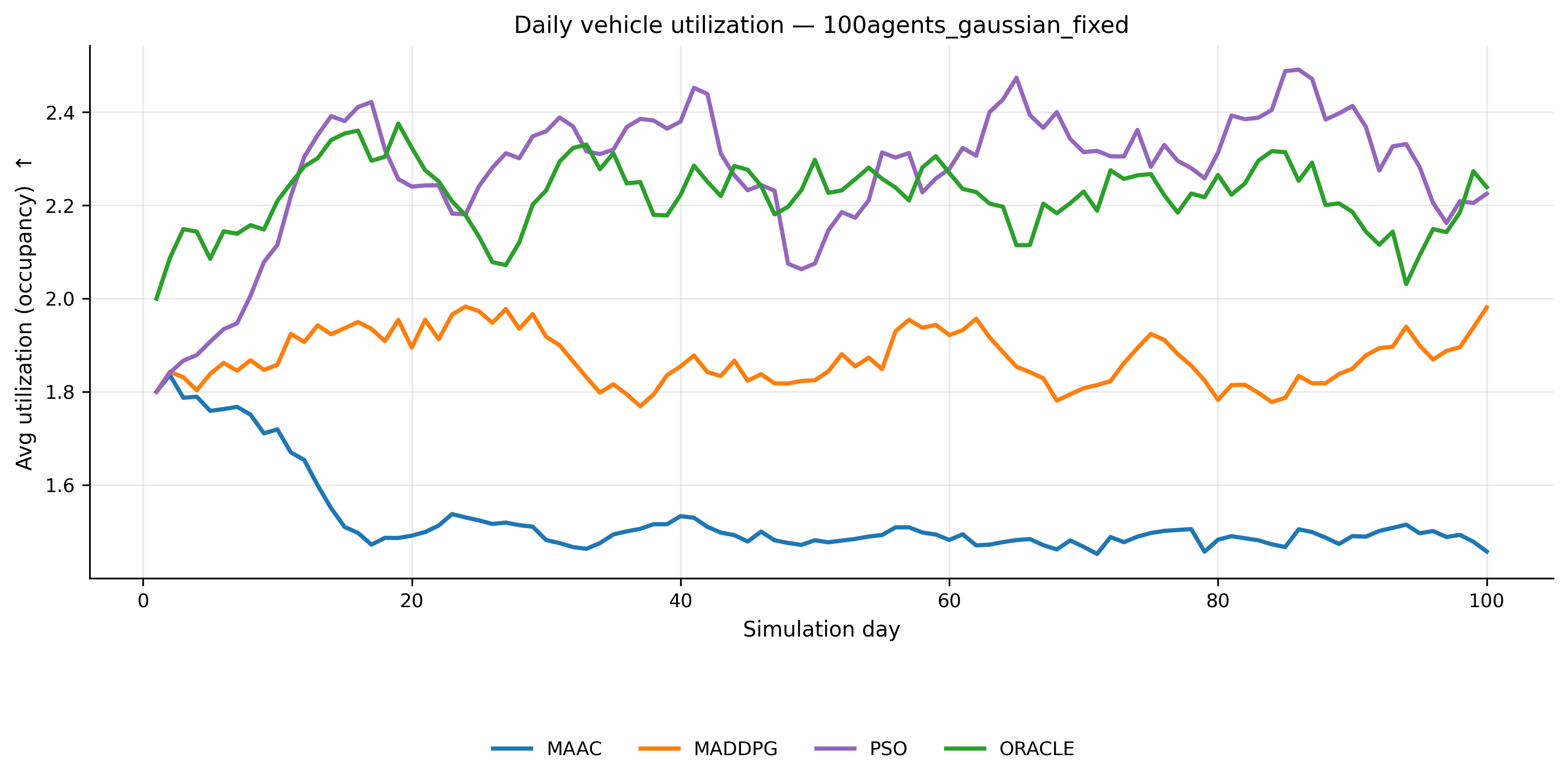}}
\caption{Gaussian allocation; no entry and exit of agents}
\end{subfigure}

\vspace{0.5em}

\begin{subfigure}{0.48\textwidth}
\centering
\includegraphics[width=\linewidth]{{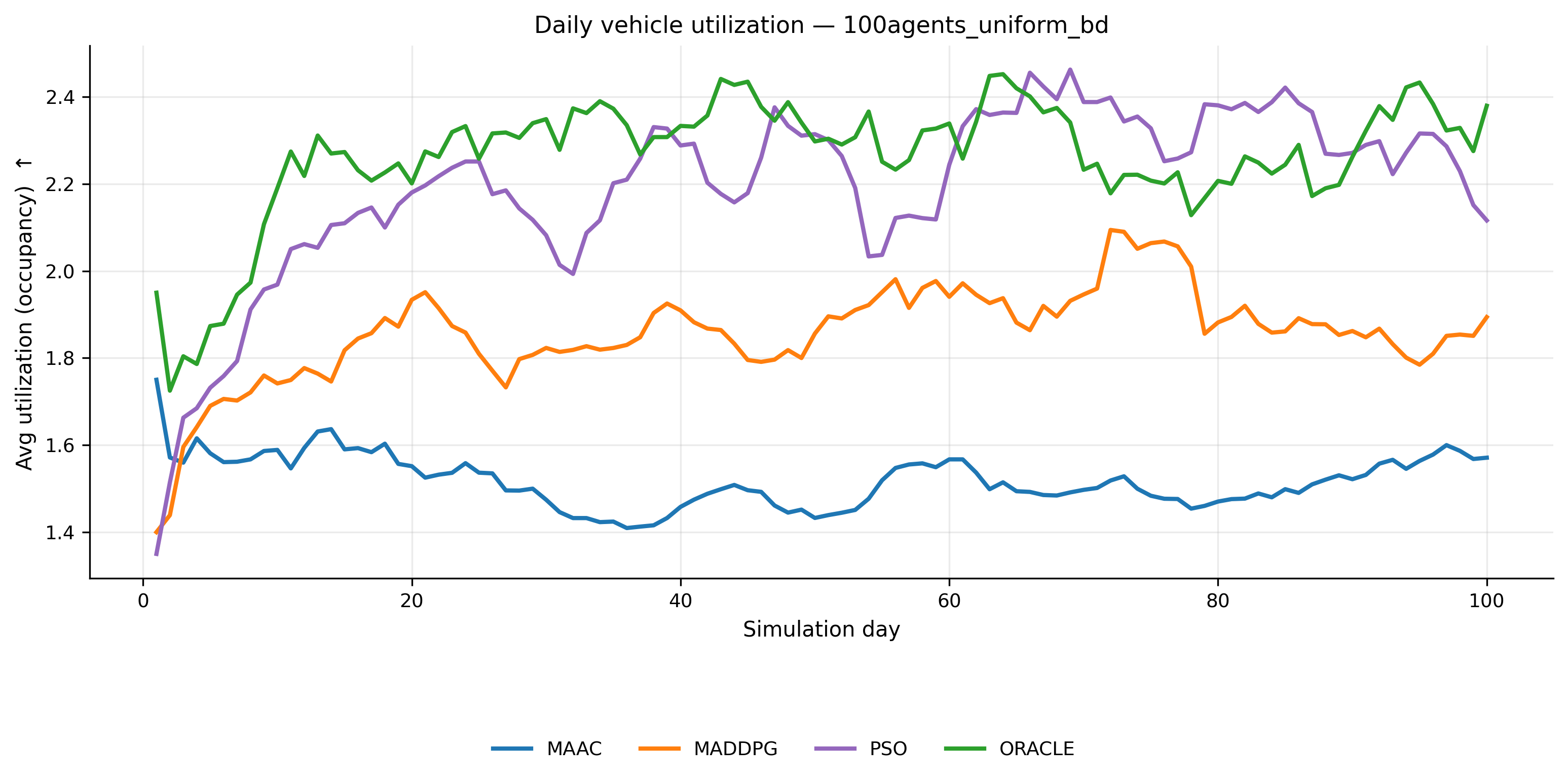}}
\caption{Uniform allocation; agents may enter and exit (Birth–Death)}
\end{subfigure}
\hfill
\begin{subfigure}{0.48\textwidth}
\centering
\includegraphics[width=\linewidth]{{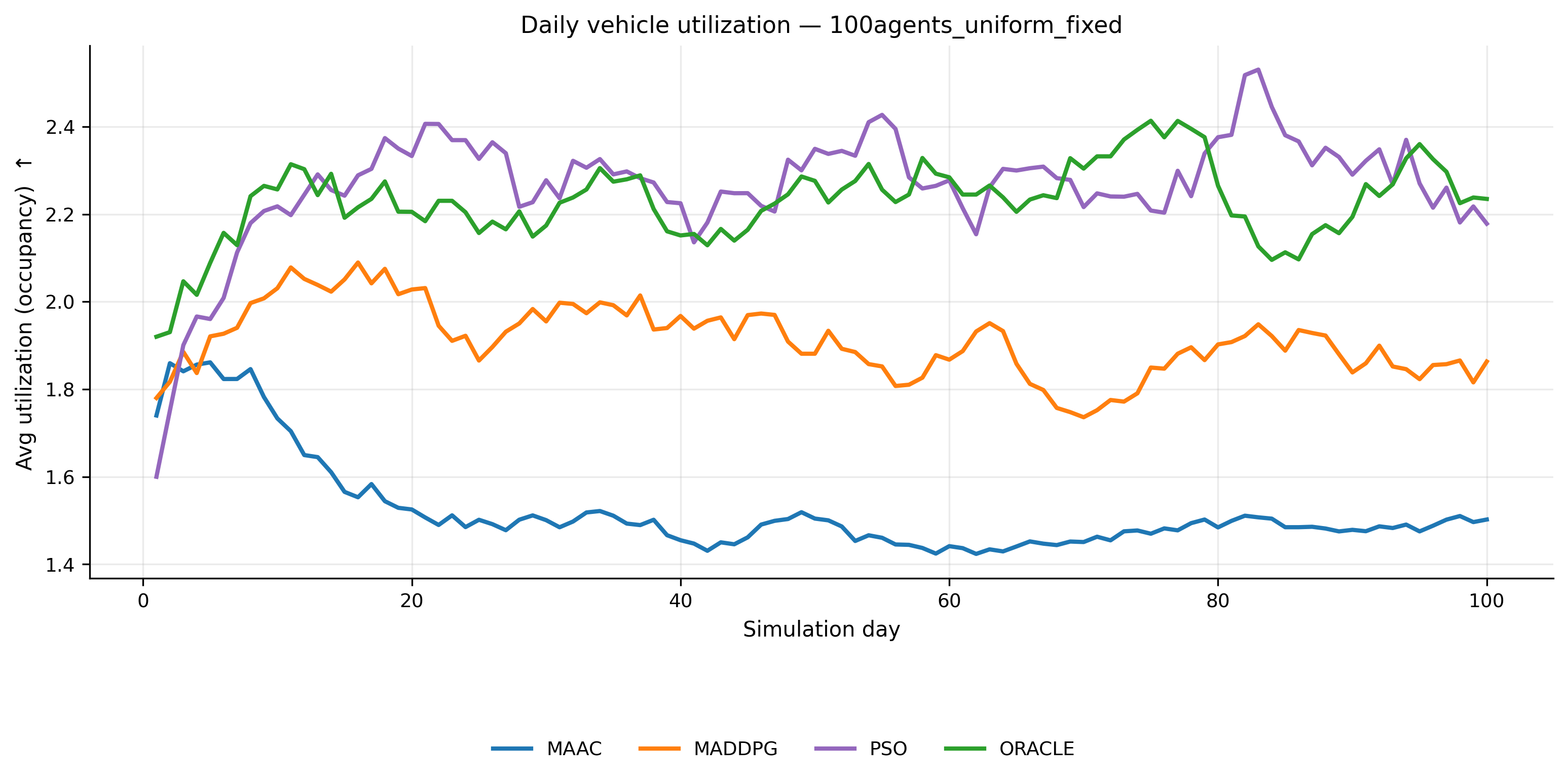}}
\caption{Uniform allocation; no entry and exit of agents}
\end{subfigure}

\caption{Daily vehicle utilization with 100 agents.}
\end{figure}

\begin{figure}[H]
\centering
\begin{subfigure}{0.48\textwidth}
\centering
\includegraphics[width=\linewidth]{{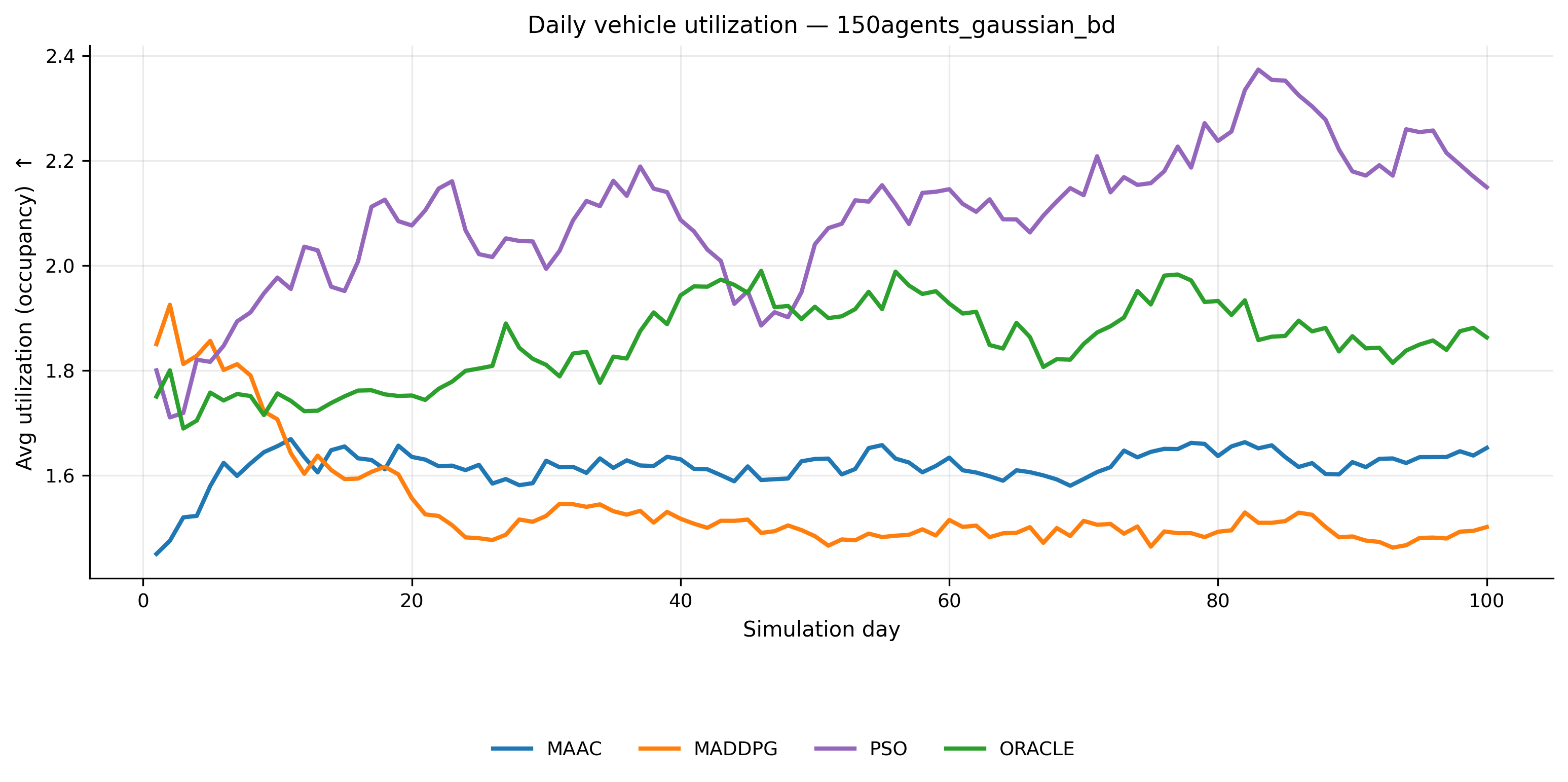}}
\caption{Gaussian allocation; agents may enter and exit (Birth–Death)}
\end{subfigure}
\hfill
\begin{subfigure}{0.48\textwidth}
\centering
\includegraphics[width=\linewidth]{{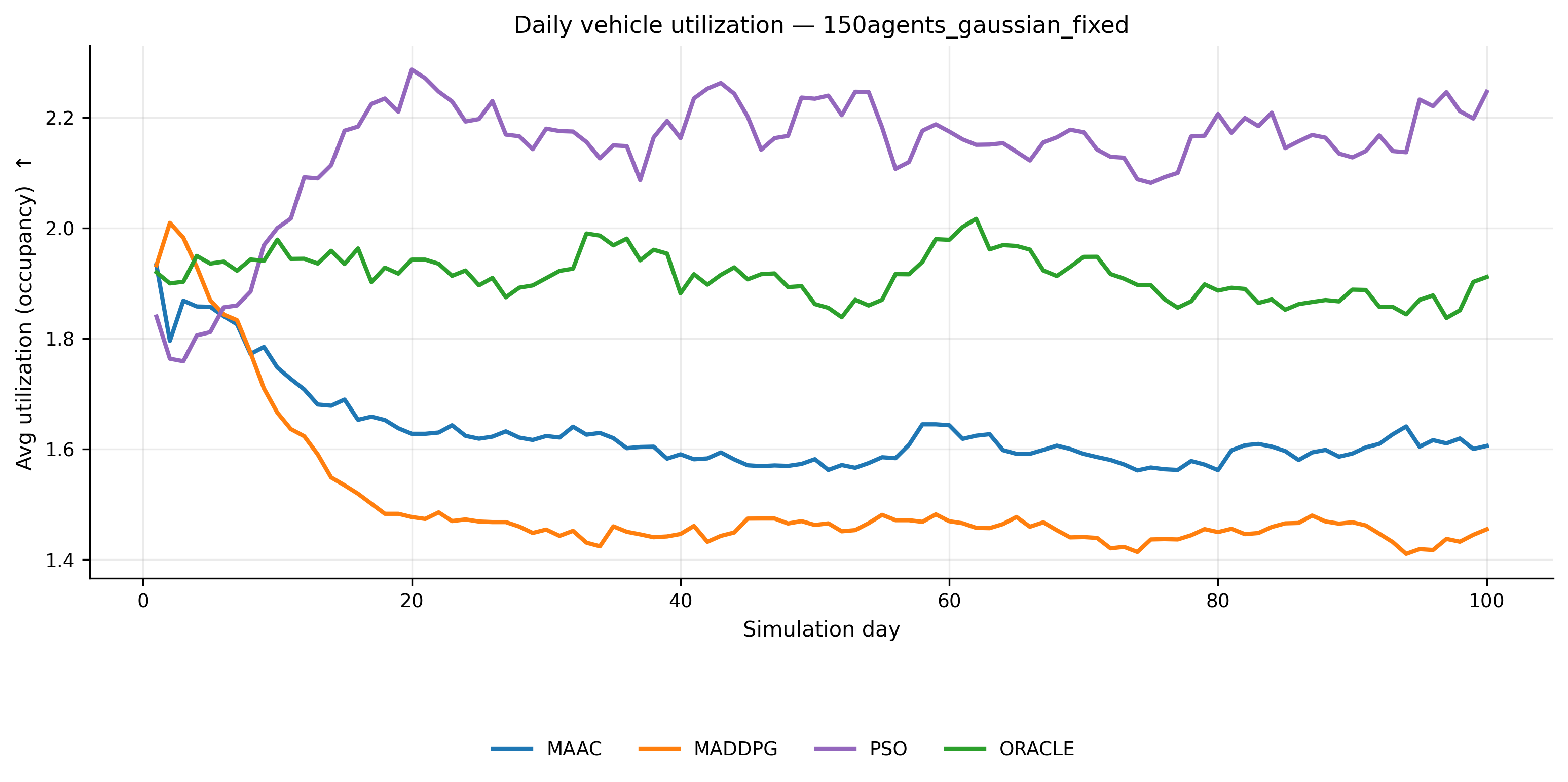}}
\caption{Gaussian allocation; no entry and exit of agents}
\end{subfigure}

\vspace{0.5em}

\begin{subfigure}{0.48\textwidth}
\centering
\includegraphics[width=\linewidth]{{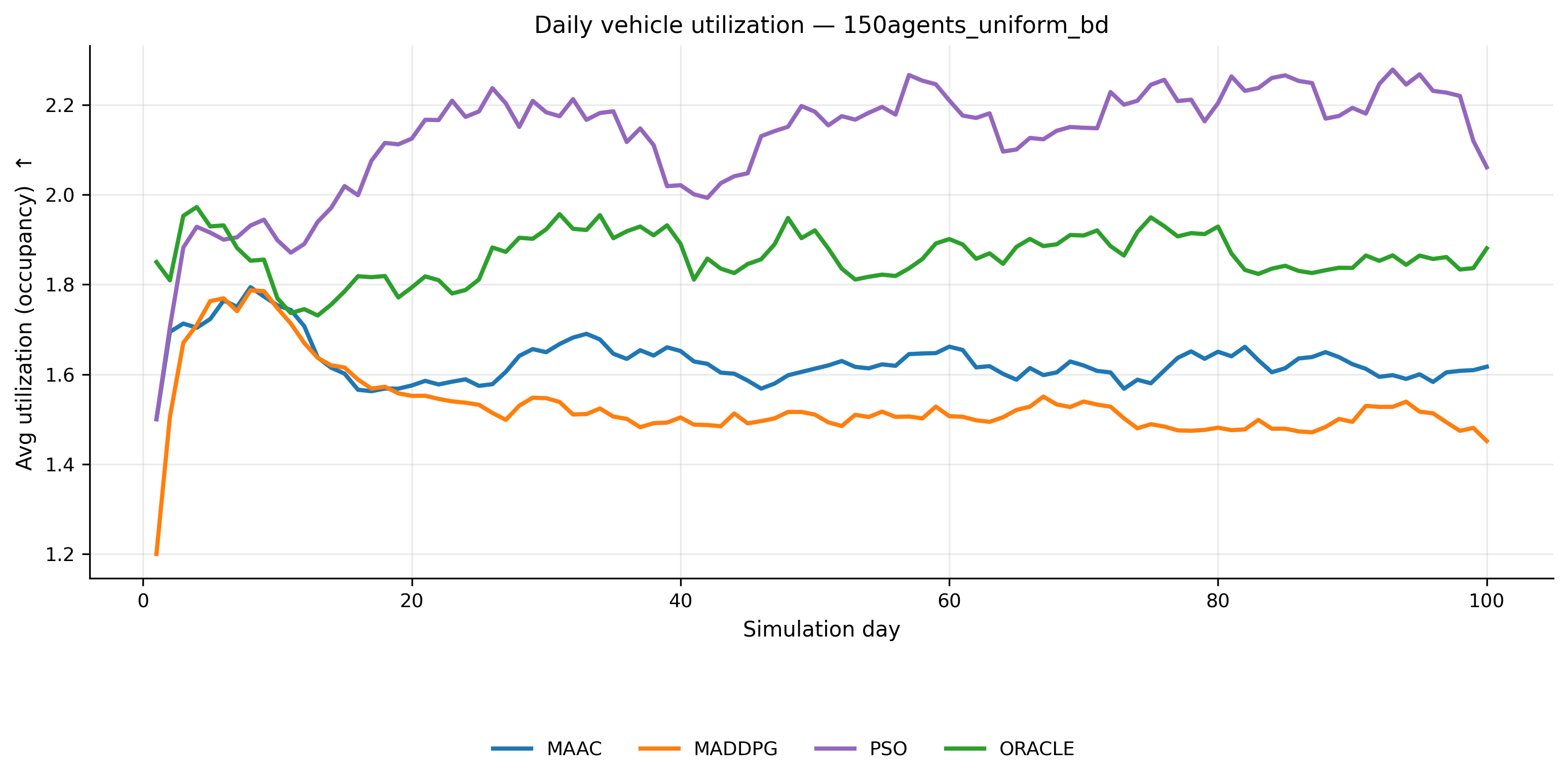}}
\caption{Uniform allocation; agents may enter and exit (Birth–Death)}
\end{subfigure}
\hfill
\begin{subfigure}{0.48\textwidth}
\centering
\includegraphics[width=\linewidth]{{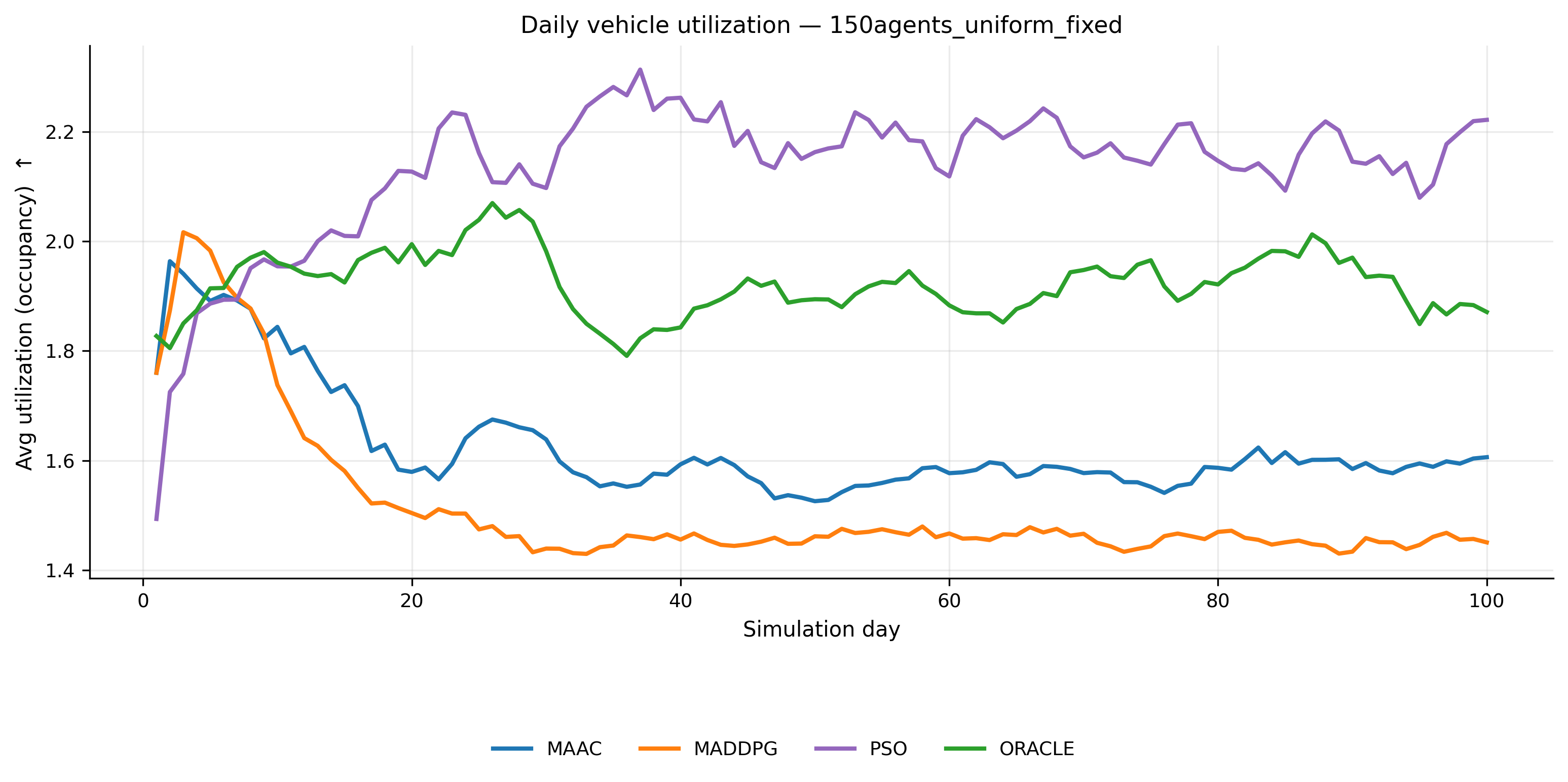}}
\caption{Uniform allocation; no entry and exit of agents}
\end{subfigure}

\caption{Daily vehicle utilization with 150 agents.}
\end{figure}

\begin{figure}[H]
\centering
\begin{subfigure}{0.48\textwidth}
\centering
\includegraphics[width=\linewidth]{{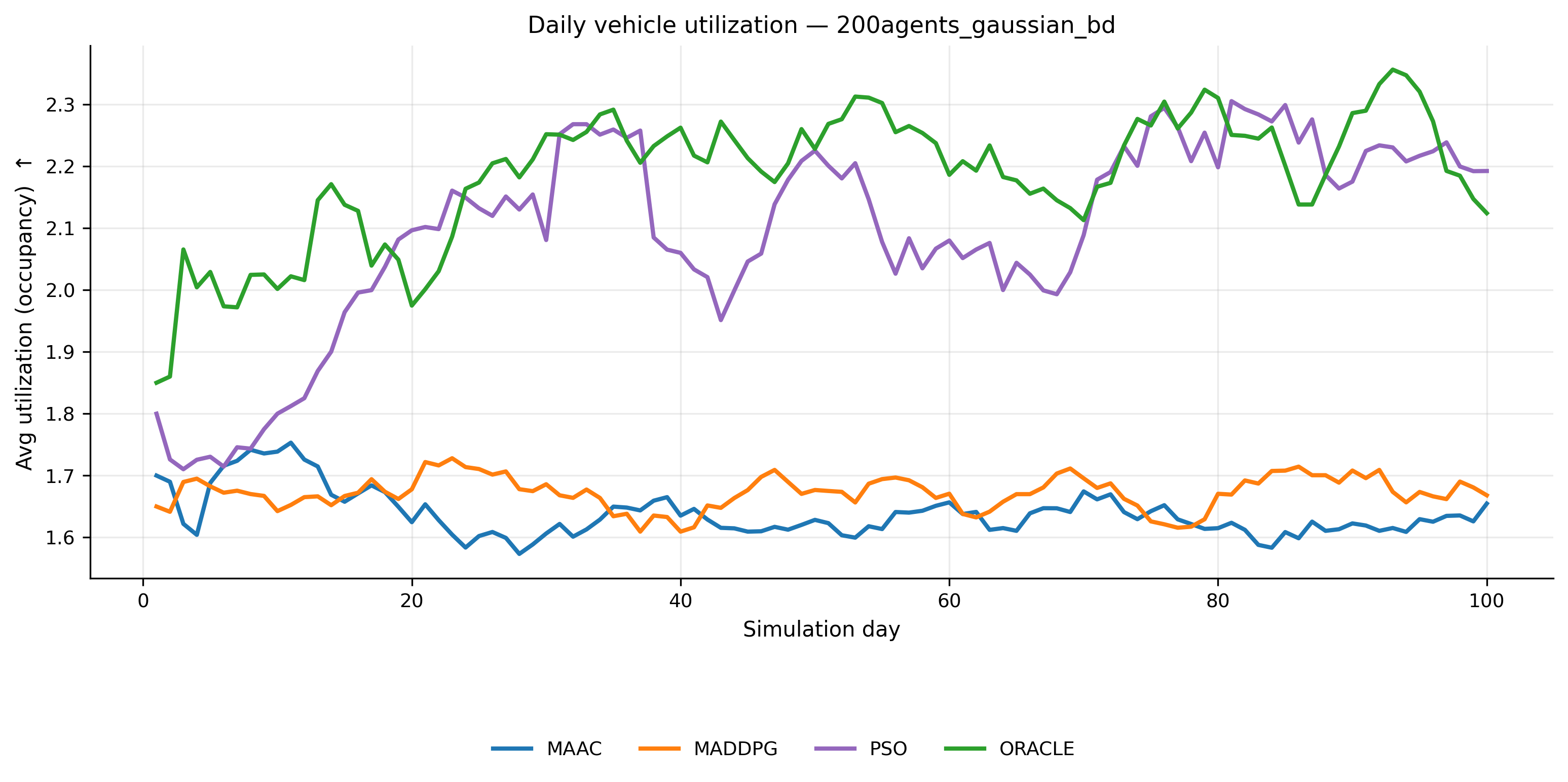}}
\caption{Gaussian allocation; agents may enter and exit (Birth–Death)}
\end{subfigure}
\hfill
\begin{subfigure}{0.48\textwidth}
\centering
\includegraphics[width=\linewidth]{{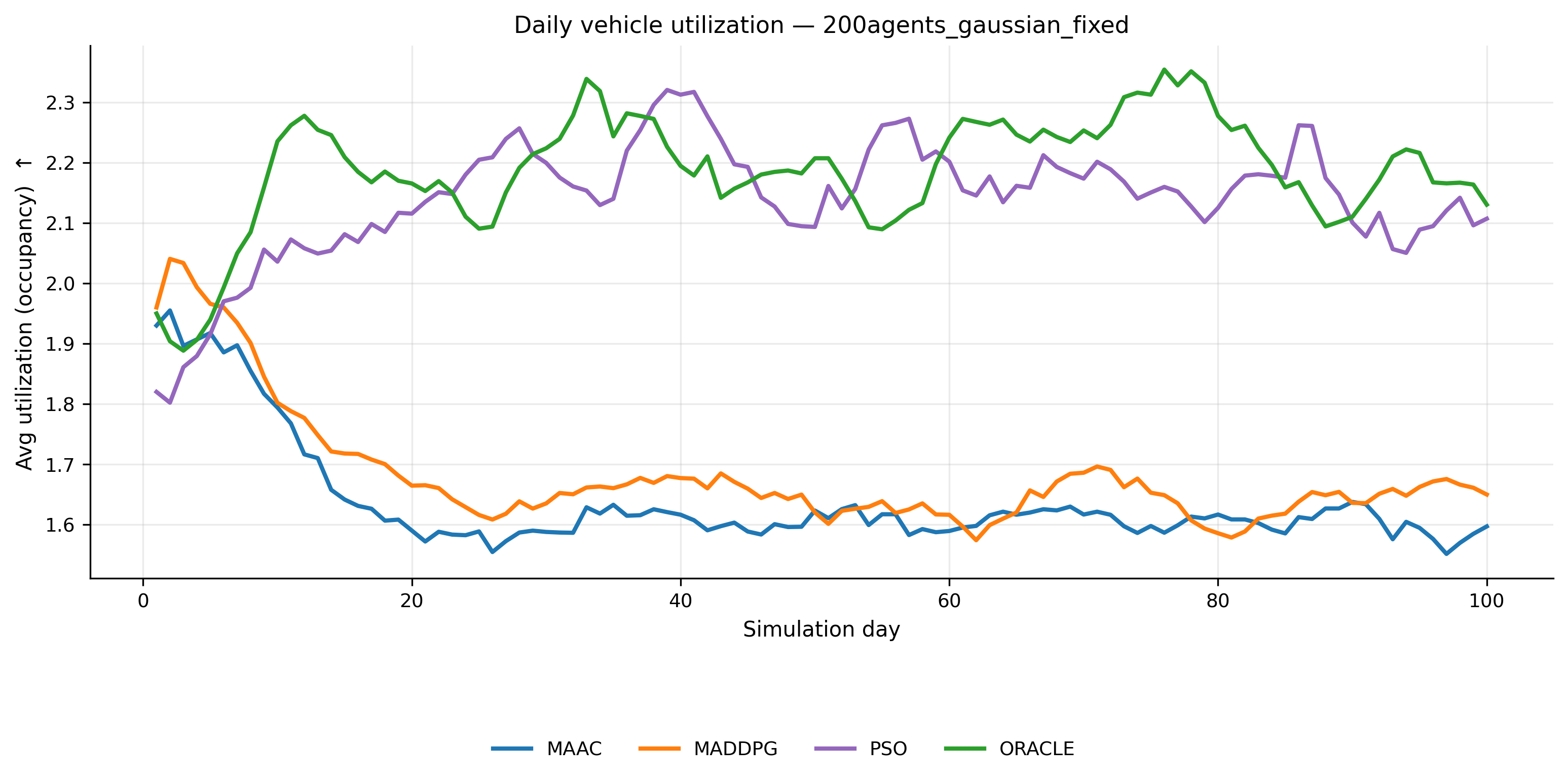}}
\caption{Gaussian allocation; no entry and exit of agents}
\end{subfigure}

\vspace{0.5em}

\begin{subfigure}{0.48\textwidth}
\centering
\includegraphics[width=\linewidth]{{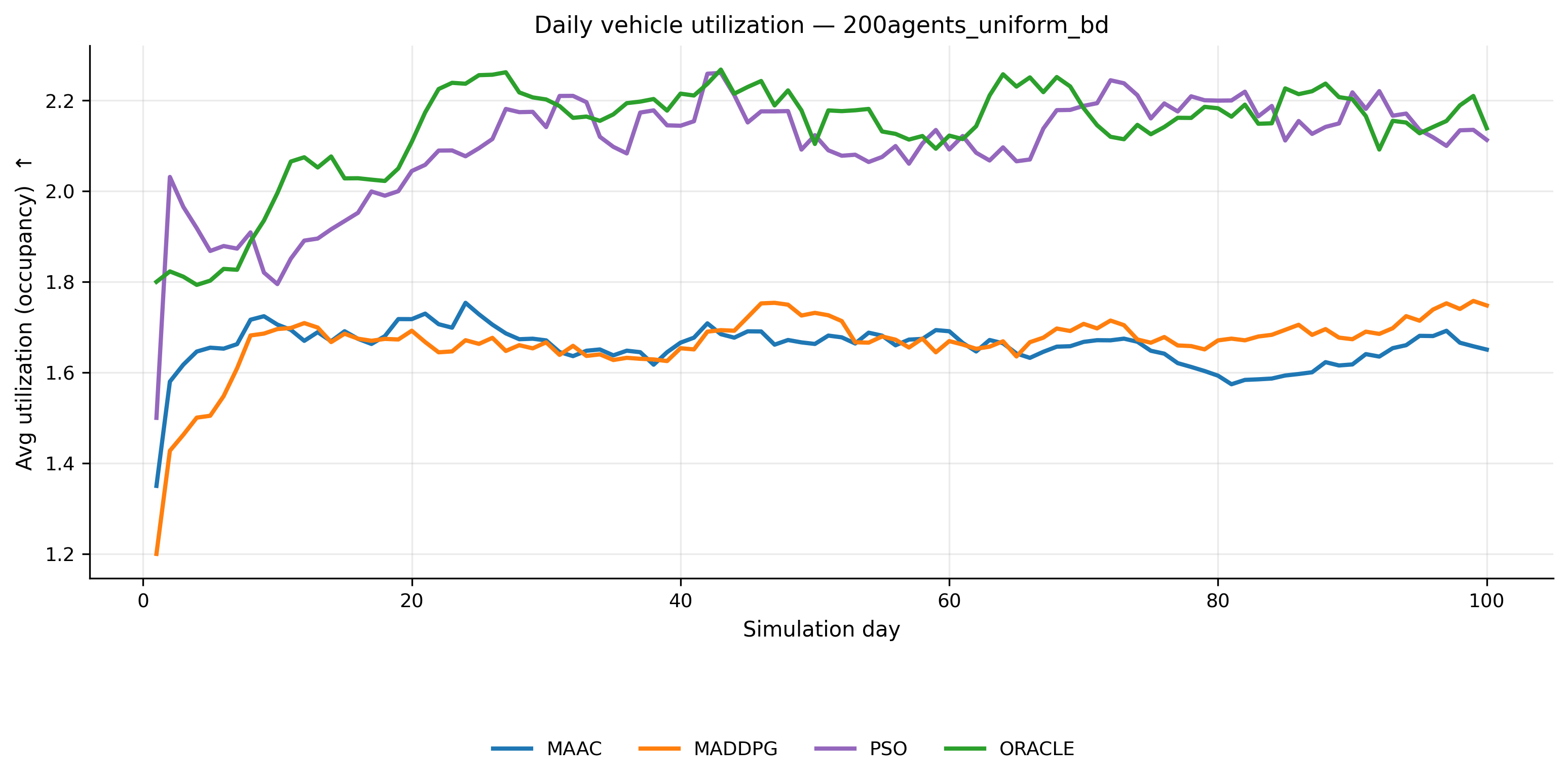}}
\caption{Uniform allocation; agents may enter and exit (Birth–Death)}
\end{subfigure}
\hfill
\begin{subfigure}{0.48\textwidth}
\centering
\includegraphics[width=\linewidth]{{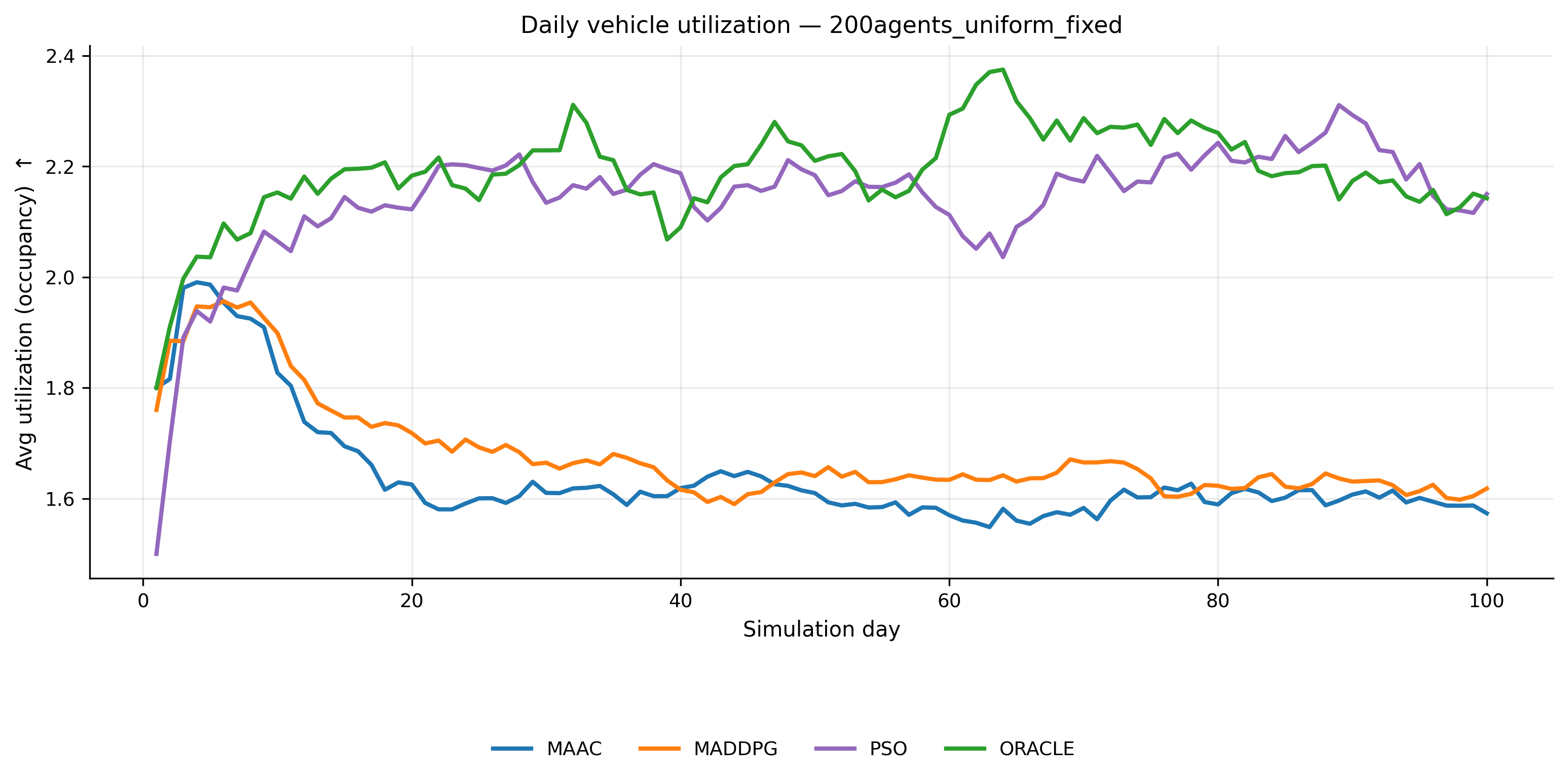}}
\caption{Uniform allocation; no entry and exit of agents}
\end{subfigure}

\caption{Daily vehicle utilization with 200 agents.}
\end{figure}

\begin{figure}[H]
\centering
\includegraphics[width=0.8\linewidth]{{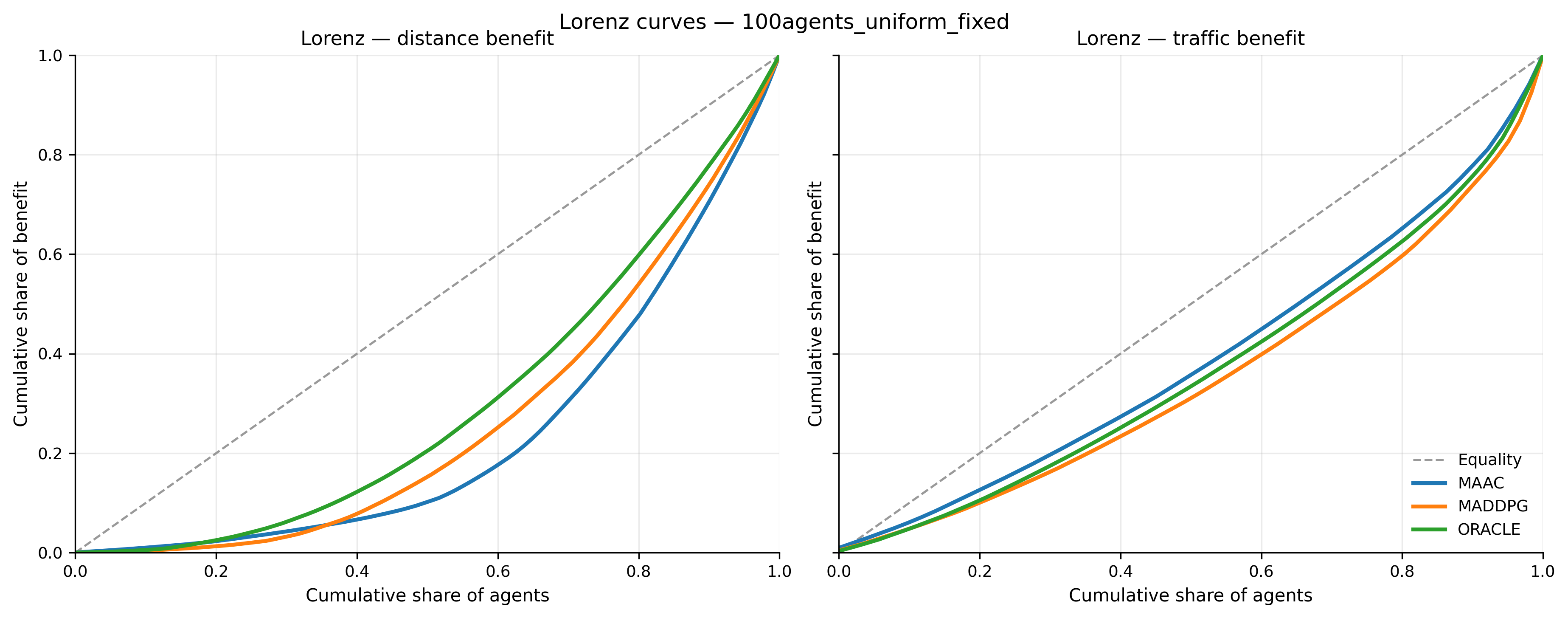}}
\caption{Lorenz curve overlay illustrating distributional inequality in outcomes for the 100-agent setting with uniform altruism allocation and no entry or exit of agents.}
\end{figure}

%% ===================================================================
\begin{figure}[H]
    \centering
    \includegraphics[width=0.6\textwidth]{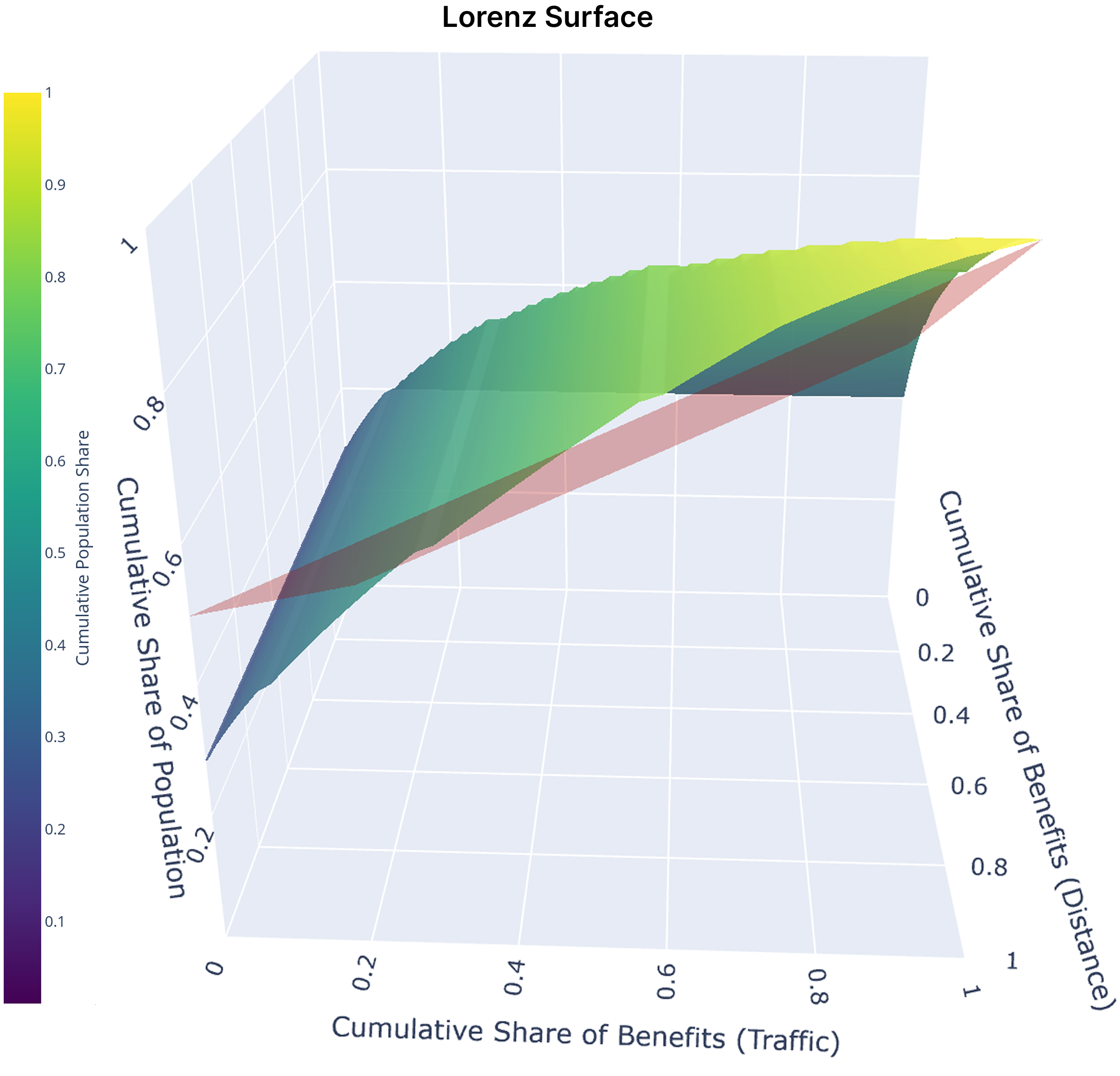}
    \vspace{5pt}
    \caption {Three-dimensional Lorenz surface representing joint inequality in cumulative shares of population, distance benefits, and traffic benefits from ride-sharing. The surface deviates from the diagonal plane (perfect equality), highlighting heterogeneity in benefit distribution.
Positive distance benefits Gini $\approx 0.471$, indicating moderate inequality in distance savings.
Traffic Benefits Gini $\approx$ 0.235, indicating relatively more equitable distribution of congestion-related gains.
A Granger-causality analysis reveals that past altruism significantly predicts subsequent benefits (p = 0.0038 at the system level), providing evidence that heterogeneity in altruism contributes to the observed concentration of benefits.}
\end{figure}

\begin{figure}[H]
\centering
\includegraphics[width=0.8\linewidth]{{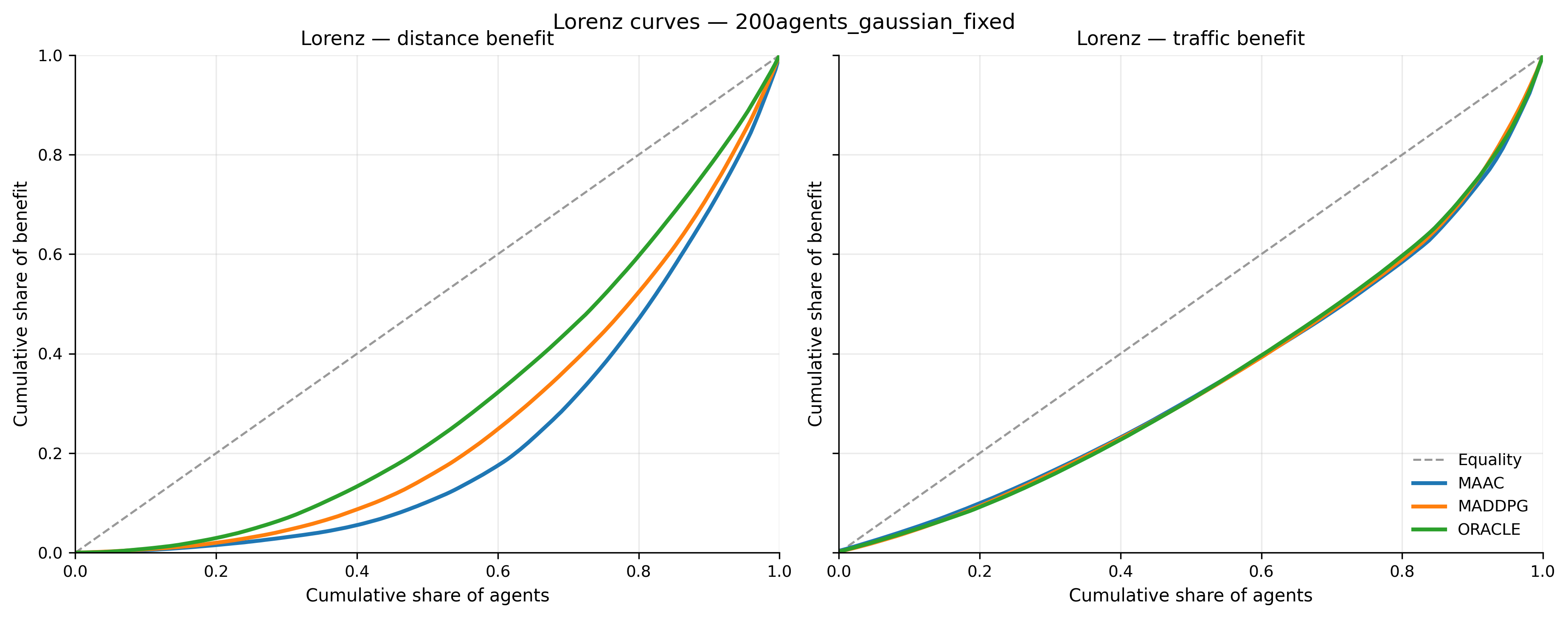}}
\caption{Lorenz curve overlay illustrating inequality in system outcomes for the 200-agent scenario with Gaussian altruism allocation and no entry or exit of agents.}
\end{figure}

% --- Interaction analyses ---
\begin{figure}[H]
\centering
\includegraphics[width=0.8\linewidth]{{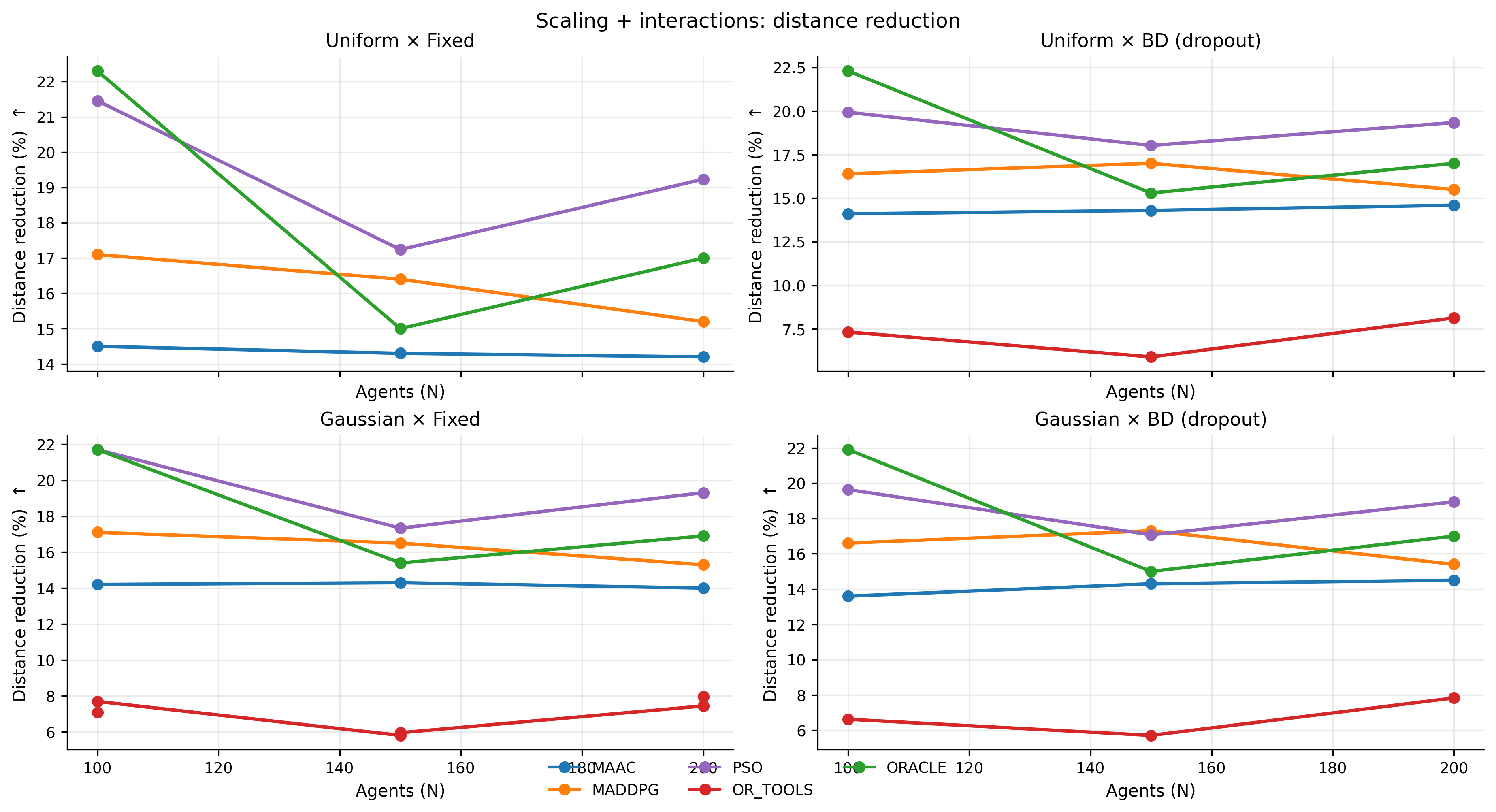}}
\caption{Interaction analysis showing the relationship between altruism mechanisms and overall distance reduction in the transportation system.}
\end{figure}

\begin{figure}[H]
\centering
\includegraphics[width=0.8\linewidth]{{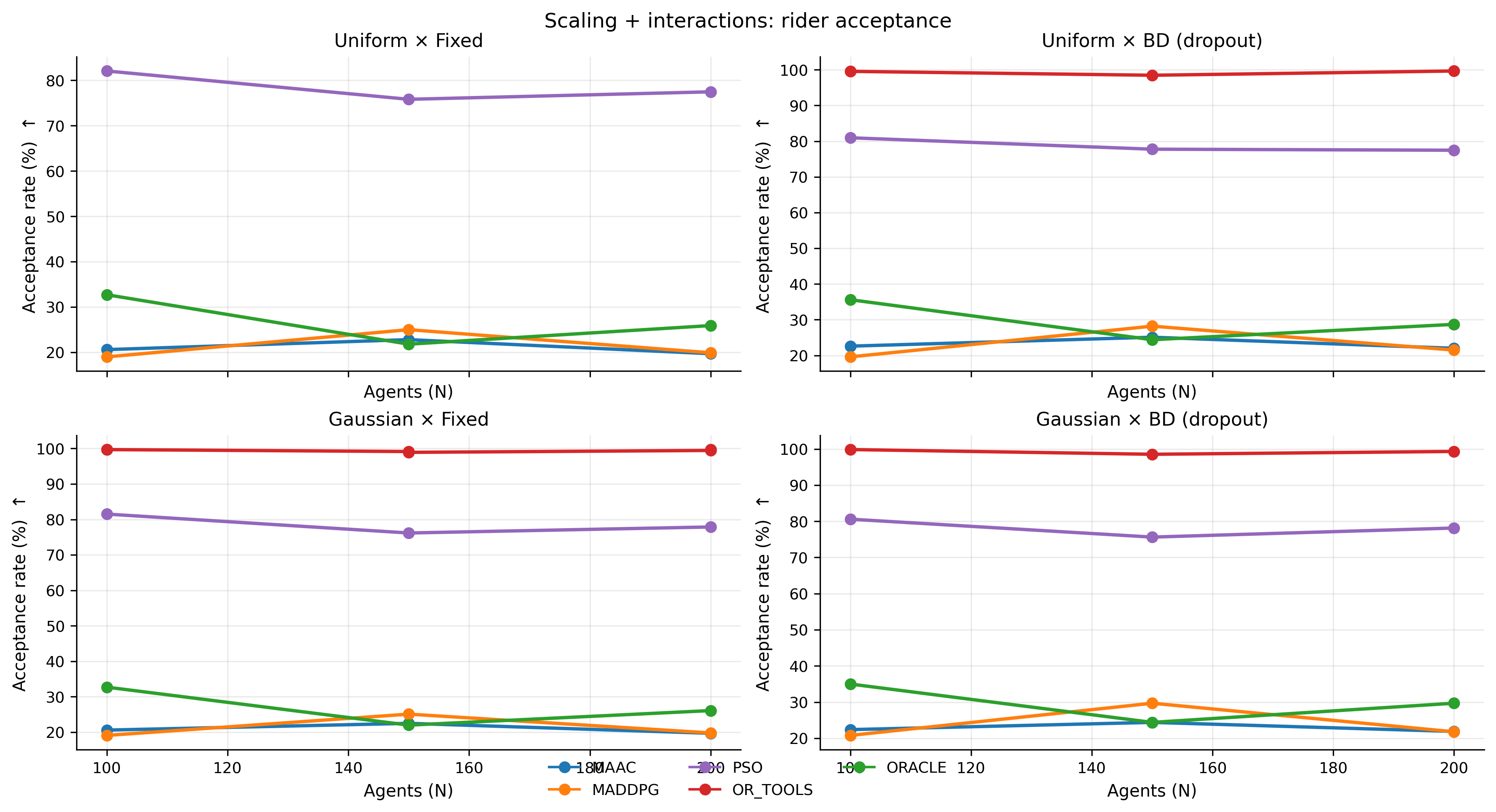}}
\caption{Interaction effects illustrating how altruism dynamics influence ride acceptance rates across agents.}
\end{figure}

\begin{figure}[H]
\centering
\includegraphics[width=0.8\linewidth]{{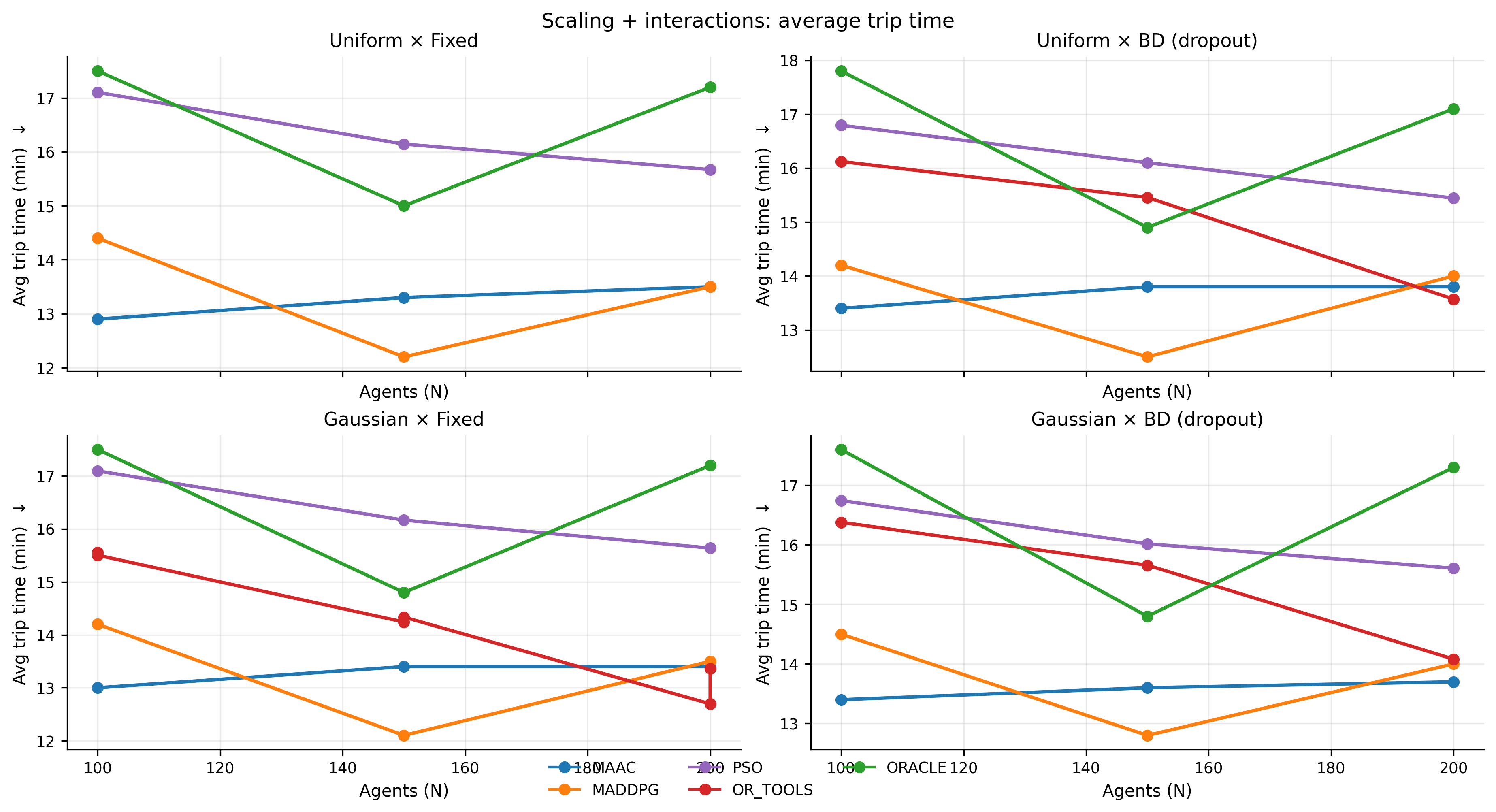}}
\caption{Interaction effects between system parameters and average trip time outcomes.}
\end{figure}

\begin{figure}[H]
\centering
\includegraphics[width=0.8\linewidth]{{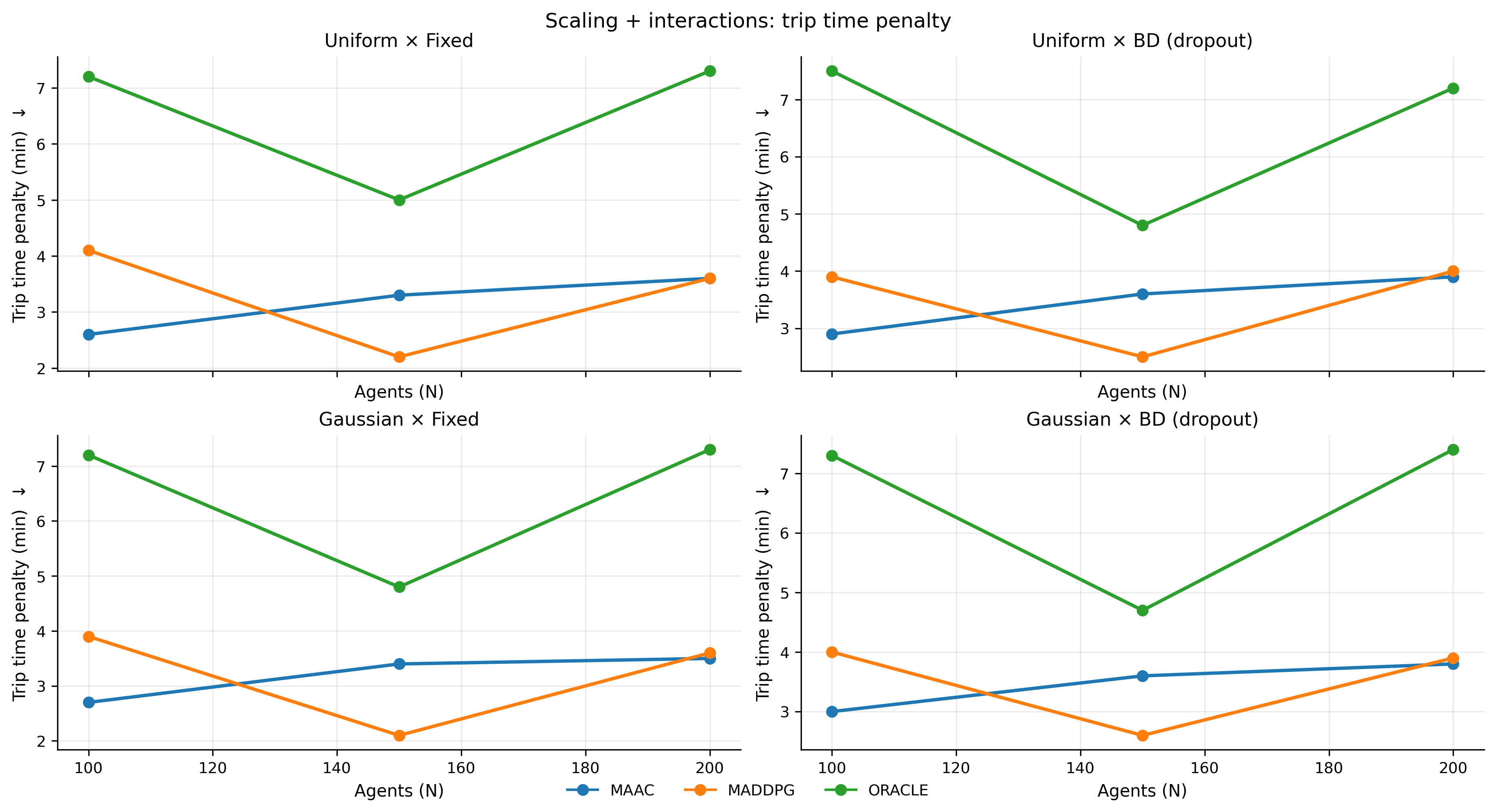}}
\caption{Interaction analysis showing changes in trip time relative to baseline system configurations.}
\end{figure}

\begin{figure}[H]
\centering
\includegraphics[width=0.8\linewidth]{{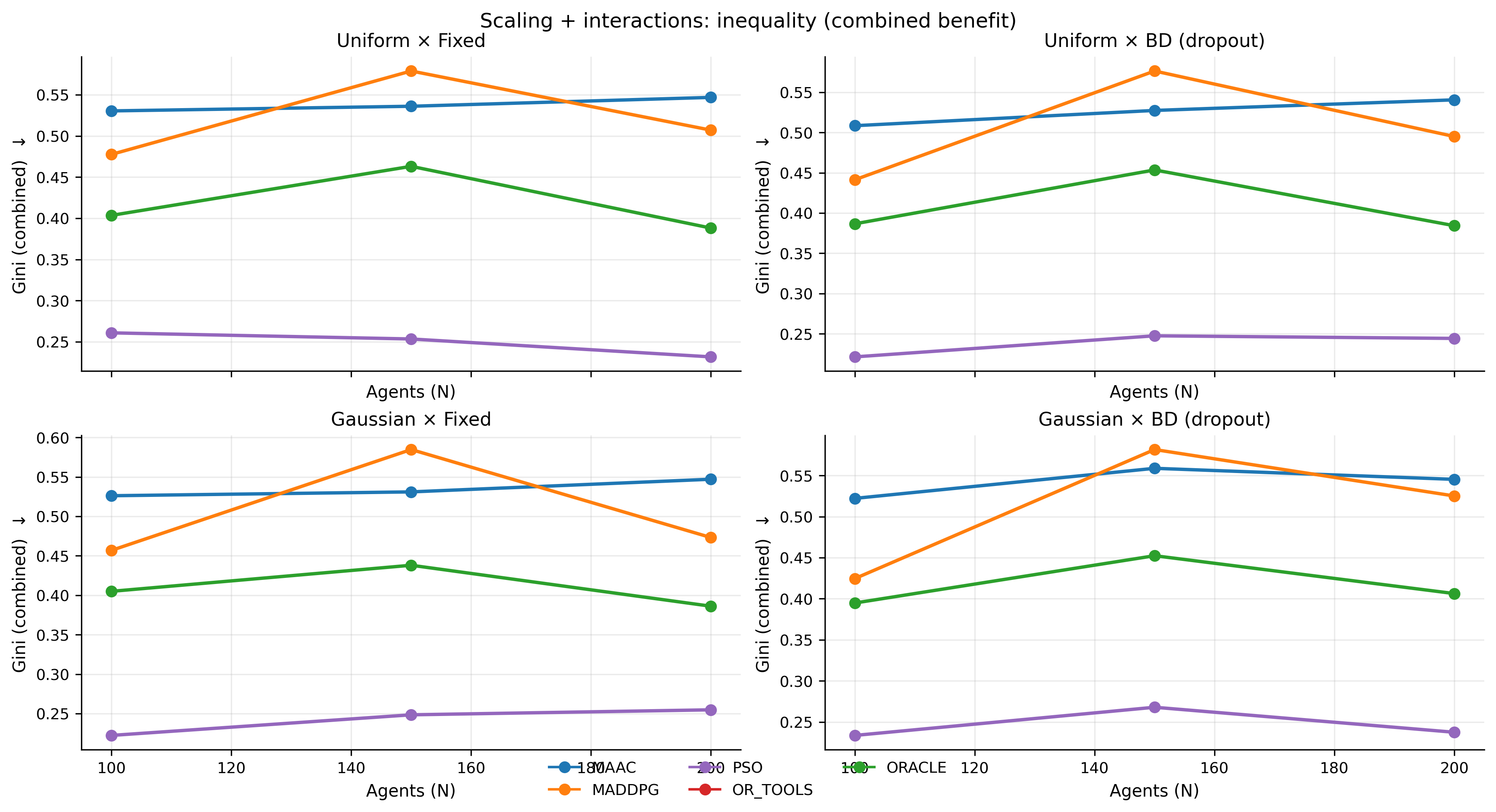}}
\caption{Interaction effects on the Gini coefficient, measuring inequality in benefits among participating agents.}
\end{figure}

\begin{figure}[H]
\centering
\includegraphics[width=0.8\linewidth]{{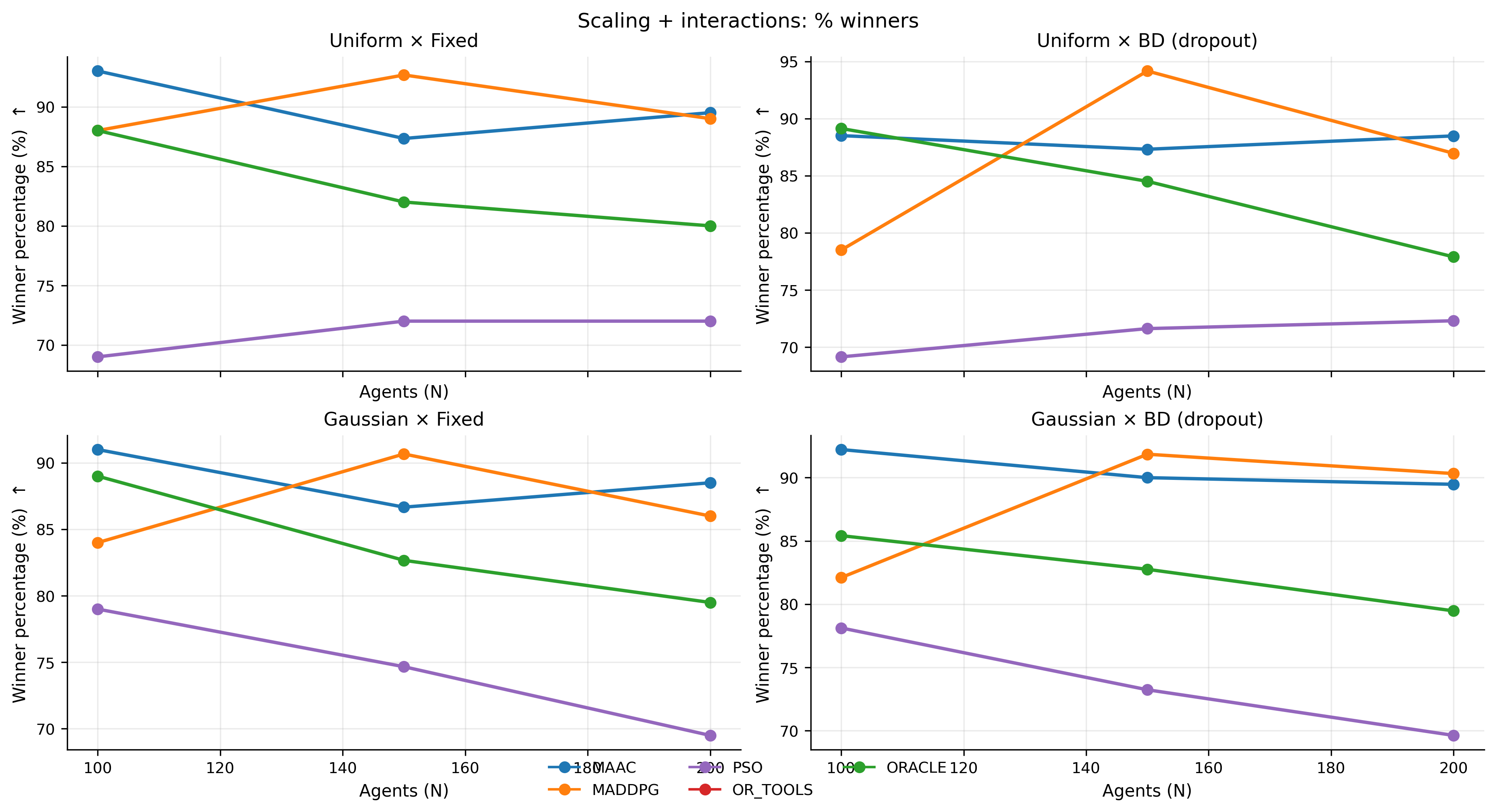}}
\caption{Analysis of agents benefiting most under different altruism-based allocation mechanisms.}
\end{figure}

\begin{figure}[H]
\centering

\begin{subfigure}{0.48\textwidth}
\centering
\includegraphics[width=\linewidth]{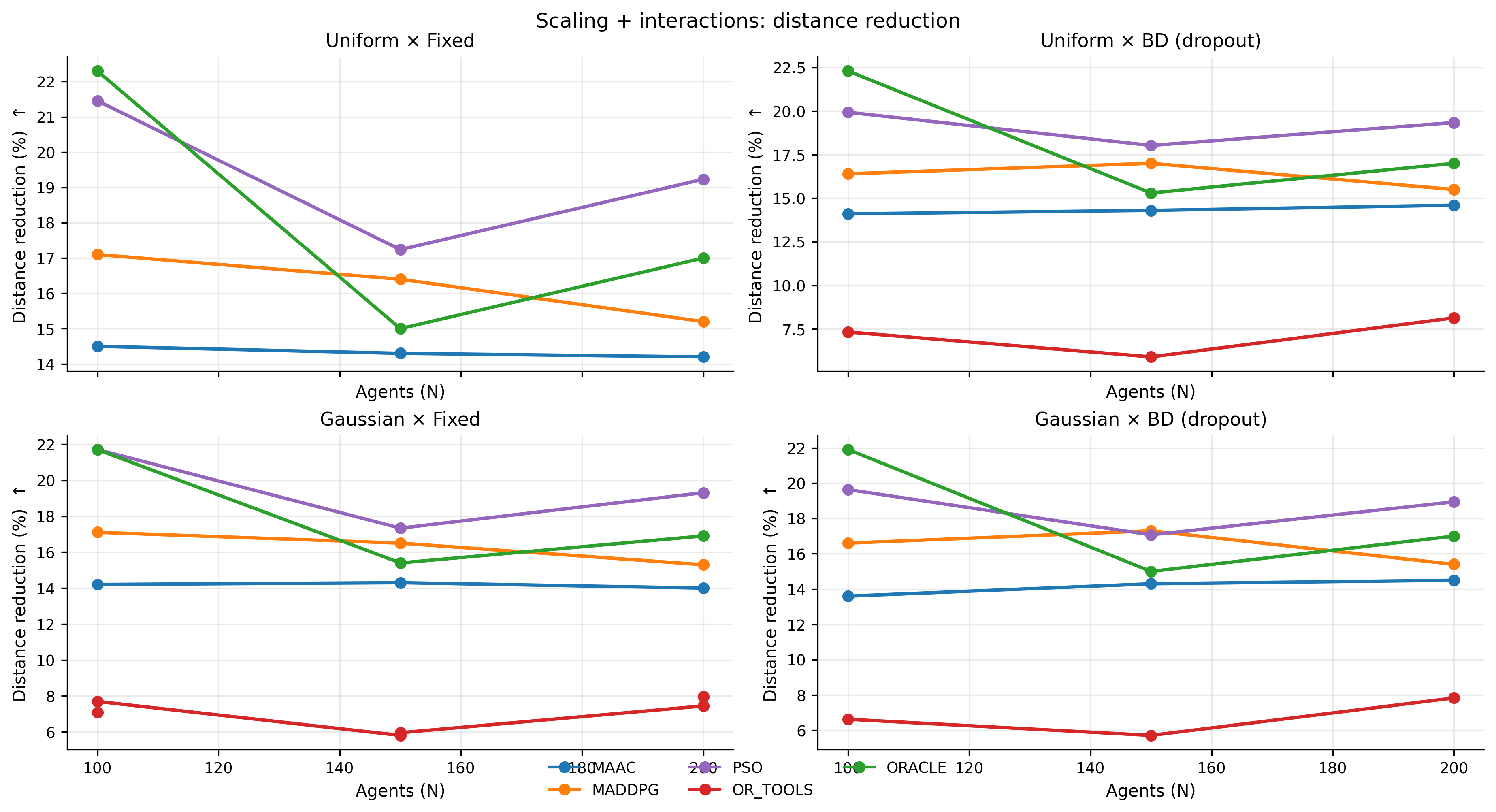}
\caption{Impact of altruism mechanisms on overall distance reduction.}
\end{subfigure}
\hfill
\begin{subfigure}{0.48\textwidth}
\centering
\includegraphics[width=\linewidth]{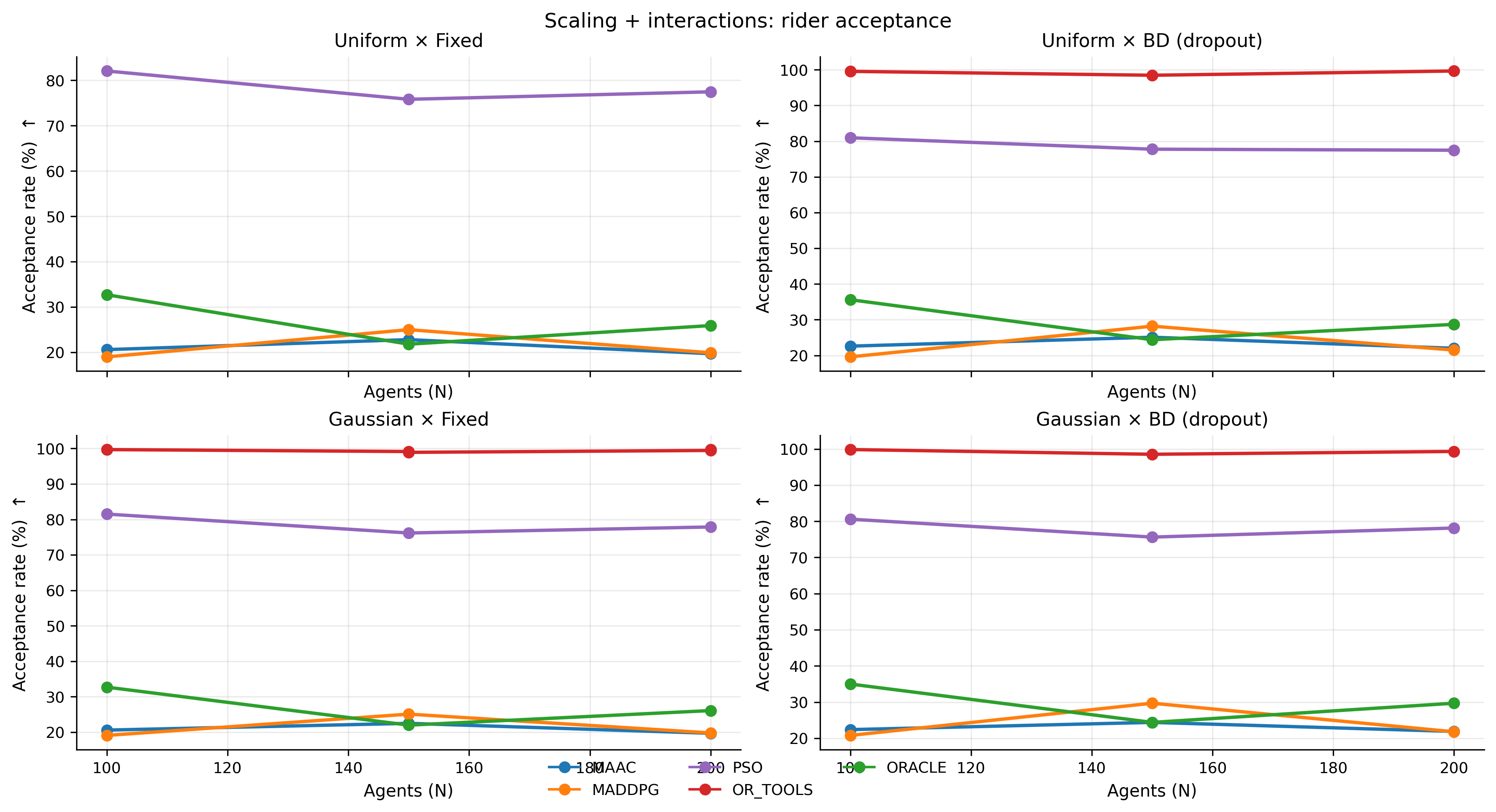}
\caption{Impact of altruism dynamics on ride acceptance rates.}
\end{subfigure}

\caption{Interaction analyses showing how altruism mechanisms affect key transportation metrics.}
\end{figure}

% --- Distribution and family analyses ---
\begin{figure}[H]
\centering
\includegraphics[width=0.8\linewidth]{{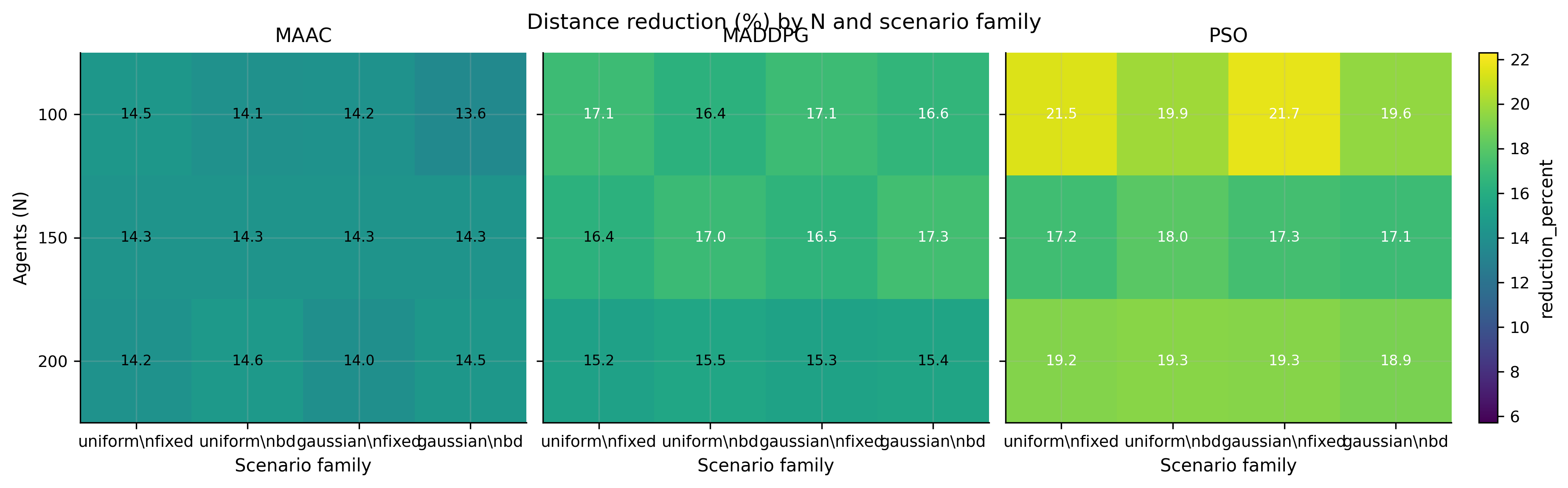}}
\caption{Heatmap illustrating distance reduction across families of experimental configurations.}
\end{figure}

\begin{figure}[H]
\centering
\includegraphics[width=0.8\linewidth]{{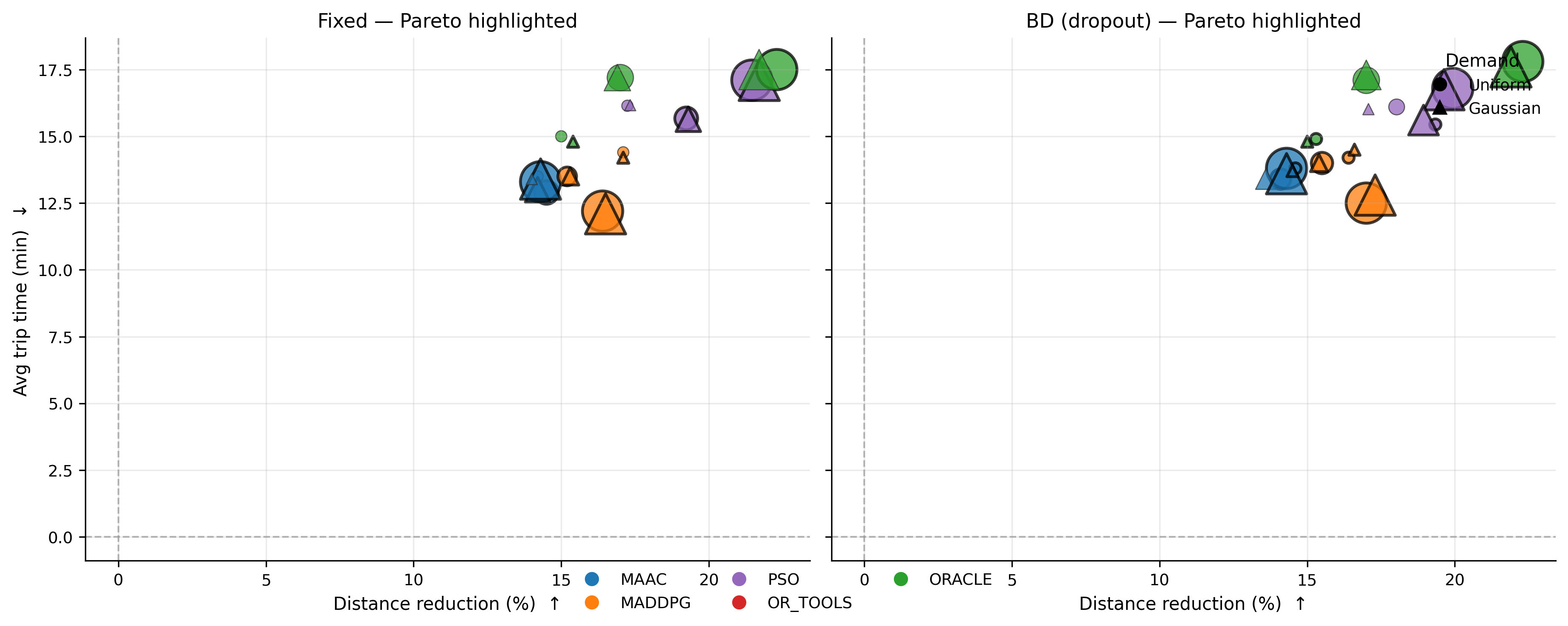}}
\caption{Pareto frontier illustrating trade-offs between efficiency improvements and equity outcomes across system configurations.}
\end{figure}

% --- Correlation analyses ---
% -------- MAAC + MADDPG --------
\begin{figure}[H]
\centering

\begin{subfigure}{0.48\textwidth}
\centering
\includegraphics[width=\linewidth]{{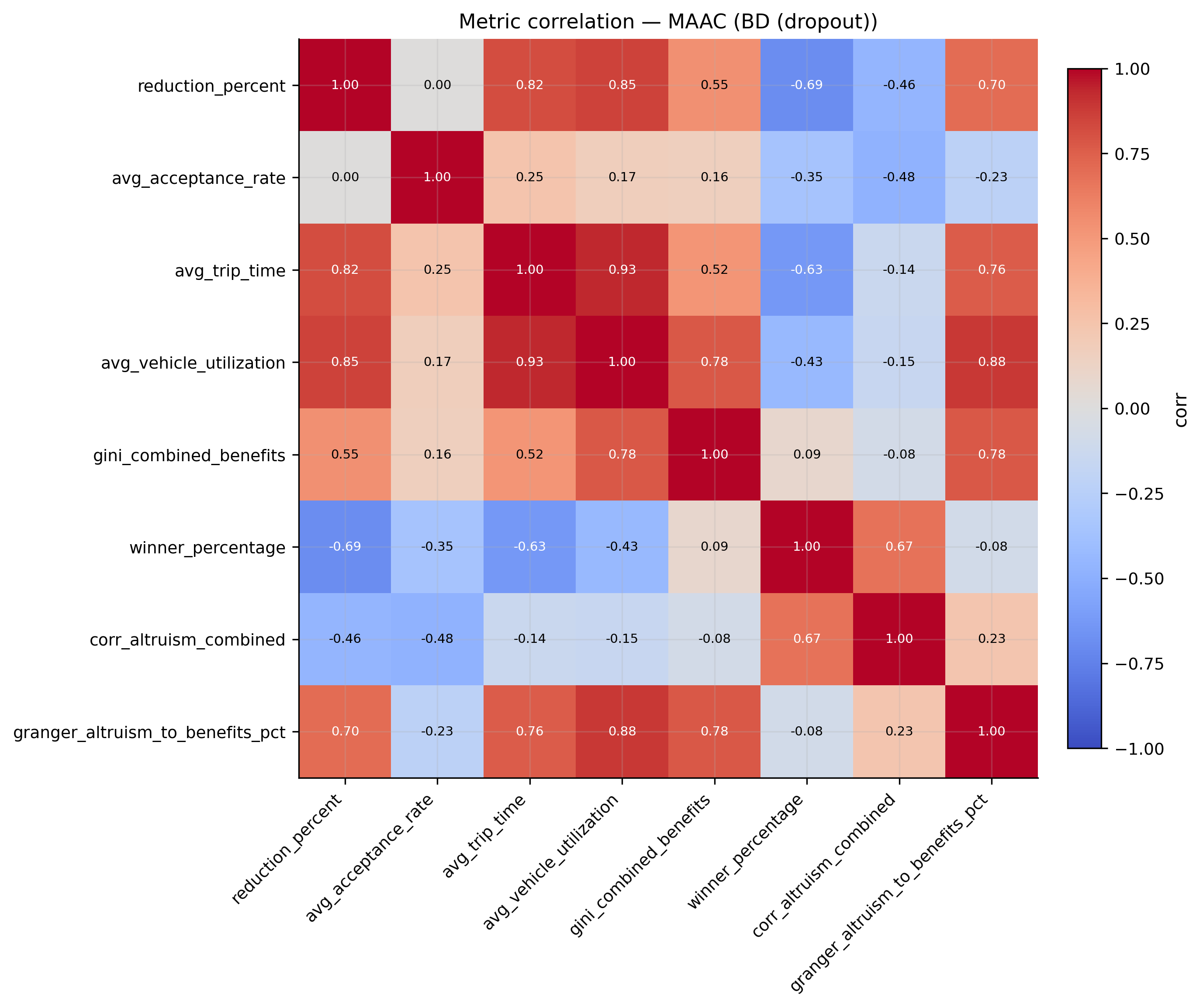}}
\caption{MAAC — agents may enter and exit (Birth–Death)}
\end{subfigure}
\hfill
\begin{subfigure}{0.48\textwidth}
\centering
\includegraphics[width=\linewidth]{{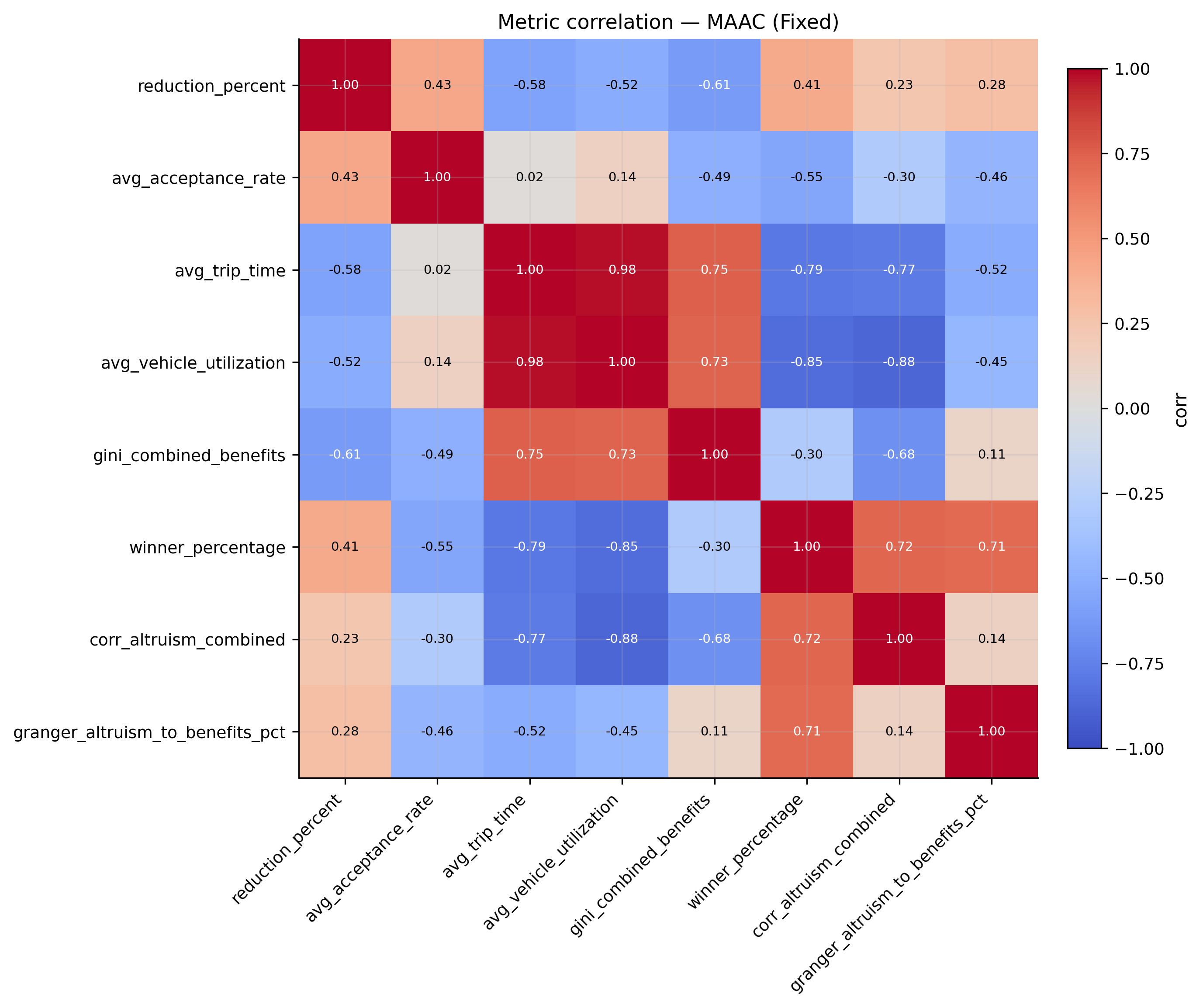}}
\caption{MAAC — no entry and exit of agents}
\end{subfigure}

\vspace{0.5em}

\begin{subfigure}{0.48\textwidth}
\centering
\includegraphics[width=\linewidth]{{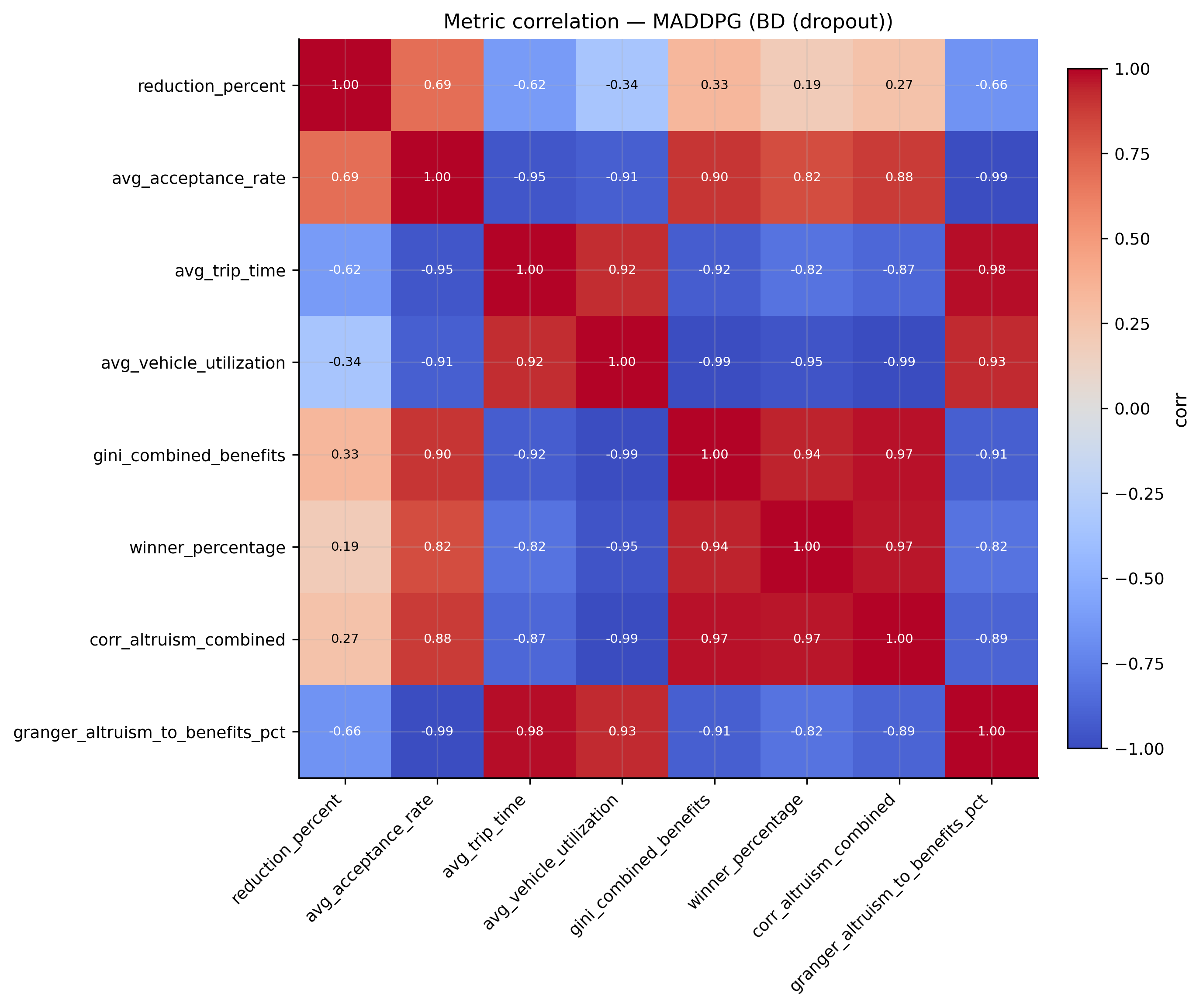}}
\caption{MADDPG — agents may enter and exit (Birth–Death)}
\end{subfigure}
\hfill
\begin{subfigure}{0.48\textwidth}
\centering
\includegraphics[width=\linewidth]{{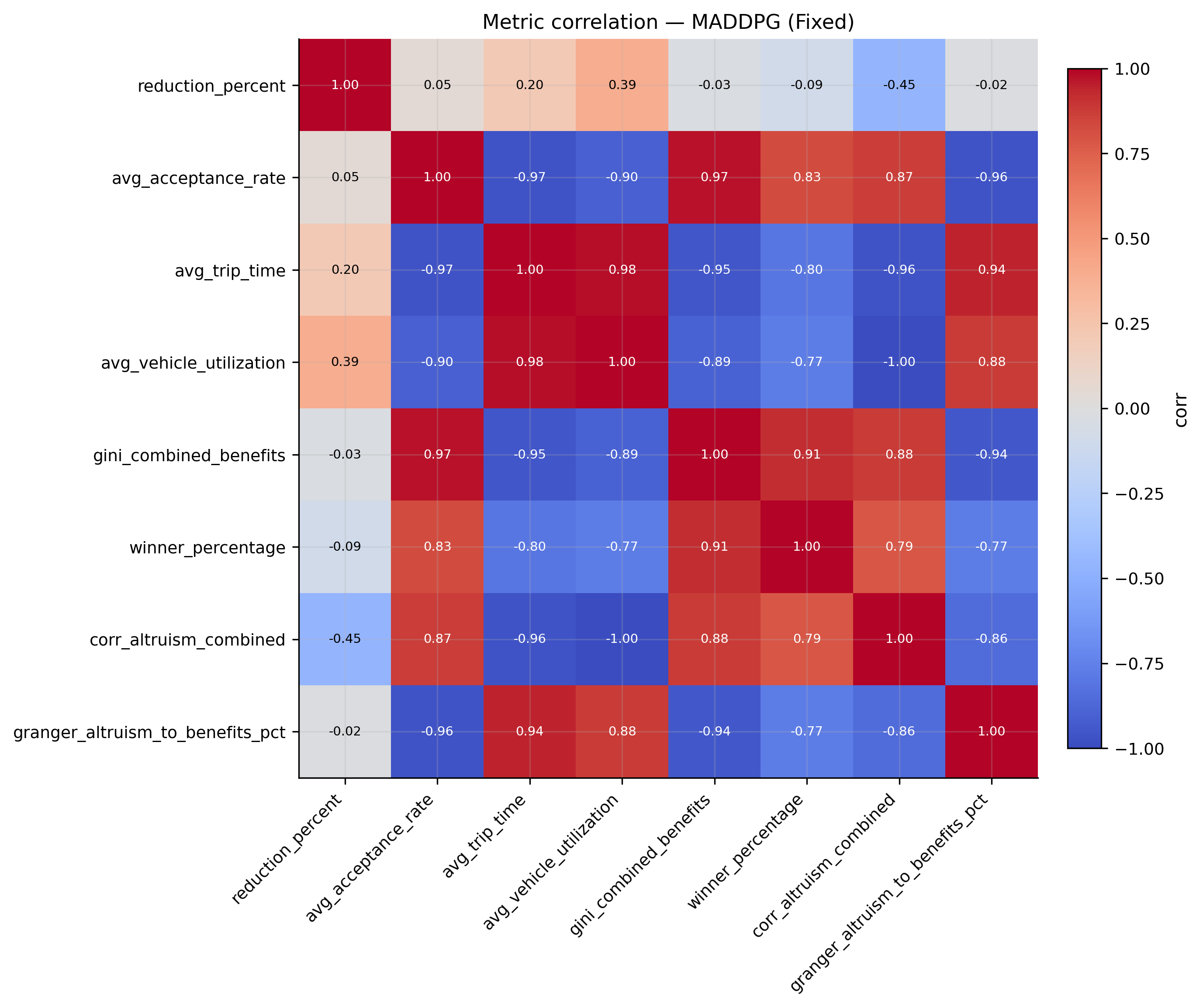}}
\caption{MADDPG — no entry and exit of agents}
\end{subfigure}

\caption{Correlation matrices of key system metrics for multi-agent reinforcement learning approaches under different population dynamics.}
\end{figure}

\begin{figure}[H]
\centering
\includegraphics[width=0.3\linewidth]{{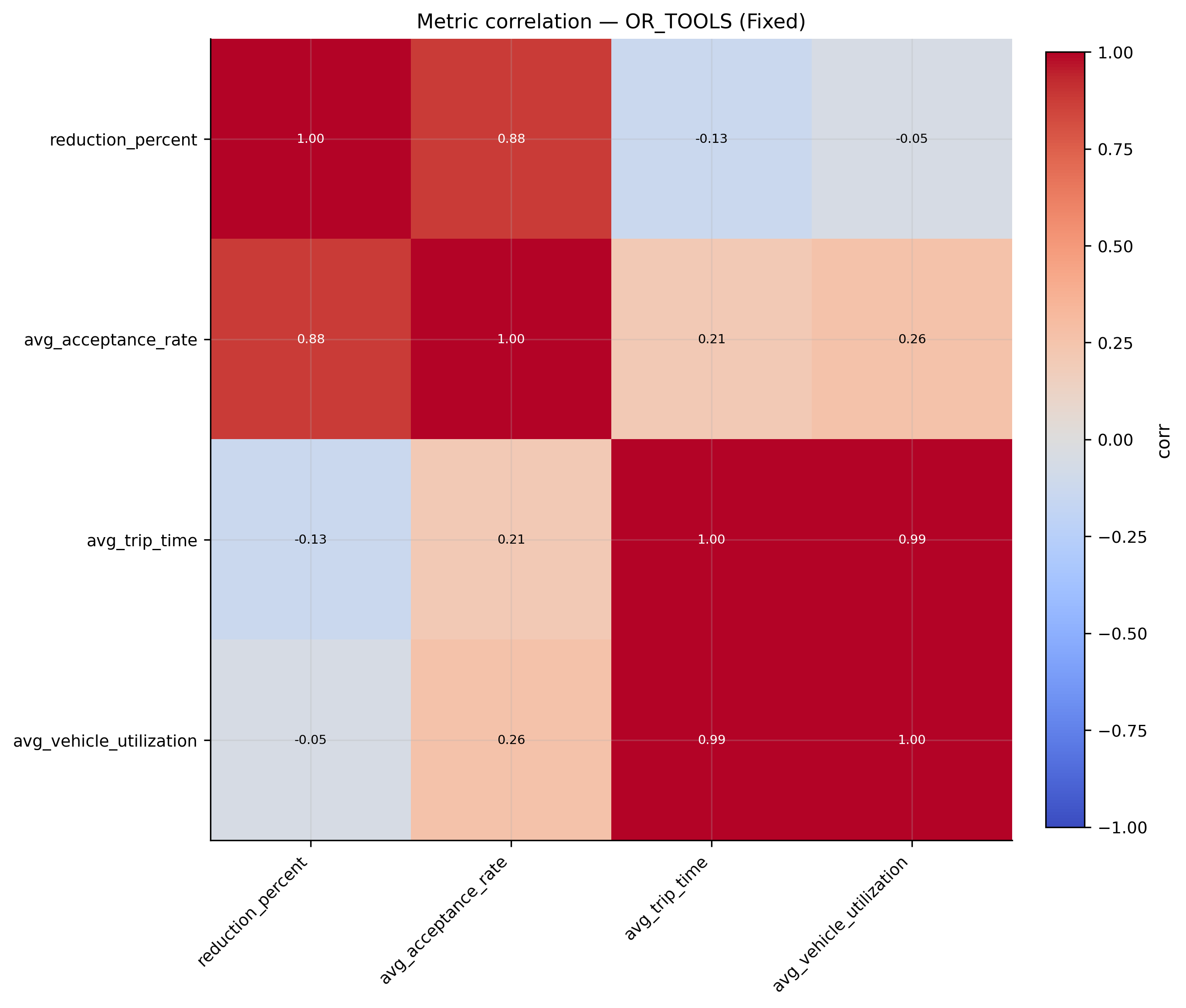}}
\caption{Correlation matrix of key performance metrics for the OR-Tools baseline with fixed agents.}
\end{figure}

% -------- Baselines --------
\begin{figure}[H]
\centering

\begin{subfigure}{0.48\textwidth}
\centering
\includegraphics[width=\linewidth]{{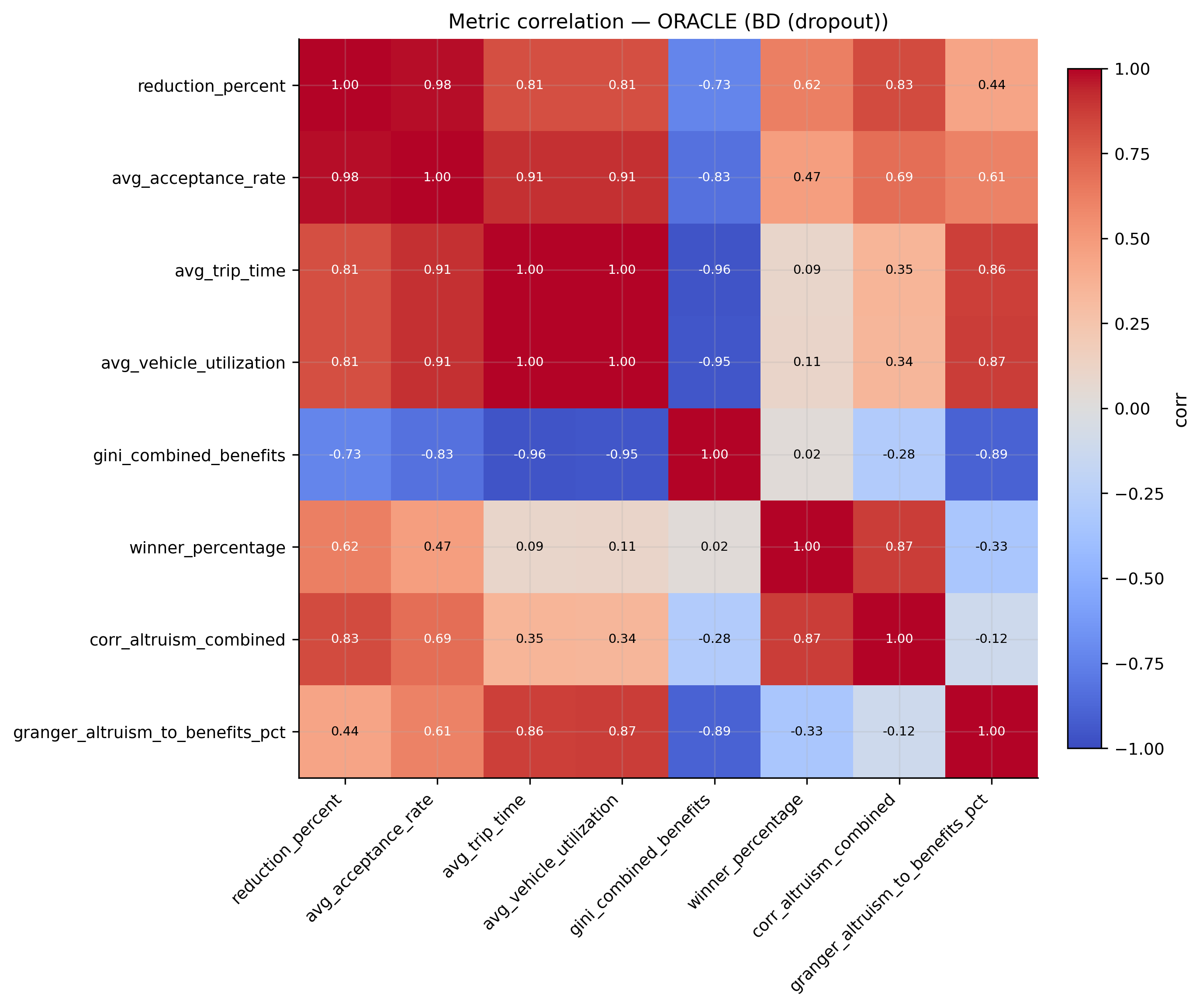}}
\caption{Oracle baseline — agents may enter and exit (Birth–Death)}
\end{subfigure}
\hfill
\begin{subfigure}{0.48\textwidth}
\centering
\includegraphics[width=\linewidth]{{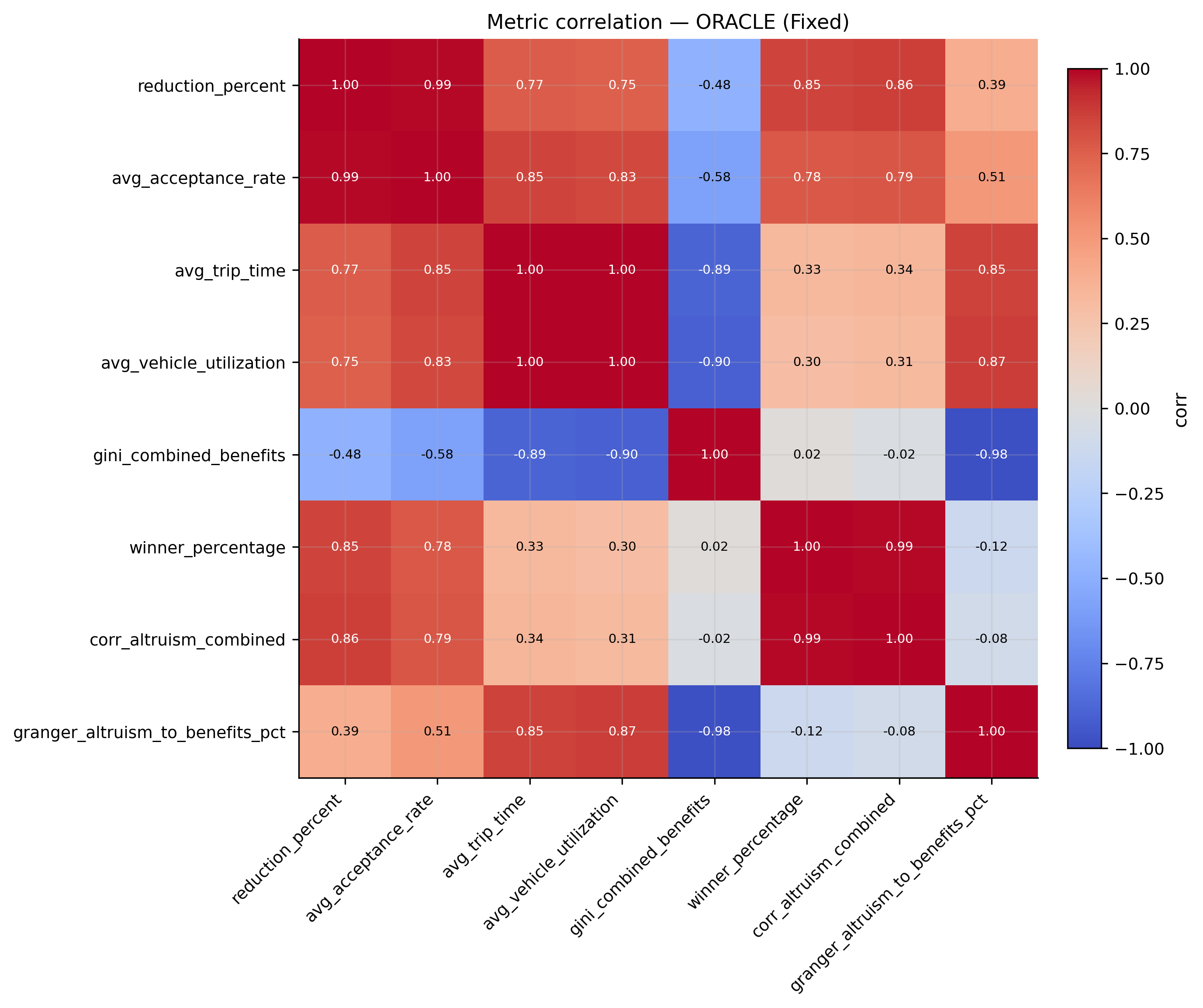}}
\caption{Oracle baseline — no entry and exit of agents}
\end{subfigure}

\vspace{0.5em}

\begin{subfigure}{0.48\textwidth}
\centering
\includegraphics[width=\linewidth]{{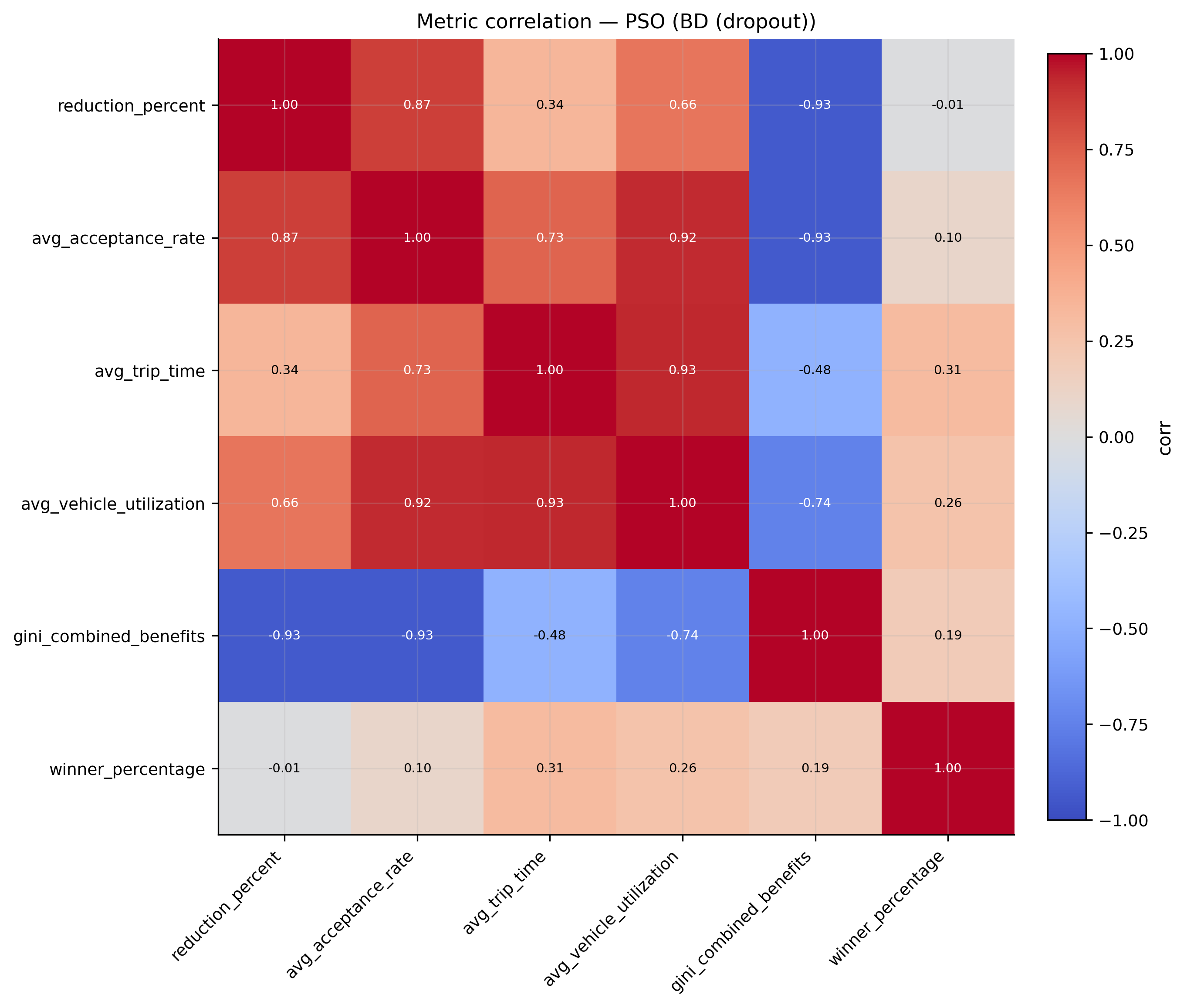}}
\caption{PSO baseline — agents may enter and exit (Birth–Death)}
\end{subfigure}
\hfill
\begin{subfigure}{0.48\textwidth}
\centering
\includegraphics[width=\linewidth]{{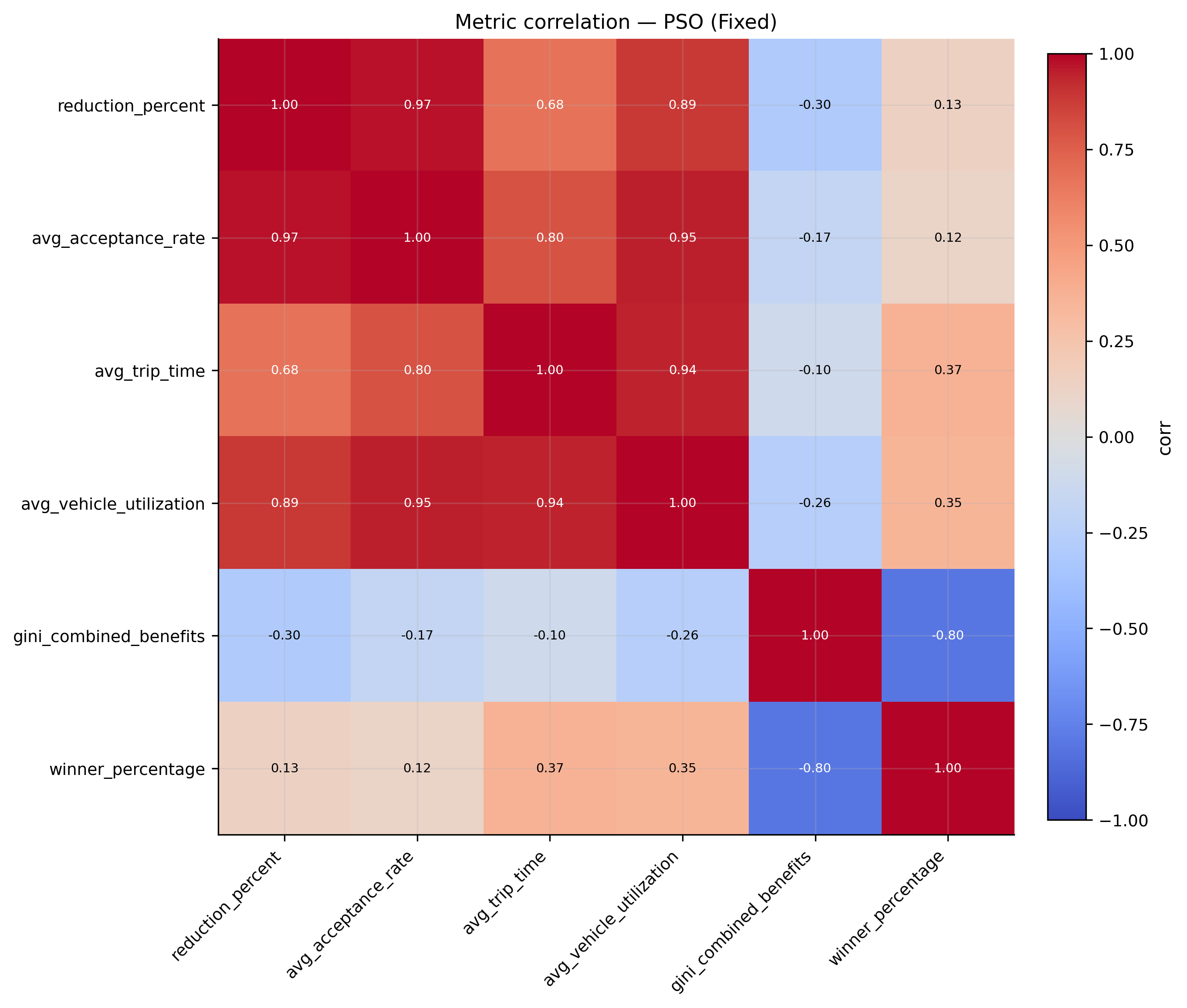}}
\caption{PSO baseline — no entry and exit of agents}
\end{subfigure}

\caption{Correlation matrices for optimization-based baselines comparing metric relationships under fixed populations and Birth–Death agent dynamics.}
\end{figure}

% --- Distribution plots ---
\begin{figure}[H]
\centering
\includegraphics[width=0.8\linewidth]{{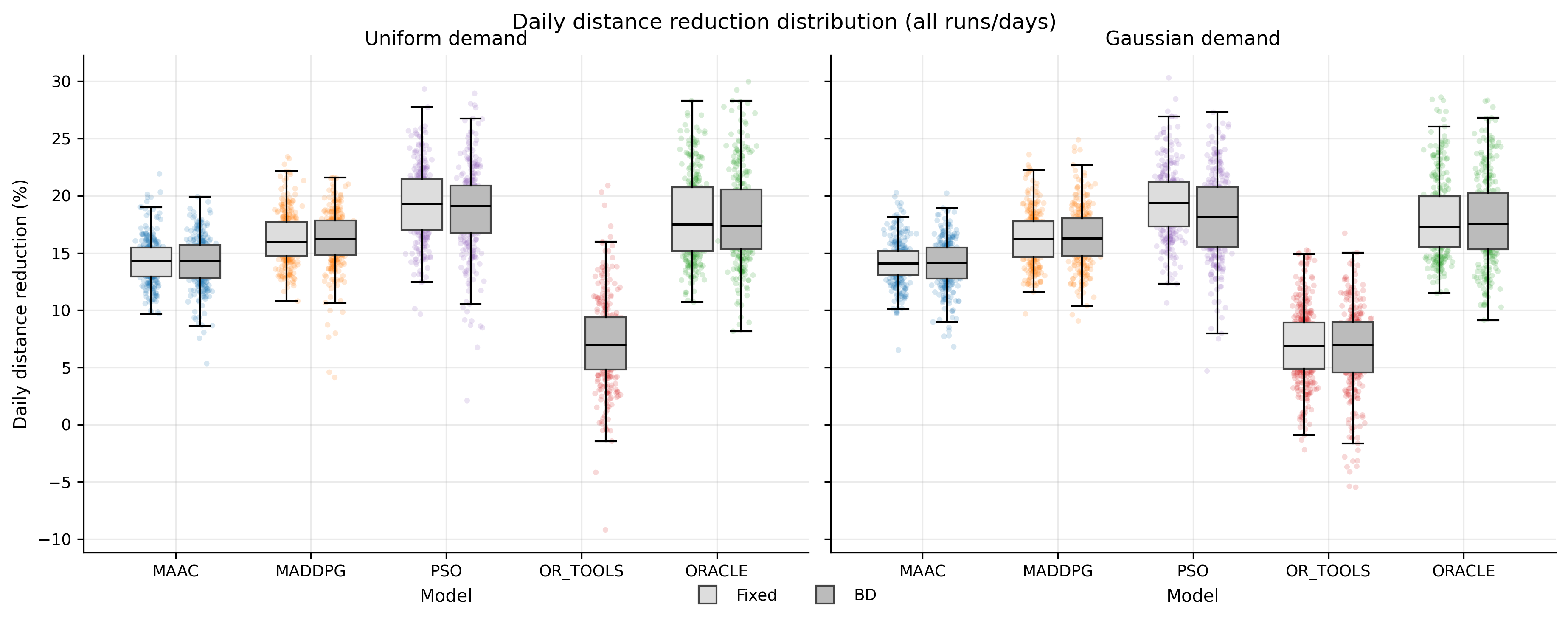}}
\caption{Distribution of daily distance reduction across experimental runs.}
\end{figure}

\begin{figure}[H]
\centering
\includegraphics[width=0.8\linewidth]{{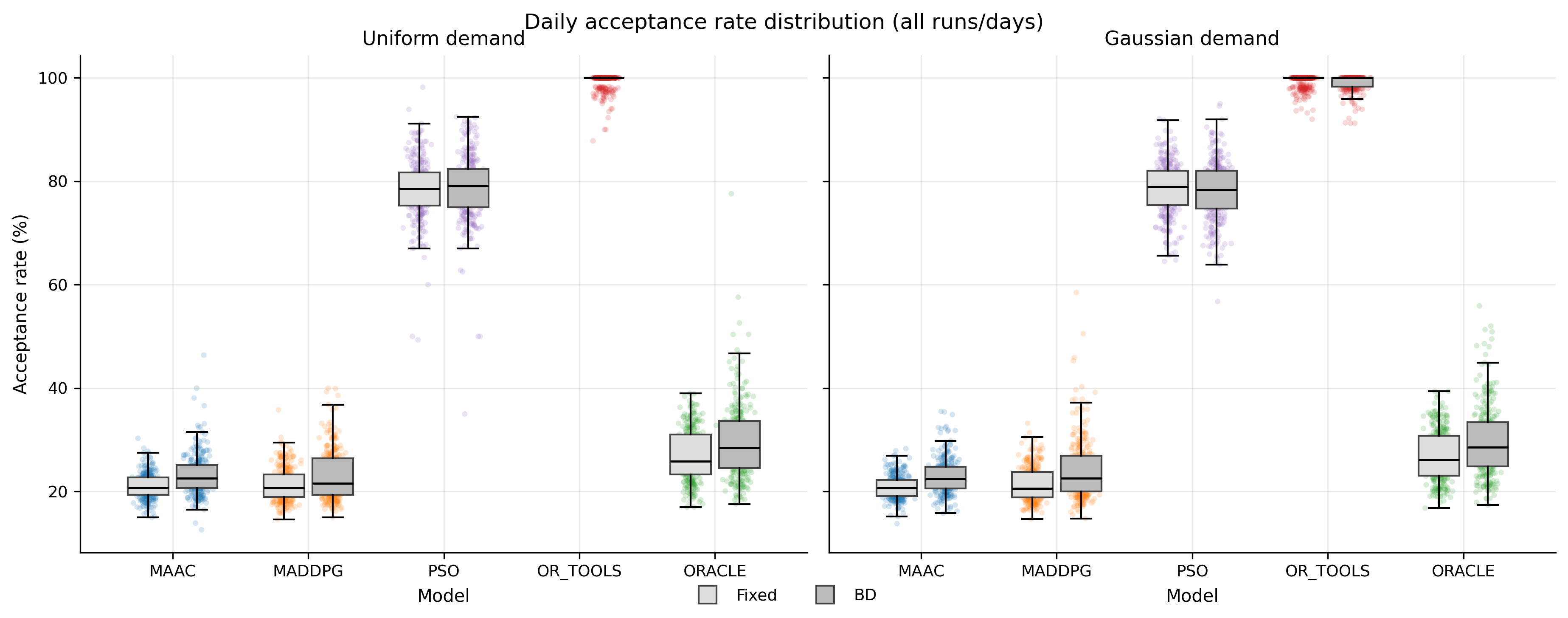}}
\caption{Distribution of ride acceptance rates across system configurations.}
\end{figure}

% --- Benefit distribution ---
\begin{figure}[H]
\centering
\includegraphics[width=0.8\linewidth]{{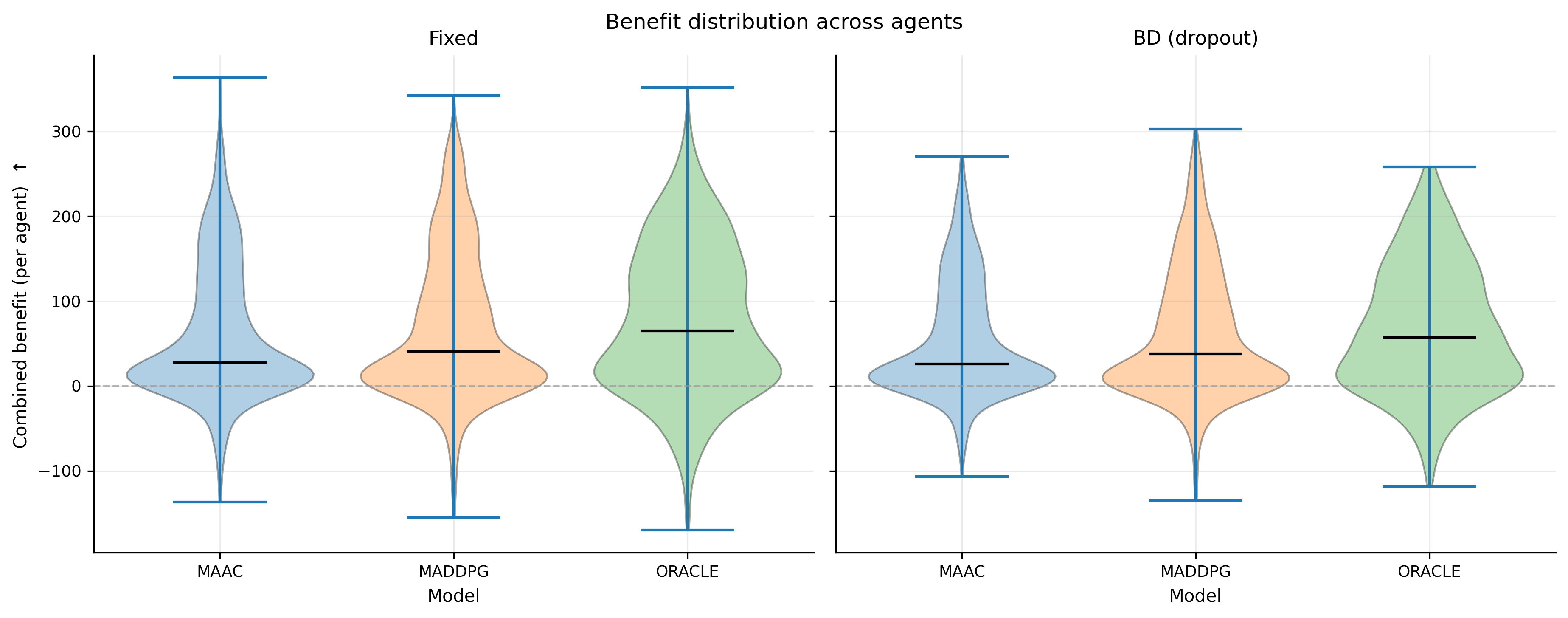}}
\caption{Violin plot showing distribution of agent-level benefits under different altruism allocation mechanisms.}
\end{figure}

\begin{figure}[H]
\centering
\includegraphics[width=0.8\linewidth]{{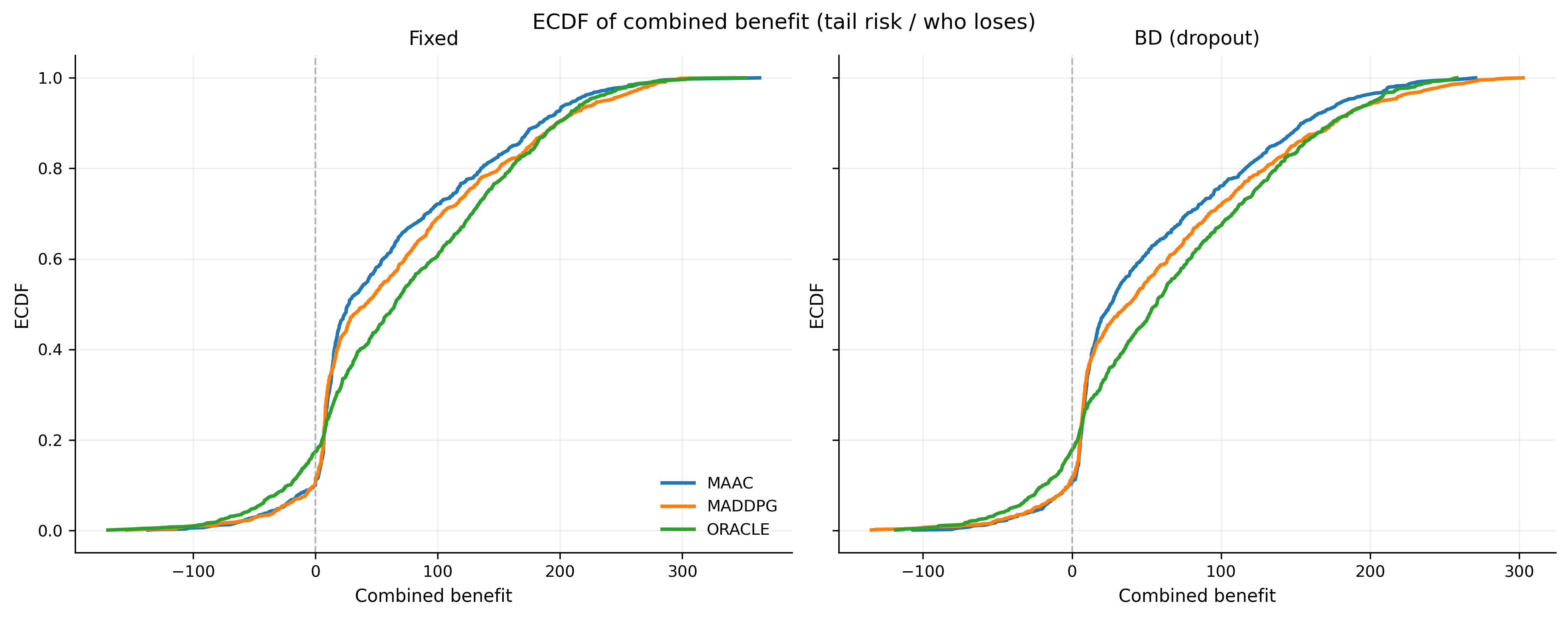}}
\caption{Empirical cumulative distribution of agent-level benefits across all experimental configurations.}
\end{figure}

\begin{figure}[H]
\centering
\includegraphics[width=0.8\linewidth]{{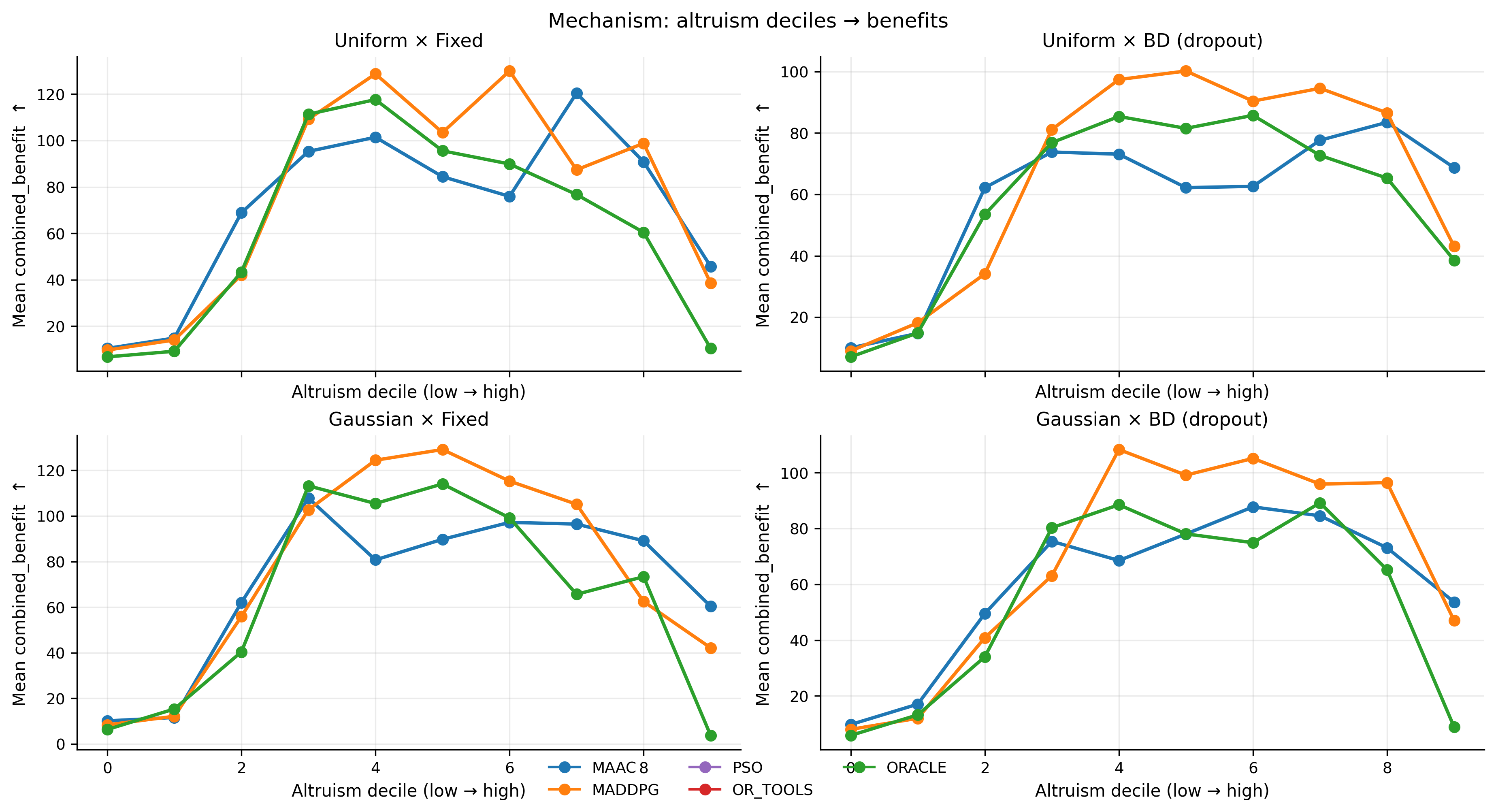}}
\caption{Benefit outcomes across altruism deciles illustrating how system gains vary with agent altruism levels.}
\end{figure}

% --- Main summary figures ---
\begin{figure}[H]
\centering
\includegraphics[width=0.8\linewidth]{{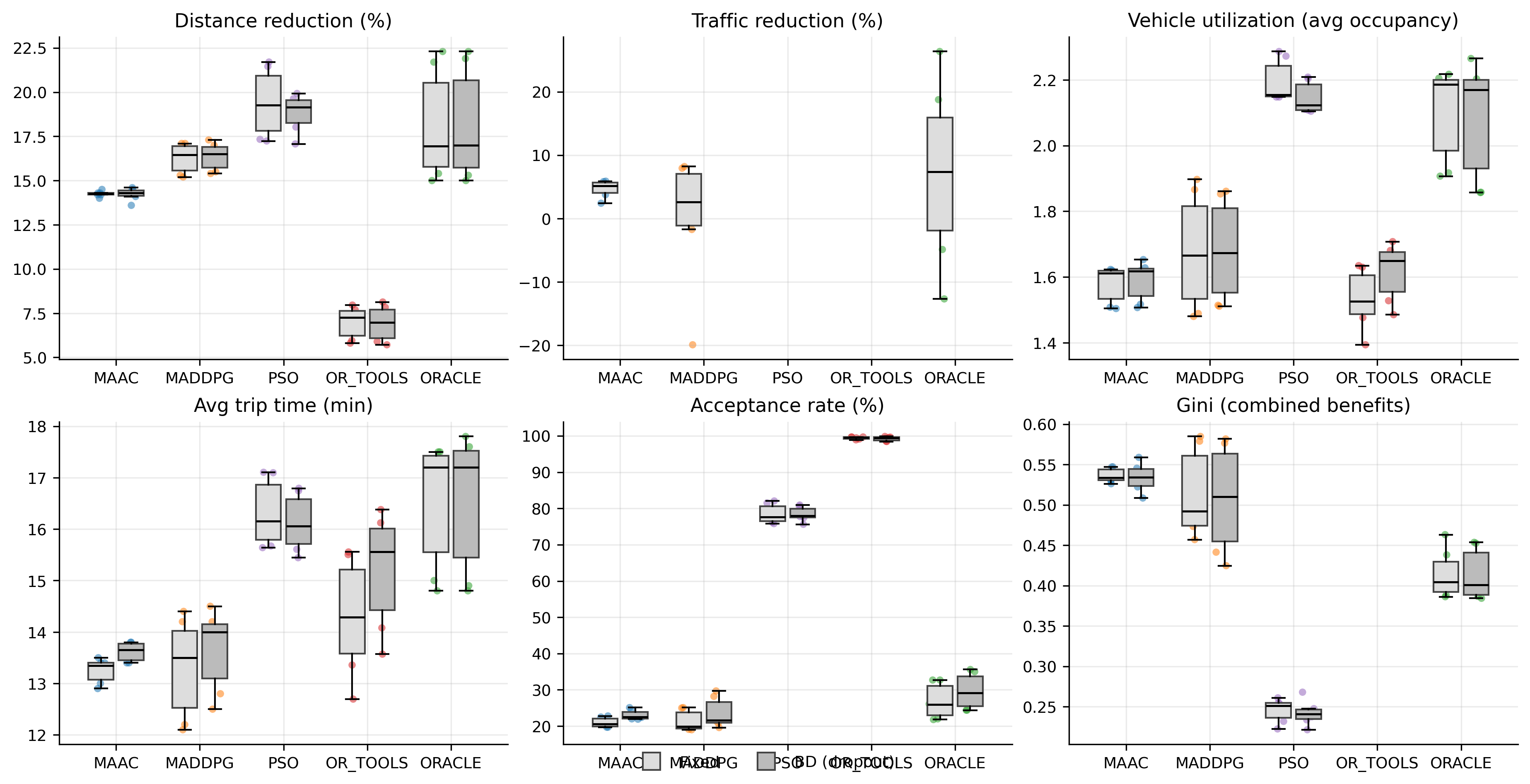}}
\caption{Boxplots summarizing the distribution of core system performance metrics across all experimental scenarios.}
\end{figure}

\begin{figure}[H]
\centering
\includegraphics[width=0.8\linewidth]{{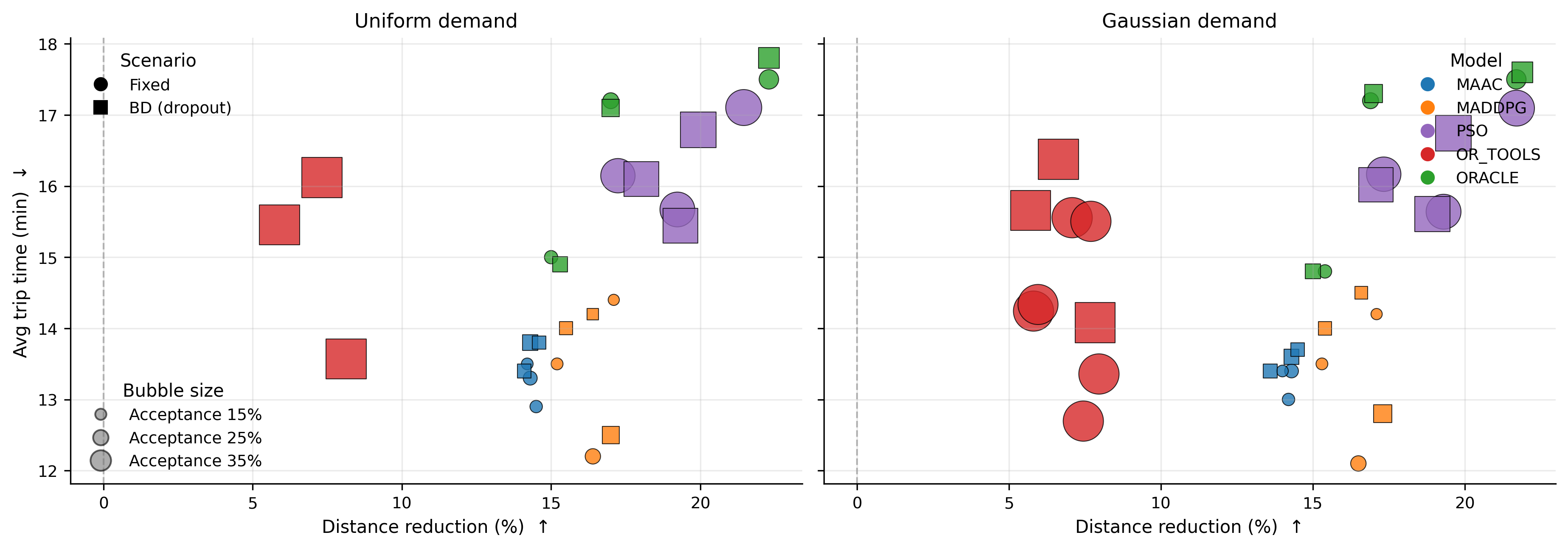}}
\caption{Trade-off analysis illustrating relationships between efficiency improvements and equity outcomes.}
\end{figure}

\begin{figure}[H]
\centering
\includegraphics[width=0.8\linewidth]{{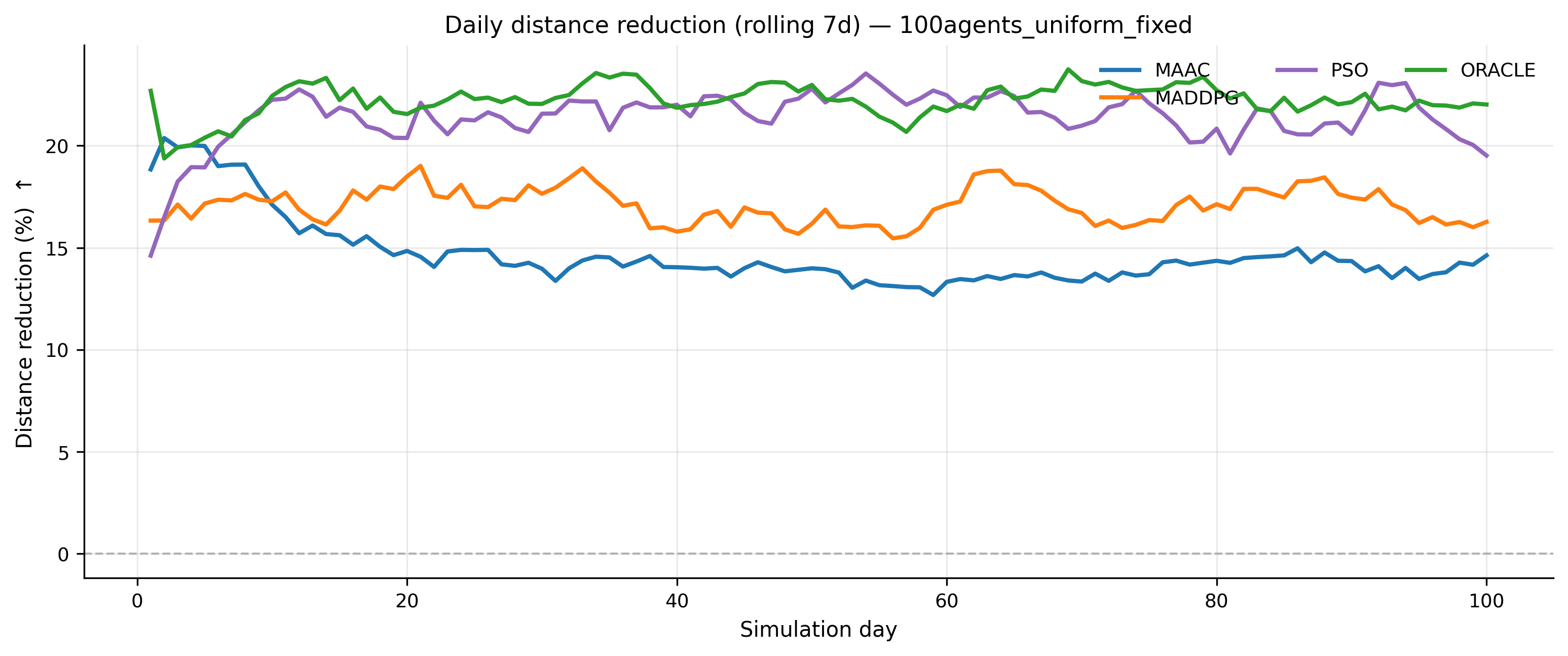}}
\caption{Daily distance reduction over time for the 100-agent scenario with uniform altruism allocation and no entry or exit of agents.}
\end{figure}

\begin{figure}[H]
\centering
\includegraphics[width=0.8\linewidth]{{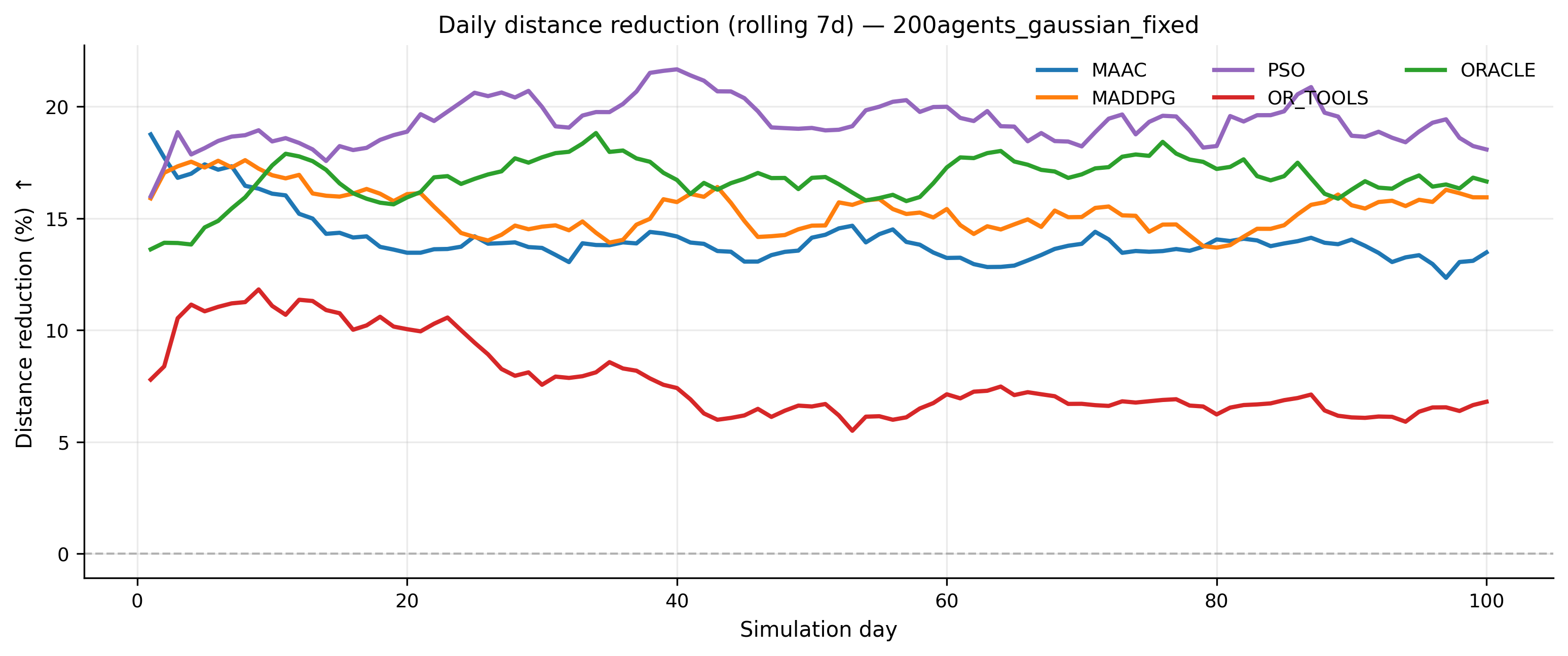}}
\caption{Daily distance reduction over time for the 200-agent scenario with Gaussian altruism allocation and no entry or exit of agents.}
\end{figure}

\end{document}